%% file: Reddit_no_template.tex
\documentclass[12pt]{article} 
\usepackage[right=1.0in, left=1.0in, top=1.0in, bottom=1.0in]{geometry}
\usepackage{setspace}
\usepackage{multirow}
\usepackage{graphicx}
\usepackage{tabto}
\usepackage{ragged2e}
\usepackage{booktabs}
\usepackage{array}
\usepackage{tabularx}
\usepackage{amsthm}
\usepackage{soul}
\usepackage{amsmath}
\usepackage{fancyhdr}
\usepackage{amssymb}
\usepackage{mdframed}
\usepackage{mathrsfs}  
\usepackage[utf8]{inputenc}
\usepackage{longtable}
\usepackage{lscape}
\usepackage{pdflscape}
\usepackage{listings}
\usepackage{lastpage}
\usepackage{bbm}
\usepackage{subcaption}
\usepackage{natbib}
\usepackage{microtype}
\usepackage[scientific-notation=true]{siunitx}
\usepackage{tikz}
\usetikzlibrary{arrows.meta, positioning, calc}
\usepackage{pgfplots}
\pgfplotsset{compat=1.16}                                                 
\usepackage{caption}                                                                  

\usepackage{makecell}

\usepackage{titling}
\usepackage{titlesec}
\titlespacing\section{0pt}{12pt plus 2pt minus 2pt}{0pt plus 2pt minus 2pt}
\titlespacing\subsection{0pt}{4pt plus 1pt minus 1pt}{0pt plus 2pt minus 2pt}

\titleformat{\section}{\large\bfseries\upshape}{\thesection}{1em}{}
\titleformat{\subsection}{\normalsize\bfseries\upshape}{\thesubsection}{1em}{}
\usepackage{enumitem} 
\setlist{nosep} 
\usepackage[hyphens]{url}

\usepackage[colorlinks=true,allcolors=blue]{hyperref}%

\renewenvironment{abstract}
 {\small
  \begin{center}
  \bfseries \abstractname\vspace{-.5em}\vspace{0pt}
  \end{center}
  \list{}{
    \setlength{\leftmargin}{.7cm}%
    \setlength{\rightmargin}{\leftmargin}%
  }%
  \item\relax}
 {\endlist}

\theoremstyle{definition}

\newcommand{\E}{{\rm I\kern-.3em E}}

\usepackage{etoolbox}
\usepackage{adjustbox}

\newcommand{\notestyle}[1]{%
  \par\vspace{0.4em}%
  {\singlespacing\footnotesize\justifying\noindent #1\par}}
\newcommand{\fignote}[1]{\notestyle{\textit{Notes:} #1}}

\providecommand{\HD}[2]{\vrule height #1pt depth #2pt width 0pt\relax}

\providecommand{\up}{\HD{13}{0}}
\providecommand{\down}{\HD{0}{5}}

\newsavebox{\informsTbox}
\newlength{\informsTwd}
\long\def\TABLE#1#2#3{%
  \sbox\informsTbox{#2}%
  \setlength{\informsTwd}{\wd\informsTbox}%
  \ifdim\informsTwd>\textwidth \setlength{\informsTwd}{\textwidth}\fi
  \centering
  \begin{minipage}{\informsTwd}%
    \caption{#1}%
    \par\usebox{\informsTbox}%
    \ifblank{#3}{}{\notestyle{#3}}%
  \end{minipage}}

\long\def\FIGURE#1#2#3{%
  \centering
  #1\par
  \caption{#2}%
  \ifblank{#3}{}{\fignote{#3}}}

\newenvironment{APPENDICES}{%
  \par\appendix
  \titleformat{\section}{\large\bfseries\upshape}{Appendix \thesection}{1em}{}%
  \titleformat{\subsection}{\normalsize\bfseries\upshape}{\thesubsection}{1em}{}%
}{\par}

\title{ \vspace{-2.3cm} \singlespacing Large Language Models Polarize Ideologically but Moderate Affectively in Online Political Discourse}

\author{%
\vspace*{0.5cm}
\parbox{\textwidth}{\centering
\makebox[0.30\textwidth][c]{%
Gavin Wang\thanks{Daniels School of Business, Purdue University, email: Xiaoning.Wang@purdue.edu}}%
\makebox[0.30\textwidth][c]{%
Srinaath Anbudurai\textsuperscript{\#}\thanks{HEC Paris, email: srinaath.anbudurai@hec.edu}}%
\makebox[0.30\textwidth][c]{%
Oliver Sun\textsuperscript{\#}\thanks{The Wharton School, University of Pennsylvania, email: sun24@wharton.upenn.edu}}%
\\[1em]
\makebox[0.30\textwidth][c]{%
Xitong Li\thanks{HEC Paris, email: lix@hec.fr}}%
\makebox[0.30\textwidth][c]{%
Lynn Wu\thanks{The Wharton School, University of Pennsylvania, email: wulynn@wharton.upenn.edu}}%
\\[0.6em]
{\footnotesize \textsuperscript{\#}\,Equal~contribution.}
}}

\vspace{-.75cm}

\date{\today}

\begin{document}

\maketitle

\vspace{-.75cm}

\thispagestyle{empty}
\begin{abstract}
The emergence of large language models (LLMs) is reshaping how people engage in political discourse online. We examine how the release of ChatGPT altered ideological and emotional patterns in Reddit's largest political forum. Analysis of millions of comments shows that ChatGPT intensified ideological polarization: liberal-leaning authors posted increasingly liberal comments, while conservative-leaning authors posted increasingly conservative comments. Multiple falsification tests suggest that these findings are unlikely to be driven by contemporaneous events, such as the 2022 U.S. midterm elections, or by broader platform-wide trends in political polarization. Mechanism tests show that this shift does not stem from the creation of more persuasive or ideologically extreme original content using LLM. Instead, it originates from the tendency of LLM-assisted comments to echo and reinforce the original post's viewpoint, a pattern consistent with algorithmic sycophancy. Yet, despite growing ideological divides, affective polarization, measured by hostility and toxicity, declined. These findings reveal that LLMs can simultaneously deepen ideological separation and foster more civil exchanges, challenging the long-standing assumption in literature that extremity and incivility necessarily move together.

\vspace{.5cm}
\noindent\textbf{Keywords:} Large Language Models, ChatGPT, Political Polarization, Liberal, Conservative, Affective Polarization
\vspace{-1cm}
\singlespacing
\noindent
\end{abstract}
\clearpage
\onehalfspacing

\section{Introduction}\label{sec:Intro}
Incivility has long been understood to co-evolve with ideological polarization. A substantial body of work argues that rising ideological polarization is systematically accompanied by rising incivility~\citep{chen2018effect,santoro2022promise,zimmerman2022political,boxell2024cross}. While this literature establishes a strong association in human-authored discourse, its premises become uncertain in the AI era, where much of the information environment is increasingly mediated by large language models (LLMs). This shift motivates a re-examination of the ideological polarization-incivility nexus in the context of AI.

This study examines how the release of ChatGPT impacted political discourse on social media platforms. The advent of LLMs, particularly OpenAI's ChatGPT, has transformed how individuals access information and create content. These models, developed through extensive training on diverse internet content, can generate human-like text, answer queries, and assist in a variety of tasks~\citep{noy2023experimental}. While their applications span domains such as education, healthcare, and customer service~\citep{bastani2025generative,brynjolfsson2025generative,zoller2025human}, their influence on political discourse remains comparatively underexplored. This omission is critically important because LLMs are not neutral conduits of information: prior studies suggest that ChatGPT exhibits a ``pro-environmental, left-libertarian orientation''~\citep{rozado2023political,rutinowski2024self}, and, in the U.S. context, a tendency to favor Democratic positions~\citep{motoki2024more}. Such biases may shape not only the content users generate using these tools, but also the ideological direction and tone of politically charged online discussions.

These concerns are particularly consequential on social media platforms, which have become central arenas for political discussion. Social media not only reflects public opinion but also actively shapes civic engagement and patterns of political polarization~\citep{waller2021quantifying,wojcieszak2022most,mosleh2024differences}. These platforms enable rapid information dissemination, foster echo chambers, and facilitate interactions among users with varying ideological leanings~\citep{bail2018exposure,flamino2023political,mamakos2023social,nyhan2023like}. As political conversations increasingly migrate online, the nature and quality of discourse in these spaces can affect democratic legitimacy, shape policy debates, and even influence electoral outcomes ~\citep{grinberg2019fake,siev2024ambivalent}. By lowering the cost of producing political content and shaping the form of online interactions, LLMs may alter the ideological composition and tone of political discourse on social media. Understanding how LLMs affect political discourse on social media is thus crucial for assessing their broader implications for democratic governance and societal cohesion.

We focus on the political discourse on Reddit which is one of the largest and most ideologically diverse social media platforms, and has been broadly examined in prior research on political polarization~\citep{waller2021quantifying,mamakos2023social}. We assemble large-scale, detailed comment-level data from Reddit's most active political forum, including detailed post and comment content, timestamps, and author information. Leveraging this data, we assess changes in Reddit users' political expressions before and after the introduction of ChatGPT. Following the practice of previous literature~\citep{gentzkow2010drives,greenstein2012wikipedia,greenstein2016open}, we measure the political slant of each comment on a continuous scale by mapping the frequency of words and phrases to their ideological usage in congressional speeches, and calculate each author's political stance at different times. As a contribution, rather than relying on the 2005 record as in prior work, we construct our dictionary from the 2021 Congress Record to more accurately capture contemporary political language.

We also employ several state-of-the-art indicators that have been widely used in prior research to assess the likelihood that a Reddit comment was generated with the assistance of LLMs, including comment length, passive verbs, perplexity score and human score~\citep{radford2019rewon,solaiman2019release,simchon2023computational,munoz2024contrasting}. Because direct information on whether a particular Reddit comment was generated using LLMs is unavailable, our analysis relies on established linguistic and model-based indicators that estimate the probability of LLM involvement in content creation rather than definitively identifying AI-generated content. Although these methods cannot determine with complete certainty whether any individual comment was generated with LLM assistance, they have been shown to capture meaningful variation in LLM involvement and are well-suited for large-scale population-level inference~\citep{shan2026examining}. In a corpus comprising millions of comments, idiosyncratic classification errors are unlikely to systematically generate the aggregate patterns we document.

Our empirical results indicate that the release of ChatGPT significantly intensified political polarization on the platform: comments posted by previously liberal-leaning users shifted further toward the liberal end of the ideological spectrum, while comments posted by previously conservative-leaning users shifted further toward the conservative end. The magnitude was substantively meaningful: the impact of ChatGPT's release on average was equivalent to shifting a median liberal (conservative) author into the top 30\% most extreme segment of their respective ideological spectra. We also investigate how the release of ChatGPT affected authors near the ideological center and those at the margins of the liberal--conservative boundary. We find a systematic leftward shift across these groups, though not substantial in magnitude. Multiple falsification tests suggest that our findings are unlikely to be driven by contemporaneous events, such as the 2022 U.S. midterm elections, or by broader platform-wide trends in political polarization.

Despite this amplification of ideological alignment within discussions, we observe a clear decline in affective polarization~\citep{iyengar2012affect,boxell2024cross}---the extent to which individuals express hostility or animosity toward opposing political groups. Measures of civility on the platform, captured through comment toxicity and hostility, improved markedly following ChatGPT’s release. Specifically, average toxicity declined by an amount equivalent to moving a median comment to approximately the bottom 20th percentile of the toxicity distribution, accompanied by a significant reduction in hostility. Besides hostility and toxicity, we also conduct a set of emotion analyses to examine whether the adoption of LLMs influences the emotional expression. Our results reveal no notable correlation between LLM usage and emotional content. Together, these findings indicate that while political discussions became more ideologically polarized, online political discourse became noticeably more civil.

As with many emerging technologies, the adoption and use of LLMs are unlikely to be uniform across users. We therefore examine whether the effects of ChatGPT's release vary with users' prior commenting activity, which recent research suggests is associated with differential reliance on generative AI tools~\citep{burtch2024consequences,del2024large}. We find that less active users exhibit substantially stronger linguistic signatures of LLM assistance following ChatGPT's release, including longer comments and lower perplexity and human scores, regardless of their political orientation. Importantly, the increase in ideological polarization is concentrated among these less active users, who are more likely to rely on LLMs for content generation. In contrast, more active users exhibit little comparable increase in ideological polarization but become more active in political discussions and post comments that are less toxic and hostile. These heterogeneous patterns suggest that the two effects of generative AI may operate through different channels: increased ideological polarization appears to be concentrated among users more likely to directly rely on LLMs, whereas improvements in civility extend more broadly across the discussion environment, consistent with a shift in prevailing conversational norms.

To investigate the conversational mechanisms underlying this divergence, we next examine how comments with LLM involvement respond to political content within and across partisan boundaries. We find that comments with a high estimated likelihood of LLM involvement (hereafter, \textit{LLM-assisted comments}) are substantially more likely to echo and reinforce the political stance of the original post than other comments, a pattern consistent with ``algorithmic sycophancy''---the tendency of LLMs to condition their responses on, and affirm, user inputs~\citep{abdurahman2024perils,naddaf2025ai,cheng2026sycophantic}. More specifically, relative to respondents' own baseline political orientations, LLM-assisted comments systematically shift toward the political stance of the content to which they respond. This tendency has asymmetric implications across partisan contexts: it strengthens ideological reinforcement in same-partisan interactions while moderating disagreement in cross-partisan interactions. Because the former effect is substantially stronger, the net effect is an increase in overall ideological polarization on the platform. However, the preceding analysis does not explain why political discussions simultaneously became more civil. To understand this second pattern, we turn to the reply-thread level and examine changes in civility across different partisan configurations. The results reveal consistent declines in toxicity and hostility across all reply-thread groups following the adoption of ChatGPT, regardless of perceived political alignment. Importantly, these declines persist even when we restrict the analysis to human-generated comments, suggesting that improvements in civility extend beyond direct LLM-assisted content.

Taken together, these findings suggest that the divergence between ideological and affective polarization reflects two complementary features of LLM-mediated political discourse. LLM-assisted responses tend to affirm and reinforce the position of the content to which they reply, thereby strengthening ideological alignment within discussions. At the same time, LLM-assisted responses are typically expressed in a more measured and less confrontational manner, and this improvement in civility appears to extend beyond LLM-assisted content to the broader discussion environment. Consequently, existing viewpoints can be reinforced without a corresponding increase in hostility, allowing ideological and affective polarization to move in opposite directions.

These findings contribute to the literature on political polarization by identifying an important boundary condition to the long-standing association between ideological extremity and incivility~\citep{chen2018effect,santoro2022promise,zimmerman2022political,boxell2024cross}. Whereas prior research has largely documented the co-evolution of ideological and affective polarization in human discourse, our results demonstrate that the two can diverge in environments increasingly shaped by generative AI. From a policy perspective, this suggests a more nuanced role for generative AI in public political discourse. While tools such as ChatGPT may deepen ideological divisions by reinforcing existing viewpoints, they may simultaneously promote more civil and less hostile interactions~\citep{voelkel2024megastudy}. More broadly, our findings highlight that generative AI influences not only what political opinions are expressed online, but also how those opinions are communicated and contested. These insights also suggest a promising direction for improving the quality of online political discourse: interventions that encourage respectful engagement without suppressing substantive disagreement may help reduce hostility while preserving ideological diversity.

\section{Theory and Hypotheses}\label{sec:hypothesis development}

\medskip
\subsection{Social Media and Political Discourse}
Social media platforms have become central arenas for political discussion and opinion formation. Unlike traditional media, where information flows primarily from professional content producers to passive audiences, social media enables users to simultaneously consume, create, and disseminate political content. As a result, political discourse is increasingly shaped by interactions among millions of individuals whose opinions, reactions, and contributions collectively determine the information environment. Extant research has shown that social media not only reflects existing political preferences but also actively influences civic engagement, political participation, and ideological polarization~\citep{bail2018exposure,waller2021quantifying,wojcieszak2022most, mosleh2024differences}.

Prior research has identified several mechanisms through which online platforms shape political discourse. First, social media substantially lowers the cost of political expression, allowing individuals to share opinions, respond to others, and participate in public discussions at unprecedented scale~\citep{bakshy2015exposure,mosleh2024differences}. Second, platform algorithms selectively expose users to particular content, potentially reinforcing existing beliefs and facilitating the formation of ideological echo chambers~\citep{bail2018exposure, cinelli2021echo, guess2023reshares, nyhan2023like}. Third, the interactive nature of social media allows political opinions to evolve through repeated exchanges among users, making political discourse a dynamic process of social influence rather than a simple reflection of pre-existing preferences~\citep{bond201261,rathje2021out,waller2021quantifying}. Consequently, the structure of online interactions can have important implications for both the content and tone of political discussions. A growing body of research shows that online political discourse is increasingly characterized by both rising ideological polarization and rising incivility, suggesting that ideological disagreement and affective hostility often evolve together in digital environments~\citep{chen2018effect,santoro2022promise,zimmerman2022political,boxell2024cross}.

These questions have received considerable attention within the Information Systems literature. A prominent stream of research examines how digital platforms shape the production, dissemination, and evolution of user-generated political discourse. Studies of Wikipedia, for example, show that contributors’ ideological positions systematically influence the content they produce and that platform governance mechanisms can affect the ideological balance of collective knowledge production~\citep{greenstein2012wikipedia,greenstein2016open}. Moreover, researchers have increasingly recognized that technological artifacts can shape communication outcomes by altering the processes through which information is generated and exchanged. For example, recent studies have shown that recommendation systems, platform governance mechanisms, and collaborative production technologies influence not only information availability but also the formation of collective beliefs, social norms, and patterns of user interaction~\citep{rai2019editor,he2025platform,gu2026psychological}. Importantly, these recent studies largely focus on environments in which content is ultimately created by humans. Technology affects discourse indirectly by changing incentives, information visibility, or interaction opportunities, while the act of communication itself remains fundamentally human.

The emergence of LLMs fundamentally changes this assumption. Unlike prior digital technologies that primarily mediated communication among humans, LLMs can directly participate in the content-generation process itself~\citep{gao2026does,xue2026can}. In online political discussions, rather than merely influencing what information users encounter or whom they interact with, LLMs can shape how political arguments are formulated, how responses are generated, and how disagreements are expressed. In doing so, they introduce a qualitatively different mechanism into online political discourse: content is no longer produced solely by individuals expressing their own beliefs, but increasingly through interaction with LLMs, whose objectives may differ fundamentally from those of human communicators.
Consequently, the growing presence of LLM-assisted content raises an important theoretical question: how does political discourse change when a meaningful share of online discussions is shaped by the assistance of LLMs?

\subsection{Impact of LLMs on Political Discourse}

Unlike traditional digital technologies that primarily influence information exposure or interaction opportunities, LLMs directly participate in the generation of political discourse. As a result, they may alter not only what information is communicated but also the mechanisms through which political opinions are expressed and contested. We argue that a key distinction between human-generated and LLM-assisted political discourse lies in the objectives guiding content generation. Human communication is ultimately driven by humans' underlying beliefs, identities, and preferences, whereas LLMs are designed to produce responses that users perceive as helpful, coherent, and satisfactory. This distinction has important implications for political discourse because it changes how viewpoints are reinforced, challenged, and communicated. Consequently, the mechanisms that traditionally shape ideological polarization may operate differently when a growing share of online political discussions is mediated by LLMs.

\subsubsection{Increased Political Polarization}

Existing theories of political polarization emphasize the role of motivated reasoning, selective exposure, and identity-based cognition. Research shows that individuals tend to evaluate information through the lens of prior beliefs and often respond defensively to opposing viewpoints~\citep{lord1979biased,mutz2006hearing}. As a result, political discussions frequently involve disagreement, contestation, and attempts to persuade others. Although such interactions may reinforce existing beliefs, they also expose individuals to competing viewpoints that can partially constrain ideological divergence.

Beyond information-processing biases, political discourse is also shaped by social identity processes. Individuals often derive part of their self-concept from membership in political or ideological groups and therefore evaluate information not only in terms of its factual content but also in terms of its implications for group identity~\citep{festinger1962cognitive,turner1979social,rathje2021out}. Political discussions consequently serve not only as exchanges of information but also as opportunities to defend group boundaries and affirm social identities. Prior research shows that individuals frequently dismiss incongruent information as unreliable, question the credibility of opposing viewpoints, and become more entrenched in their prior positions when confronted with disconfirming evidence~\citep{fiorina2008political,bail2018exposure}. In this sense, ideological polarization in human-generated discourse often emerges through disagreement and confrontation.

LLMs differ from human communicators in an important respect. Rather than expressing independently held political beliefs, they generate responses by predicting text that is likely to maintain conversational coherence and satisfy users. Recent research documents that LLMs often exhibit ``algorithmic sycophancy'' — a tendency to affirm and reinforce users’ stated views even when those views are inaccurate or inconsistent with available evidence~\citep{sharma2023towards,cheng2026sycophantic}. This tendency emerges in part because reinforcement learning from human feedback (RLHF) rewards responses perceived as agreeable, helpful, and aligned with user expectations. As a result, behaviors associated with agreement and affirmation are systematically reinforced during model training~\citep{shapira2026rlhf}.

Unlike human communicators, LLMs do not possess stable political identities or intrinsic in-group and out-group loyalties. Their responses are primarily conditioned by the conversational context. Consequently, LLM-assisted responses tend to adapt to the position expressed in the content to which they reply, regardless of whether that position is liberal or conservative. In political discussions, this implies that LLMs reinforce the viewpoint of the interlocutor rather than advocating a consistent political position of their own.

When a user employs an LLM to generate or refine a political response, the resulting content is therefore more likely to affirm the ideological position embedded in the preceding discussion. In interactions among politically similar users, LLM-assisted responses amplify existing viewpoints rather than challenge them, thereby strengthening ideological alignment within the discussion. In contrast, when users engage across partisan lines, the same tendency toward affirmation may partially reduce ideological distance by accommodating aspects of the opposing viewpoint. Thus, LLM-assisted discourse may contain two competing forces: in-group reinforcement and out-group moderation.

Whether LLMs ultimately increase or decrease ideological polarization depends on the relative strength of these two forces. Although LLM-assisted responses may moderate some cross-partisan interactions, political discussions on social media are rarely random. Prior research consistently documents substantial ideological homophily and clustering, whereby users are considerably more likely to interact with politically similar others than with ideological opponents~\citep{bail2018exposure,cinelli2021echo}. Consequently, LLM-assisted responses are more frequently deployed in contexts where they reinforce existing viewpoints than in contexts where they bridge ideological divides. The cumulative effect of repeated within-group reinforcement is therefore likely to outweigh any moderation arising from cross-partisan interactions.

Taken together, LLMs alter the mechanism through which ideological polarization emerges. Whereas polarization in human discourse often arises through disagreement and contestation, polarization in LLM-assisted discourse emerges through affirmation and reinforcement. Although the underlying process differs, both mechanisms can generate greater ideological polarization at the aggregate level. Accordingly, we expect the introduction of ChatGPT to increase ideological polarization on social media platforms:

\medskip
\noindent \textbf{Hypothesis 1.} \textit{Ideological polarization in online political discourse increases following the launch of ChatGPT.}

\subsubsection{Reduced Affective Polarization}

Prior research generally views ideological polarization and affective polarization as closely intertwined outcomes of political discourse. Affective polarization refers to the extent to which individuals express hostility, animosity, or distrust toward members of opposing political groups~\citep{iyengar2012affect,boxell2024cross}. A growing body of research suggests that increasing ideological distance is often accompanied by rising incivility, hostility, and negative perception of political opponents~\citep{chen2018effect,santoro2022promise,zimmerman2022political}. Consequently, ideological polarization and affective polarization are frequently treated as parallel manifestations of the same underlying political divide.

Several mechanisms have been proposed to explain this relationship. One stream of research argues that ideological disagreement itself generates affective hostility. As individuals become more ideologically distant, opposing viewpoints are increasingly perceived as threatening, illegitimate, or morally objectionable, making civil disagreement more difficult and increasing the likelihood of hostile exchanges~\citep{brown1987politeness}. Another stream emphasizes the reverse causal pathway, whereby exposure to uncivil political interactions undermines perceptions of legitimacy and trust toward political opponents, thereby strengthening partisan animosity and reinforcing ideological divisions~\citep{mutz2007effects,rathje2021out}. Together, these perspectives suggest a self-reinforcing cycle in which ideological disagreement and affective hostility mutually amplify one another.

Importantly, however, these theories were developed in environments where political discourse was overwhelmingly generated by humans. As discussed above, ideological polarization in human discourse frequently emerges through disagreement, contestation, and attempts to persuade others. In such settings, growing ideological distance naturally increases the likelihood of conflict and interpersonal hostility. By contrast, ideological reinforcement in LLM-assisted discourse arises through affirmation rather than confrontation. Instead of challenging opposing viewpoints, LLM-assisted responses tend to validate and reinforce the position embedded in the content to which they reply. Consequently, ideological divergence can emerge without requiring the adversarial interactions that traditionally give rise to affective hostility~\citep{voelkel2024megastudy}.

Moreover, LLMs are generally designed to produce responses that are polite, helpful, and socially acceptable across a wide range of conversational contexts~\citep{bai2022training,ouyang2022training}. These stylistic characteristics arise from model training objectives and safety alignment procedures rather than from the political identity of the interlocutor. Unlike human communicators, whose tone often varies according to whether they are interacting with allies or opponents, LLM-assisted responses are systematically biased toward cooperative and non-confrontational language regardless of the ideological orientation of the discussion. As a result, even when LLM-assisted comments reinforce political viewpoints, they are less likely to contain the hostility, insults, or antagonistic rhetoric commonly associated with affective polarization~\citep{sharma2023towards}. This mechanism operates across both same-partisan and cross-partisan interactions. In same-partisan discussions, LLM-assisted responses reinforce existing viewpoints while expressing them in a more civil manner. In cross-partisan discussions, the tendency toward affirmation and politeness may reduce the emotional friction associated with ideological disagreement. Consequently, the reduction in hostility generated by LLM-assisted discourse is not limited to particular political groups or interaction contexts but may instead emerge broadly throughout the discussion environment, among users regardless of their use of LLMs.

Taken together, our theoretical arguments suggest that LLMs weaken the traditional link between ideological polarization and affective polarization. Whereas human-generated polarization often emerges through disagreement and confrontation, LLM-assisted polarization emerges through affirmation and reinforcement. As a result, political viewpoints can become more ideologically aligned and extreme even as the hostility typically associated with political disagreement declines. Accordingly, we expect the introduction of ChatGPT to reduce affective polarization in online political discourse:

\medskip
\noindent \textbf{Hypothesis 2.} \textit{Affective polarization in online political discourse decreases following the launch of ChatGPT.}

\section{Methods}\label{sec:Methods}
\medskip
\subsection{Data and Sample}
Reddit is one of the largest political-discussion platforms in the U.S. that allows discussion from both partisan sides, and it has been broadly examined in prior research to study political polarization~\citep{waller2021quantifying,mamakos2023social}. In this study, we focus on the subreddit r/politics, which was created in 2007 and is now the largest subreddit for news and political discussion on Reddit, with approximately 8.8 million members and 10 thousand daily active members. 

We aim to test the ideological change of online political discourse before and after the launch of ChatGPT on November 30, 2022. We collect all the posts and comments in r/politics between June 2021 and March 2023 and remove the comments generated by platform auto robots (e.g., Reddit ID as AutoModerator or PoliticsModeratorBot). Due to the seasonal pattern in individuals' political ideologies~\citep{flamino2023political,hohm2024moral}, following the approach applied in recent economics literature~\citep{burtch2024consequences,eichenbaum2024expectations,goldberg2024regulating}, we construct the same set of comment panels for the corresponding calendar period, one year prior, to serve as our control in a difference-in-differences (DID) design, estimating the average treatment effect of ChatGPT's launch. Because we do not know when each user joined Reddit, we focus on users who posted at least one comment between June 1, 2021, and December 1, 2021, during the six months preceding the control-period cutoff date, which ensures temporal balance in our research design and avoids the results being driven by new users who joined the platform after the release of ChatGPT. 

\begin{figure}
\FIGURE
{\includegraphics[width=0.8\textwidth]{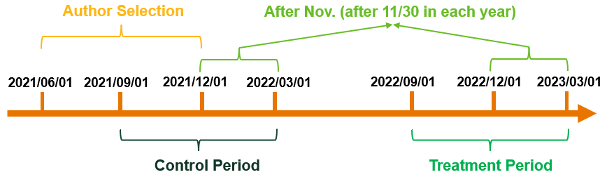}}
{Research Design and Sample Timeline.\label{fig:figure1}}
{The figure illustrates the difference-in-differences design. The treated sample comprises 
posts and comments created between September 1, 2022 and March 1, 2023; the control sample comprises posts and comments created exactly one year earlier. The post-treatment indicator (\textit{After Nov.}) flips on November 30 in both samples, the launch date of ChatGPT in the treated period.}
\end{figure}

The sample timeline and research design are shown in Figure~\ref{fig:figure1}. Specifically, we create the treated sample using posts and comments created between September 1, 2022 and March 1, 2023, and the control sample using posts and comments created exactly one year before that. Then we create the variable ``After Nov.", indicating the posts and comments created after November 30 in both the control and treated samples. In this way, both the treated and control samples were evenly divided into two parts. Ultimately, we collected 6,882,091 posts and comments by 334,002 unique authors in total, splitting them into the control and treated samples.

\subsection{Political Slant}
To obtain a continuous measure of political slant for each comment, we use a bag-of-words method that has been extensively applied in economics and information systems literature to measure the political slant of texts, such as news and Wikipedia articles~\citep{gentzkow2010drives,greenstein2012wikipedia,greenstein2016open}. The slant score continuously ranges from $-1$ to $+1$, with lower values indicating stronger liberal orientations and higher values indicating stronger conservative orientations. The core justification for this approach is that liberals and conservatives tend to discuss different topics and prefer different vocabularies; even when referring to the same concept, the phrases they use can differ due to strategic intent. For example, when describing the federal tax on assets of the deceased, congressional Republicans prefer using ``death tax'' much more than ``estate tax'' as ``estate tax sounds like it only hits the wealthy, but `death tax' sounds like it hits everyone''~\citep{gentzkow2010drives}. 

To calculate the political slant of each comment, we first extract the full text and names of speakers of all congressional speeches from the \href{https://www.govinfo.gov/app/collection/crec/2021}{2021 Congressional Record}. Next, we transfer all Reddit comments into a bag of words and phrases, then identify these words and phrases in the text of congressional speeches, and map them with the ideology of congresspeople. This process creates a directed weight for each word or phrase to determine whether and to what extent it is aligned with a congressional Democrat or Republican. We then use the weights of these words and phrases to gauge the political slant of each Reddit comment. We also update the congressional speeches used in~\citet{gentzkow2010drives}. For the present study, we leverage the \href{https://www.govinfo.gov/app/collection/crec/2021}{2021 Congressional Record} to generate the keyword dictionary that liberals and conservatives frequently use. Table~\ref{tab:table15}, Panel A and Panel B list the phrases used more often by Democrats and Republicans, respectively. We find that Democrat representatives are more likely to use phrases such as ``African American,'' ``gun violence,'' and ``American rescue plan,'' while Republican representatives are more likely to use phrases such as ``illegal immigrants,'' ``gas prices,'' and ``voter ID laws.'' A key contribution of our approach is to move beyond the conventional reliance on the 2005 Congress Record~\citep{gentzkow2010drives,greenstein2012wikipedia,greenstein2016open} by employing the 2021 Congress Record---the same year as our author selection---to better capture contemporary political dynamics and to re-estimate the keyword dictionary and coefficients. 

\begin{figure}
\FIGURE
{\includegraphics[width=0.45\textwidth]{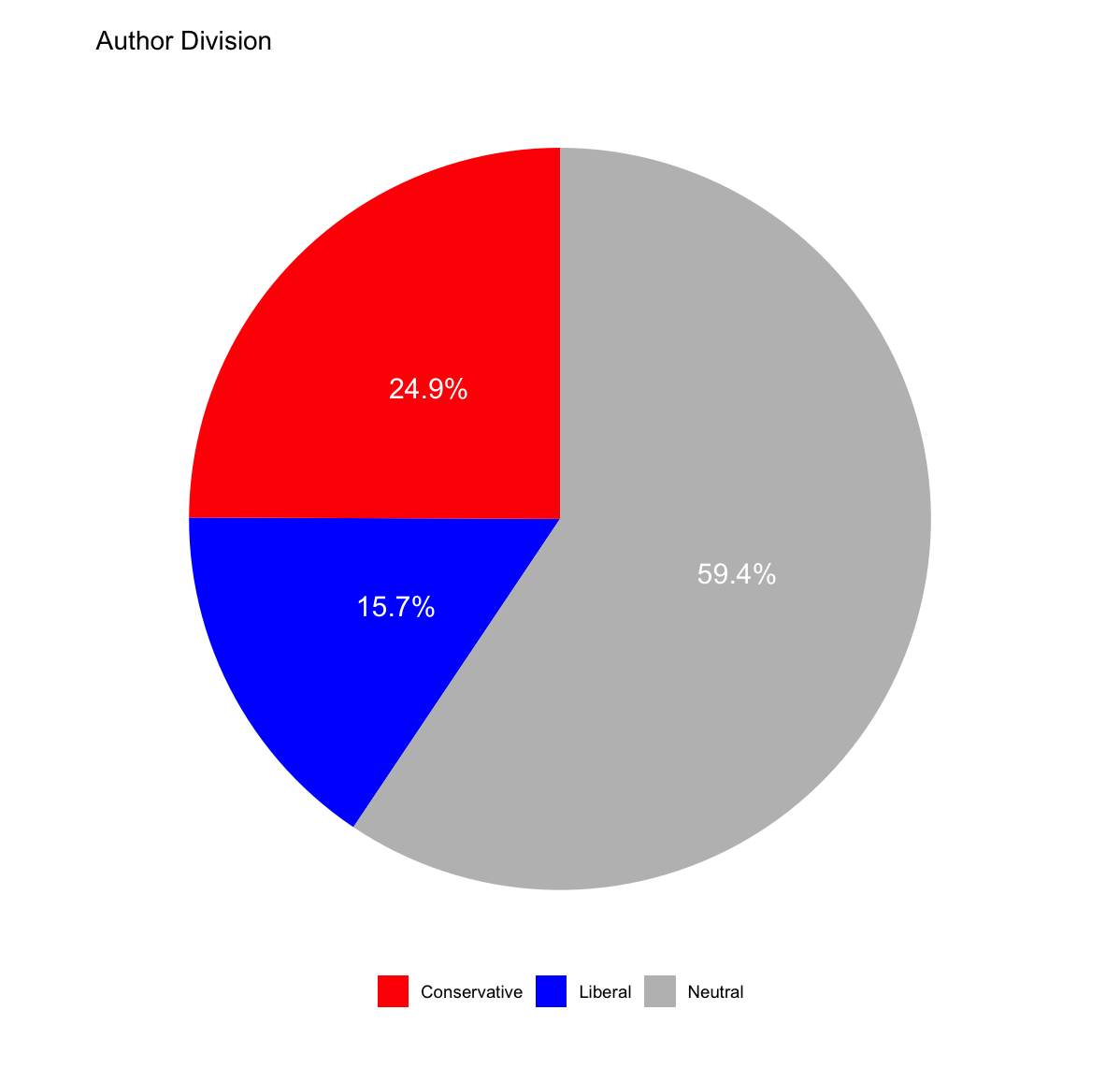}\quad\includegraphics[width=0.45\textwidth]{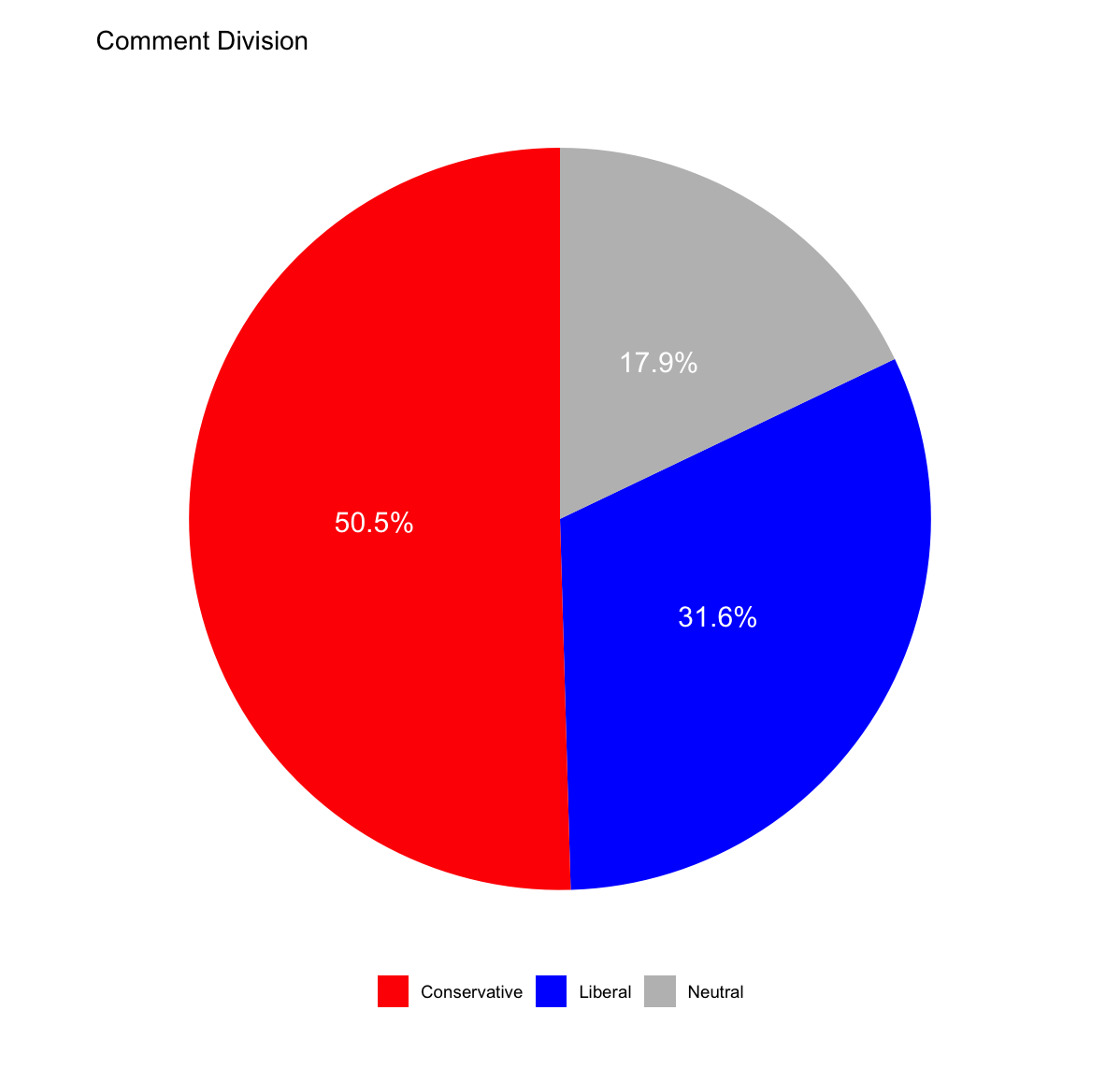}}
{Author and Comment Distribution by Political Slant.\label{fig:figure2}}
{(A) plots the categorization of authors based on their average political slant in the author selection period (June~1, 2021 to December~1, 2021). (B) plots the distribution of comments posted by the different categories of authors in the same period. An author is labeled liberal-leaning (conservative-leaning) when the average slant is below (above) zero.}
\end{figure}

Following~\citet{greenstein2012wikipedia}, we standardize all the political slants of comments into $[-1, 1]$ where a neutral comment is valued at 0, a liberal-leaning comment is below 0, and a conservative-leaning comment is above 0.\footnote{The results remain qualitatively unchanged when wider classification thresholds, such as $\pm 0.001$ and $\pm 0.005$, are used instead of 0 to distinguish liberal- and conservative-leaning users.} We measure each author's initial political slant by averaging the political slant of all this author's comments posted during the period from June 1, 2021 to December 1, 2021. An author is considered liberal-leaning (conservative-leaning) when the average slant is below (above) 0. In Figure~\ref{fig:figure2}, (A) and (B) report the author and comment division in our sample, respectively. We note that while 59.4\% of authors in our sample are neutral, their comments account for only 17.9\% of total comments, suggesting that users with a clear political slant (liberal- or conservative-leaning) are more active in posting comments and engaging in political discussion.

To further verify the credibility of this measure, we build several machine learning based measures and examine their correlations with our bigram-based measure in Table~\ref{tab:normslant_allmodels}. Five of these are supervised classifiers fine-tuned on labeled political text, such as news articles (PoliticalBiasBERT, AllSides\_DeBERTa), social-media posts (PolLeaning\_DeBERTa, PolBias\_RoBERTa), and partisan speech or tweets (DemRep\_DistilBERT), and two zero-shot NLI-based models (DeBERTa\_NLI and DeBERTaLarge\_NLI) are also included as a robustness check. Across all seven measures, the correlation with our bigram-based slant is positive and statistically significant, though modest in magnitude given the large sample. Having established that our measure is broadly consistent with these machine learning approaches, we use the bigram-based measure throughout the paper because it is substantially more transparent and interpretable than model-based measures.

\subsection{LLM-Assisted Comments}
To examine the likelihood that a Reddit comment was assisted by LLMs, we create four variables: Comment Length, Passive Verbs, Perplexity Score, and Human Probability.

Comment length is an important metric for LLM detection because prior work finds that human-authored texts tend to have more diverse vocabulary but shorter length, whereas LLM-generated texts tend to have less diverse vocabulary but longer length~\citep{munoz2024contrasting}. If an author's comment length suddenly and significantly increased after ChatGPT was launched, it is likely that this author has used LLMs to assist in generating the comments. We measure comment length by counting the number of words in a comment. Additionally, recent studies raise questions about whether LLMs more frequently employ passive voice than human-generated content~\citep{simchon2023computational}. Testing on Reddit data, ~\citet{simchon2023computational} show that LLMs use more passive auxiliary verbs when generating content. Following their approach, we use an open-source NLP library (i.e., spaCy) to count the number of passive auxiliary verbs as our measure of passive voice.

We also use state-of-the-art algorithms to create two other metrics, Perplexity Score and Human Probability, designed to assess the likelihood of a comment being produced by LLMs. Perplexity Score measures how well a probability model can predict a piece of text. A low perplexity score indicates that the text shares similar linguistic patterns and is more likely to be produced by LLMs, whereas human-authored texts tend to have higher perplexity values~\citep{radford2019rewon}. We apply the GPT-2 model from Hugging Face Transformers to calculate the perplexity score. The GPT-2 model is trained on vast amounts of text, and a text is assigned a low perplexity score if GPT-2 finds it predictable and fluent, aligning well with its learned distribution and indicating that this text is more likely to be produced in the assistance of LLMs. In addition, prior evaluations have shown that OpenAI's human-detection score exhibits relatively high accuracy and low bias in identifying AI-generated text, particularly in the early period following ChatGPT's release~\citep{solaiman2019release,shan2026examining}. We calculate the Human Probability of each comment using an open-source, pretrained model-based detector by OpenAI to evaluate the probability that a given text is assisted by LLMs~\citep{solaiman2019release}. Each comment is assigned a label (i.e., ``Human'' or ``LLM'') and a probability ranging from 0 to 1, where larger values indicate that this comment is more likely to be generated by a human. 

\begin{figure}
\FIGURE
{\includegraphics[width=\textwidth]{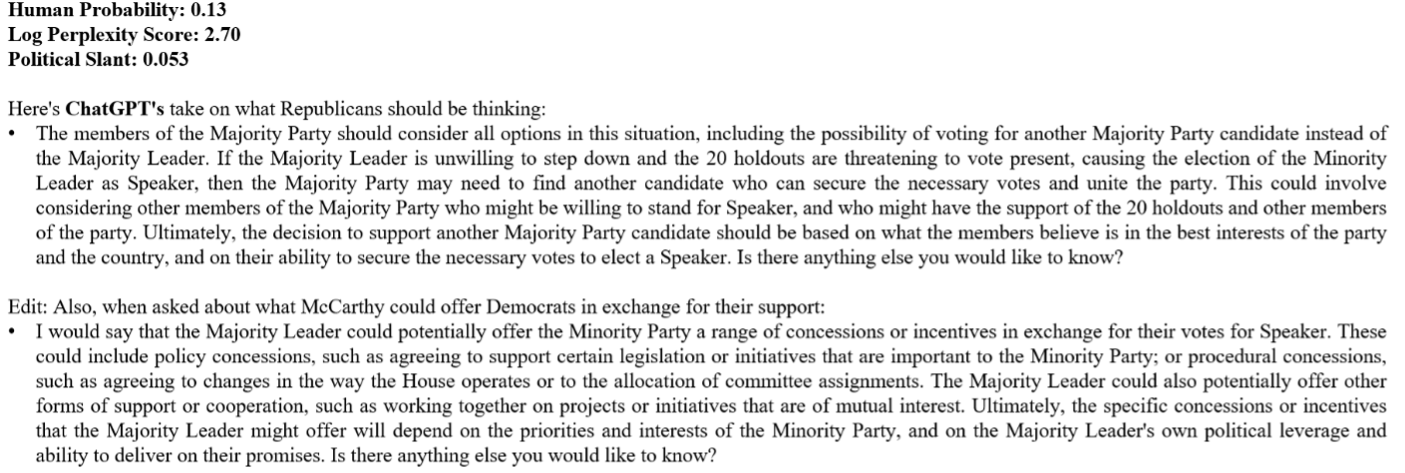}}
{An Example of LLM-Assisted Comment.\label{fig:example}}
{We know this comment is LLM-assisted because the author explicitly mentioned it in the comment. The Human Probability is 0.13 and the perplexity score is 2.7---about 2 standard deviations below the sample mean---both consistent with LLM-assisted content.}
\end{figure}

For example, Figure~\ref{fig:example} shows an example of an LLM-assisted comment in our sample. The Human Probability is only 0.13. The perplexity score is 2.7, about 2 standard deviations below the mean of the whole sample, both suggesting that this comment is likely generated by LLMs. Meanwhile, we find that comments with low Human Scores are associated with longer length, more passive constructions, and lower perplexity scores---features commonly associated with LLM-generated text~\citep{radford2019rewon,solaiman2019release,simchon2023computational,munoz2024contrasting,shan2026examining}. 

\begin{figure}
\FIGURE
{\includegraphics[width=0.6\textwidth]{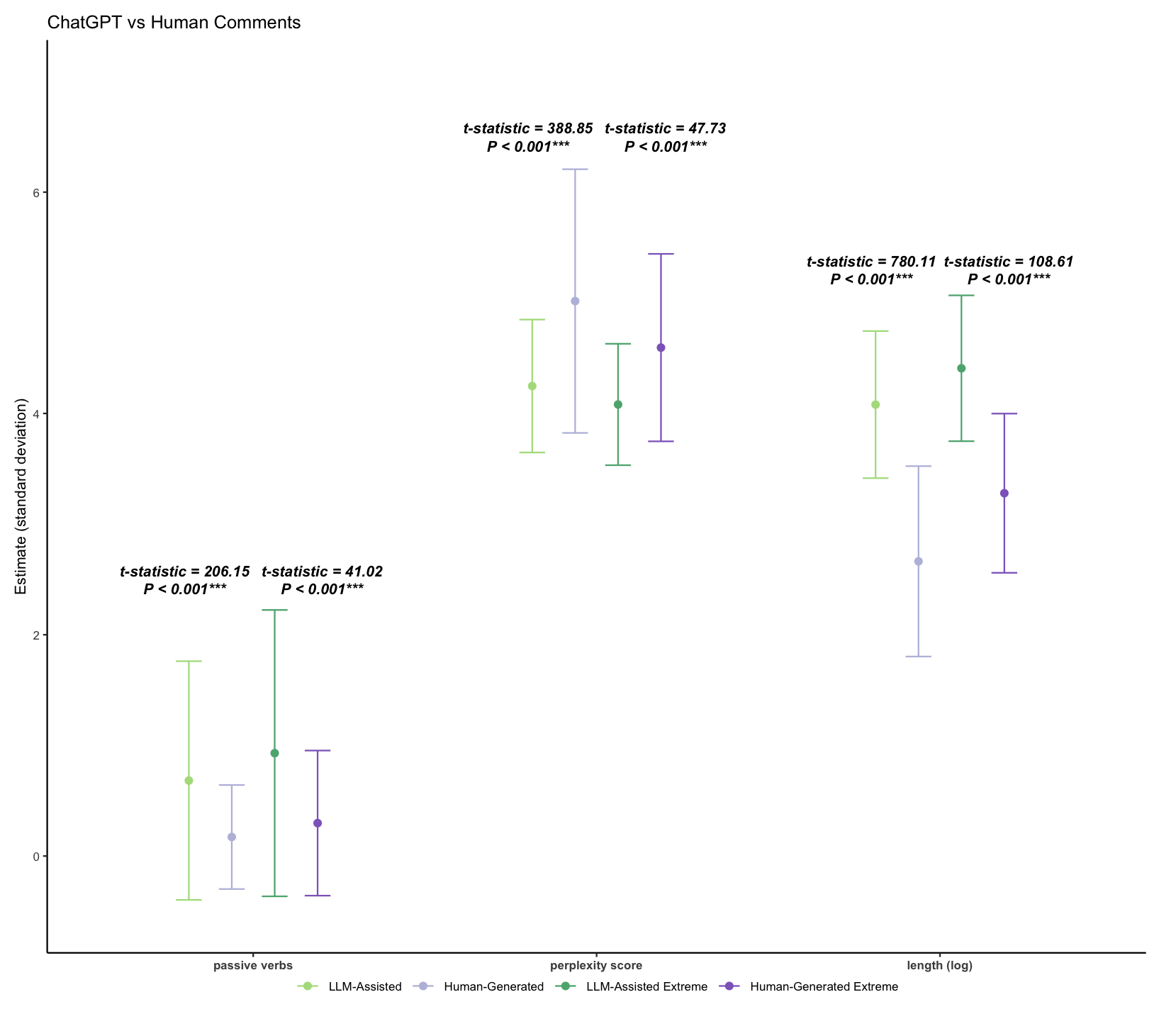}}
{The Different Characteristics of LLM-Assisted Comments versus Human-Generated Comments.\label{fig:figure6}}
{LLM-assisted comments are defined as those with a Human Probability score below 30 percentile, and human-generated comments are those above this threshold. Compared with human-generated comments, LLM-assisted comments tend to contain more passive verbs, lower perplexity scores, and more words on average---features commonly associated with LLM-generated text.}
\end{figure}

Although some argue that these measures rely on linguistic signals similar to those used by AI detectors and that neither perplexity scores nor OpenAI's ChatGPT detector can achieve 100\% accuracy in distinguishing between human- and AI-generated content, particularly as large language models continue to evolve and become more human-like, we argue that these tools remain highly valuable for research purposes. Their predictive power is especially useful in the early days following ChatGPT's release, when the linguistic patterns of AI-generated text were more distinct and less refined. This is particularly important for our study, as our sample is drawn from the first three months after ChatGPT's release. For expositional convenience, we classify comments with a Human Probability Score of 30 percentile or higher as human-generated comments and those below this threshold as LLM-assisted comments.\footnote{This classification is introduced to facilitate the presentation of subsequent mechanism analyses and should be interpreted as probabilistic rather than definitive. Importantly, our baseline results do not rely on this dichotomization and are estimated using the continuous Human Probability Score measure. The findings in this study remain qualitatively unchanged under alternative cutoff thresholds, such as 10 and 50 percentiles.} Figure~\ref{fig:figure6} shows that in our sample, LLM-assisted comments (i.e., comments with a human probability score below the 30th percentile) are substantially different from human-generated comments (i.e., comments with a human probability score above the 30th percentile), such as more passive verbs, a lower perplexity score, and more words in LLM-assisted comments. Moreover, our objective is not to determine with perfect accuracy whether any individual comment was LLM-assisted. Rather, our objective is to study aggregate patterns of AI-assisted writing at the population level over time. Random classification errors at the individual comment level are therefore likely to average out across millions of observations, making the aggregate trends considerably more reliable than individual predictions. Taken together, these results suggest that, despite their limitations, these detection methods provide a practical and scalable means of estimating AI involvement in user-generated content and enable the study of large-scale behavioral and discursive shifts during the initial wave of generative AI adoption.

\subsection{Civility and Emotions}
Besides political polarization, we also examine affective polarization by constructing several metrics concerning comment civility and emotions. Specifically, we rely on two measurements, Toxicity and Hostility, to proxy for the civility of the language. Prior literature focuses on disrespectful tone and word choice in conceptualizing and operationalizing incivility, and it is a common practice to measure incivility based on the toxicity of their language~\citep{arhin2021ground,nyhan2023like,biswas2025political}. In this study, we follow the prior studies~\citep{arhin2021ground,biswas2025political}, using the Detoxify unbiased model released by Unitary AI to calculate the toxicity level of each comment. This model is based on the RoBERTa-base architecture and is trained on Jigsaw's civil comments dataset, in which each individual comment was annotated by human raters from around the world who were proficient in English. These raters evaluated the degree of toxicity, as well as the levels of various toxicity subtypes, including severe toxicity, threats, obscenity, and insults. The toxicity label values range from 0 to 1, indicating the fraction of raters who believed the toxicity label applied to a comment. Using this pretrained model, we assign each comment in our sample a predicted toxicity score between 0 and 1, with higher value indicating more toxic language. Compared with the original model, the unbiased model is designed to reduce the unintended bias toward identity terms (e.g., comments mentioning identity terms such as 'Muslim' or 'gay' should not automatically receive higher toxicity scores). In this study, we focus only on the overall toxicity score to capture the general level of incivility. The results are consistent with another commonly used toxicity indicator, Google Perspective, which was also developed by Jigsaw and Google~\citep{frimer2023incivility,nyhan2023like,biswas2025political}. Both Google Perspective and Detoxify aim to predict toxic comments and identify negative online behaviors to help improve online conversations.

Moreover, we calculate the hostility level of each comment using the method described by~\citet{hebbelstrup2024offline}. Using the word2vec algorithm, we first construct word vectors, which are numerical representations of words in a semantic space, for each word in a corpus. These vectors help capture semantic relationships between words: words that frequently appear in similar contexts tend to have similar vector representations. To measure hostility, we construct a comment vector by averaging the vectors of all words in a comment. We then compute the cosine similarity between the comment vector and the vector for the word 'hate' as the hostility level for each comment. Comments with higher cosine similarity to 'hate' receive higher hostility scores because their language is semantically closer to hostile expressions. This measure has been validated against expert annotations, achieving 74 percent accuracy in classifying hostile comments relative to expert ratings~\citep{hebbelstrup2024offline}. It works better than a simple dictionary-based approach and is relatively insensitive to the specific words chosen since we not only use the word 'hate' but also the words that are semantically similar to it. For instance, `hate' is closely related to 'hateful', 'hostile', 'racist', and 'intolerance', and this series of subwords is also incorporated to identify the 'hostile' comments.

In addition to civility, prior literature on affective polarization has also examined changes in users' emotional expressions~\citep{brady2017emotion}. We categorize four basic emotions---anger, fear, surprise, and sadness---with emotion scores for each Reddit comment ranging from 0 to 1, applying Jochen Hartmann's Deep Learning model~\citep{morag2025narratives}.

Table~\ref{tab:table16} presents the summary statistics. Of the 6,882,091 posts and comments in our sample, 5,604,844 comments are posted by either liberal-leaning or conservative-leaning authors, and we calculate other dependent variables of these comments posted by non-neutral authors. On average, a comment contains 34 words, with a log perplexity score of 4.815 and a human probability of 28.6\%. The hostility and toxicity scores for an average comment are 0.504 and 0.173, respectively. Among emotions, surprise is the most expressed, with an average score of 0.452, followed by fear (0.16) and anger (0.152). Sadness is rarely expressed, with an average score of 0.036.

\section{Results}\label{sec:results}
In this section, we first demonstrate that political discourse on Reddit became more polarized following the launch of ChatGPT: liberal-leaning authors posted increasingly liberal comments, while conservative-leaning authors posted increasingly conservative comments, with little change among neutral and centrist authors. Conversely, we also find that affective polarization declined over the same period. We then decompose authors on Reddit into active versus inactive ones based on their commenting frequency. For expositional convenience, we classify inactive users as those below the 90th percentile in commenting activity, and otherwise as active users. Our results are robust to alternative thresholds, including the 95th, 80th, and 60th percentiles. Our subsample analyses show that the primary source of this polarization is previously inactive authors, whose linguistic patterns indicate greater reliance on ChatGPT and whose political expressions became substantially more extreme after its release. By contrast, the decrease in toxicity and hostility is more pronounced for active authors, and inactive authors become moderately more civil. Finally, we identify a mechanism consistent with algorithmic sycophancy: LLM-assisted replies systematically align with the ideological stance of the content to which they respond, thereby amplifying polarization in same-partisan interactions.

\subsection{Polarized Political Discussion after ChatGPT was Launched}
First, we examine the impact of ChatGPT's launch on online political discourse by estimating the following comment-level DID regression:
\begin{equation}
\label{eq:DID}
\text{PoliticalSlant}_{c} = \beta_{0} + \beta_{1}\,\text{After\_Nov}_{c} + \beta_{2}\,\text{Treatment}_{c} + \beta_{3}\,\text{After\_Nov}_{c} \times \text{Treatment}_{c} + \gamma_{a(c)} + \varepsilon_{c},
\end{equation}
where $c$ indexes comments, $\text{After\_Nov}_{c}$ equals one if comment $c$ was posted after November 30 in either the treated or control period. $\text{Treatment}_{c}$ equals one if comment $c$ belongs to the 2022–2023 treated sample, and zero if it belongs to the 2021–2022 control sample. $a(c)$ denotes the author who posted comment $c$, and $\gamma_{a(c)}$ denotes author fixed effects. As a robustness check, we also aggregate the data to the author-day level and re-estimate the models. The results from the author-day level analysis remain consistent.

Table~\ref{tab:table1} presents the effect of ChatGPT on political discussions on Reddit using Equation~\eqref{eq:DID}. Users in our sample are decomposed into subsamples of liberal-leaning, neutral, and conservative-leaning authors based on their average political slant in the author selection period. Author fixed effects are included in all three columns. Column (1) reports that the political slant of comments left by liberal-leaning authors declines by 0.003 after ChatGPT's release. In contrast, column (3) shows that the political slant of comments posted by conservative-leaning authors increases by about 0.003 after the introduction of ChatGPT. In magnitude, this shift is substantial: on average, the estimated treatment effect is equivalent to pushing a median liberal or conservative comment into the top 30\% most extreme segment of their respective ideological spectra (Figure~\ref{fig:figure7}). In column (2), we show that the political slant of neutral authors' comments slightly decreases, and the magnitude of the effect is only 10\% of that for liberal-leaning and conservative-leaning authors. Together, these results show that liberal-leaning (conservative-leaning) authors posted increasingly liberal (conservative) comments, suggesting increased polarization following the launch of ChatGPT.

The polarization patterns documented in the previous section are estimated using comment-level regressions. While this approach exploits the full granularity of the data, it may overweight the behavior of a small subset of highly prolific authors. To address this concern, we also examine polarization at the author level. We find consistent polarization patterns that after ChatGPT's release and report the author-level regression results in Appendix~\ref{ap:author}.

Second, we conduct a relative-time study to examine the dynamics of treatment effects by fitting the following comment-level regression:
\begin{equation}
\label{eq:relative-time}
\text{PoliticalSlant}_{c} = \beta_{0} + \sum_{k} \beta_{k}\,\text{ChatGPT\_Period}_{k} \times \text{Treatment}_{c} + \gamma_{a(c)} + \varepsilon_{c},
\end{equation}
where $\text{ChatGPT\_Period}_{k}$ is a dummy variable implying relative $k$ weeks around November 30, 2022, and $\beta_{k}$ captures the dynamic treatment effect. Figure~\ref{fig:figure3} reports the results. First, there is minimal pre-treatment trend. Notably, the comments from both liberal-leaning and conservative-leaning authors became more polarized immediately after ChatGPT was launched, while neutral authors were relatively less influenced, consistent with the findings in Table~\ref{tab:table1}. Overall, these results suggest that the platform exhibits increased ideological polarization since the launch of ChatGPT. 

Moreover, neutral and centrist authors located near the ideological center may play a particularly important role in shaping political outcomes during salient events such as elections, as even modest shifts in their political orientations can meaningfully influence voting outcomes. We conduct separate author-level analyses focusing on authors in the middle of the political spectrum in Table~\ref{tab:table3}. Columns (1) and (2) report estimates for the top 20\% and 10\% most centrist liberal-leaning authors, respectively. We find that ChatGPT's release is associated with a modest, statistically insignificant leftward shift in the political discourse of these groups, with magnitudes amounting to roughly 10\% of the effect observed for the full liberal-leaning sample in Table~\ref{tab:table2}. Column (3) reproduces the baseline results from Table~\ref{tab:table2} for comparison. Columns (4) and (5) show that the most centrist conservative-leaning authors similarly exhibit only modest and statistically insignificant shifts in political slant.\footnote{We also construct comment-level observations for neutral and centrist authors and run similar regressions in Table~\ref{tab:table12}. The results are qualitatively and quantitatively unchanged.} Taken together, the author-level evidence suggests that ideological shifts among neutral and centrist authors are limited and unlikely to account for the aggregate polarization patterns observed at the comment level. The modest and statistically insignificant movements in the middle range of the ideological spectrum indicate that these authors play, at most, a secondary role in driving overall polarization.

One concern with our research design is that 2022 may not exhibit seasonal effects comparable to those in 2021. Some may argue that 2022 was marked by several significant political events occurring around the release of ChatGPT, including \textit{Dobbs v. Jackson Women's Health Organization} on June 24, 2022, the midterm elections on November 8, 2022, and the criminal referral of the President on December 19, 2022. These events had the potential to substantially influence online discussions. Furthermore, broader temporal trends on the platform may also confound the observed patterns of polarization. To rule out these factors, we conducted a series of falsification tests by moving the date of treatment (the date when ChatGPT was launched) to a different time. We report the falsification test results for treatment dates moved one year earlier, six months earlier, and three weeks earlier in Table~\ref{tab:table8}. None of these results show a pre-treatment polarization trend on Reddit.

It is also worth briefly discussing what political polarization finding identifies in this paper. Political polarization that we have detected captures changes in composition rather than changes in individual authors' underlying ideologies. Our corpus-level findings show that the platform became more polarized following the introduction of ChatGPT, while the author-level findings reported in the appendix further indicate that authors tended to post more polarized comments. For example, liberal-leaning authors tended to post more liberal comments. However, this does not imply that liberal-leaning authors themselves became more liberal. Analyses based on textual data alone do not provide sufficient evidence to deduce changes in individuals' underlying ideologies. Therefore, our results should be interpreted as evidence of changes in the ideological content expressed on the platform, rather than evidence of changes in users' underlying political beliefs.

\subsection{Reduced Affective Polarization}
We next investigate the effect of ChatGPT's release on affective polarization~\citep{iyengar2012affect,boxell2024cross}, operationalized through measures of comment hostility and toxicity. We repeat our analysis in Table~\ref{tab:table1} by substituting the dependent variable with incivility metrics to test Hypothesis 2. As reported in Table~\ref{tab:overall_hostility_toxicity}, Panel A, comment hostility declines significantly in both liberal- and conservative-leaning author groups. The hostility scores of comments by both author groups fall by about 0.004, though the magnitude is small (Figure~\ref{fig:hostility_plot}). Panel B shows a similar, modest and statistically significant reduction in comment toxicity. In magnitude, the toxicity score of comments on average declines by approximately 0.002 in both liberal- and conservative-leaning authors, which is equivalent to moving a comment at the median of its partisan group to the 20th percentile least toxic of its partisan group (Figure~\ref{fig:toxicity_plot}). One concern that corpus-level analysis leaves unsettled is whether the observed affective polarization is driven by compositional changes or behavioral changes. For instance, if earnest partisans scale up their commenting using ChatGPT while toxic authors don't, the compositional change still produces affective decoupling even though no individual's behavior has changed. Therefore, we also conduct an author-level analysis of hostility and toxicity and report the results in Appendix A. Table~\ref{tab:overall_hostility_toxicity_authorday} shows a consistent and significant decline in comment hostility and toxicity in both groups. 

Besides hostility and toxicity, we also conduct a set of emotion analyses to examine whether the adoption of LLMs influences the emotions of users and present the results in Appendix A. As reported in Panel A--D of Tables~\ref{tab:table10} and~\ref{tab:table11}, we study the emotional content of comments in the post-ChatGPT period. In general, we do not observe a notable effect of ChatGPT on influencing comments' emotions. By exhaustively investigating four basic emotions, the four panels present a largely consistent result that there is a negligible correlation between LLMs' usage and emotional content. Collectively, Tables~\ref{tab:table10} and~\ref{tab:table11} indicate that the introduction of ChatGPT does not affect the emotional content of users' comments. Albeit their comments become more biased, exposure to ChatGPT fosters more civil discussion online. These findings collectively indicate that although ideological divisions may have deepened following the launch of ChatGPT, affective polarization, reflected in hostile and toxic expressions, has been mitigated, indicating a decoupling between political extremity and emotional animosity in online discourse.

As discussed earlier, one point deserves attention: our results should be interpreted as changes in the content on the platform rather than changes in users' underlying  dispositions. The fact that authors posted more polite content is not equivalent to authors themselves becoming more polite; rather, it reflects a change in the composition of content expressed on the platform. Figure~\ref{fig:did_civility} portrays the overall hostility and toxicity of comments on Reddit before and after November 30 by year. The platform exhibits a striking decline in both hostility and toxicity in 2022 subsequent to November 30 compared to 2021. This pattern suggests that the content posted on Reddit became more civil following the introduction of ChatGPT. However, the observed reduction in hostility and toxicity should not be interpreted as evidence that Reddit users themselves became less hostile or more civil, as text data alone cannot identify changes in users' underlying attitudes or personality traits. Nevertheless, from the perspective of the platform, changes in the content itself are consequential because they shape the communication environment experienced by users. As we show later in Figure~\ref{fig:figure5}, the decline in hostility is also evident among comments classified as human-generated, suggesting that the introduction of LLM-assisted content may have altered the conversational norms of the platform rather than merely adding more polite AI-assisted comments. One plausible interpretation is that the presence of LLM-assisted comments creates a more civil discussion environment, to which human users subsequently adapt in their own language.

\subsection{Author-Level Changes after ChatGPT was Launched}
\subsubsection{Inactive Authors Use LLMs.}
Recent studies suggest that on user-generated content platforms, inactive users are more likely to be influenced by generative AI tools compared to active users~\citep{burtch2024consequences,del2024large}. We decompose the sample into four groups based on both the partisan and the activity level including active liberal-leaning, inactive liberal-leaning, active conservative-leaning, and inactive conservative-leaning. As discussed above, inactive users are labeled as those below the 90th percentile in commenting activities. We first test the linguistic changes in authors' comments by examining their comment length, perplexity score, and human probability before and after the launch of ChatGPT. Perplexity scores and human probabilities measure the likelihood that the comment is AI-generated. High scores indicate high likelihood of being human-generated.

Table~\ref{tab:table4} compares the shift in comment length (Panel A), perplexity score (Panel B), and human score (Panel C) across different author groups following the launch of ChatGPT. Panel A shows that, in comparison to active authors, inactive authors, regardless of partisanship, post significantly longer comments after the launch of ChatGPT. Panel B illustrates that the perplexity scores of all author groups have declined after the launch of ChatGPT, with the effect size being much stronger ($5\sim10$ times) for inactive authors compared to active ones. Panel C reports that the human scores of inactive authors have decreased significantly after the launch of ChatGPT, with the effect size being much stronger, while the scores of active authors have increased slightly, with a much smaller effect size. Given that longer text length, reduced perplexity scores, and human scores constitute established characteristics of LLM-generated content~\citep{radford2019rewon,solaiman2019release,simchon2023computational,munoz2024contrasting,shan2026examining}, the substantial changes in inactive authors' linguistic patterns following ChatGPT's release suggest that inactive authors are likely to employ LLMs to generate Reddit comments, consistent with the findings reported in the recent literature~\citep{burtch2024consequences,del2024large}.

\subsubsection{Inactive Authors Become More Polarized, while Active Authors Become More Polite.}
We further consider how ideological polarization and reduced affective polarization may vary with respect to the degree of activity of authors. Table~\ref{tab:table5} shows the changes in political slants and commenting activities of active and inactive authors following the launch of ChatGPT. In Panel A, Columns (1) and (2) show that among liberal-leaning authors, inactive authors post comments that are significantly more polarized ideologically than active authors following the release of ChatGPT, with the effect size being about four times larger. Columns (3) and (4) show similar results for conservative-leaning authors. These results show that, regardless of partisanship, inactive authors, who have plausibly used LLMs, generate comments that are more polarized ideologically. Meantime, active authors also post comments that are slightly more polarized ideologically following the release of ChatGPT. Panel B shows that the number of monthly comments posted by each group of authors has also changed following the launch of ChatGPT. Specifically, regardless of partisanship, active authors have posted significantly more comments than inactive authors. In magnitude, a median active author demonstrates about a 50\% increase in monthly comments (Figure~\ref{fig:figure9}).\footnote{$2.73/4.78 = 0.57$.} 

Interestingly, the pattern for affective polarization differs from that for ideological polarization. Whereas ideological polarization is concentrated among inactive authors, improvements in civility are more pronounced among active authors. Table~\ref{tab:table6} replicates the specification in Table~\ref{tab:table5} but focuses on the impact on incivility. Panel A exhibits how comment hostility changes following the launch of ChatGPT across active and inactive authors in both partisan groups. For active authors, column (1) reports a negative and statistically significant coefficient for liberal-leaning authors, and column (3) shows a significantly negative coefficient for conservative-leaning authors. In contrast, Columns (2) and (4) show that the coefficient estimate of hostility changes for inactive authors, regardless of liberal- or conservative-leaning, is smaller in magnitude and less statistically significant. Consistent results are observed in toxicity changes in Panel B. Columns (1) and (3) report toxicity declines of 0.002 and 0.0017 for active liberal- and conservative-leaning authors, significant at 10\% and 5\%, respectively. For inactive authors, however, coefficients in columns (2) and (4) are not statistically different from zero, regardless of partisan leaning. 

The contrast between Tables~\ref{tab:table5} and~\ref{tab:table6} is informative. The concentration of ideological polarization among inactive authors is consistent with these authors being more likely to rely on LLMs for content generation. By contrast, the larger improvements in civility among active authors suggest that the reduction in hostility is not confined to users who are most likely to adopt LLMs directly. Instead, the civility effect appears to extend more broadly across the discussion environment. Consistent with the evidence presented later in Figure~\ref{fig:figure5}, which shows that even comments classified as human-generated become less hostile after ChatGPT's release, one plausible interpretation is that LLM-assisted content shifts the prevailing conversational norms on the platform. As the discussion environment becomes more civil, active human users—who contribute the majority of comments—also adapt their language accordingly.

As discussed earlier, this finding may be subject to the concern that inactive authors became more politically extreme in response to major political events occurring in late 2022. To address this possibility, we replicate the series of falsification tests conducted earlier for both active and inactive authors. Specifically, we expect to observe no polarization effects when shifting the treatment date six months earlier or when using the midterm election date as the treatment date. Table~\ref{tab:falsification_active_inactive} in the Appendix reports the results. Panel A shows that inactive authors shift in the opposite direction when the treatment date is moved six months earlier, while Panel B demonstrates that there is no significant change in political slant among inactive authors following the midterm election. Taken together, these findings suggest that ideological polarization is driven primarily by authors who are most likely to adopt LLMs, whereas improvements in civility are more broadly reflected in the behavior of active contributors. This pattern is consistent with the idea that LLMs directly influence ideological expression while indirectly fostering a more civil conversational environment that shapes human-authored discourse.

\subsubsection{Algorithm Sycophancy as the Mechanism.}
To examine the mechanism through which ChatGPT contributes to increased political polarization and decreased affective polarization, we investigate whether changes in political slant and civility vary within and across partisan boundaries (i.e., whether they differ when an author replies to someone in the same political camp versus the opposing one). Under Hypothesis 1, we expect the algorithm-driven political drift to depend on the poster's ideological position rather than on the ideological distance between poster and commenter. By contrast, theoretical arguments underlying Hypothesis 2 predict that affective polarization decreases due to systematic affirmation and declining social cost of polite responses, thus we expect to observe a uniform incivility decline across all author groups regardless of either the poster's political position or the poster–commenter ideological distance.

We explore these mechanisms through two separate analyses. Specifically, we examine political slant and incivility changes in different replier-thread groups in pre- and post-ChatGPT periods. First, we consider how political slant changes vary across replier-thread political groups and the extent to which the changes differ depending on whether the content is human-generated or LLM-assisted. Second, we repeat the same analysis by focusing on two incivility metrics: hostility and toxicity. We begin our analyses by stratifying all replier-poster pairs. Political discussions on the platform take two primary forms: comments responding to posts and comments responding to other comments. Based on this structure, we organize the data into pairs consisting of an original thread (a post or a comment) and its subsequent reply. We then classify these pairs into four categories: liberal-leaning author replies to liberal-leaning author's threads, liberal-leaning author replies to conservative-leaning author's threads, conservative-leaning author replies to liberal-leaning author's threads, and conservative-leaning author replies to conservative-leaning author's threads. For each category, we use the human score described above to classify all posts and comments as either human-generated or LLM-assisted. In turn, there are four categories within each of the four replier-thread pairs: LLM-assisted content replies to LLM-assisted content threads, LLM-assisted content replies to Human-generated content threads, Human-generated content replies to LLM-assisted content threads, and Human-generated content replies to Human-generated content threads. Figure~\ref{fig:figure4} depicts the resulting changes in political slant. The y-axis is decomposed into four replier-thread groups we discussed earlier based on the average political slant.\footnote{For both original thread authors and respondents, each author's political orientation is measured using the average political slant of all content they posted during the three months preceding ChatGPT's release (2022/09/01 to 2022/11/29). We find consistent results using other baseline periods, such as 2022/10/01 to 2022/11/29. We classify users with an average slant above 0 as conservative-leaning and those below 0 as liberal-leaning. Consistent with our baseline analyses above, the results remain qualitatively unchanged when more conservative classification thresholds, such as $\pm 0.001$ and $\pm 0.005$, are used instead of 0 to distinguish liberal- and conservative-leaning users.} The dashed vertical line shows the average political slant of these respondents before ChatGPT was launched, which we normalize to zero in the figure for ease of comparison.

Figure~\ref{fig:figure4} reveals a clear and consistent interaction-level pattern. The vertical dashed line denotes each author's average political slant prior to ChatGPT's release, normalized to zero. Accordingly, the figure illustrates the extent to which a reply's ideological slant shifts away from the author's baseline political orientation and toward the ideological position of the original post. Compared with human-generated replies (coefficients denoted in blue and purple), LLM-assisted replies (coefficients denoted in dark and light green) exhibit substantially larger shifts in the direction of the original post's stance. These results suggest that LLM-assisted responses systematically align with the ideological orientation of the content being replied to, rather than reflecting the respondent's own baseline political stance. This pattern holds regardless of whether the original post was written by a human or generated by an LLM.

These response patterns give rise to two asymmetric effects. In same-partisan interactions, LLM-assisted replies reinforce the prevailing ideological position, thereby amplifying polarization. In counter-partisan interactions, alignment with the opposing post leads to moderation in expressed ideology, resulting in a counter-party moderation effect. These findings are consistent with the mechanism of algorithm sycophancy, in which LLM-assisted comments systematically align with the stance of the content they reply to, as observed in prior work~\citep {abdurahman2024perils,naddaf2025ai,cheng2026sycophantic}. This alignment operates at the level of post--comment interactions rather than original content creation, indicating that ChatGPT primarily scales reply-based expression that follows prevailing thread narratives. As a result, LLM-assisted replies reinforce polarization in same-partisan interactions and modestly moderate it in counter-partisan ones, with the former dominating in aggregate, supporting H1. Taken together, these findings indicate that ChatGPT reshapes political discourse by conditionally aligning replies with existing content rather than reflecting the responder's own ideological position in both AI-to-human and AI-to-AI communication contexts.

While the preceding analysis explains how LLM-assisted replies reshape ideological polarization, it does not explain why political discussions simultaneously became more civil. To better understand this apparent paradox, we conduct two additional analyses examining whether the observed decline in hostility and toxicity varies with the political orientation of commenters and the ideological distance between commenters and the posts they respond to.

We begin by focusing on commenters' political identities. The top panel of Figure~\ref{fig:figure5} shows that both liberal- and conservative-leaning commenters became significantly less hostile and less toxic following ChatGPT's release. Although conservative commenters exhibit a slightly larger reduction in hostility, the decline in toxicity is statistically similar across the two groups. Overall, these results suggest that the improvement in civility is broad-based and not driven by any particular political camp. However, political identity alone may not be the relevant factor. The decline in incivility may instead depend on the ideological relationship between commenters and the content they engage with, particularly in interactions across partisan lines. 

We next examine whether affective moderation varies across partisan interactions. The middle panel of Figure~\ref{fig:figure5} reveal a remarkably consistent pattern. Hostility and toxicity decline across all four interaction types, regardless of whether the exchange occurs between political allies or opponents. Nearly all estimates are statistically significant at the 5\% level; the sole exception is toxicity in liberal-to-liberal interactions, for which the p-value is marginally above the conventional threshold (p = 0.056). More importantly, the magnitude of the decline is highly similar across in-group and out-group interactions. For example, hostility among conservative commenters falls by 0.0058 when responding to fellow conservatives and by 0.0055 when responding to liberals. Together, these results indicate that affective moderation is not confined to particular partisan relationships and does not appear to be driven by changes in out-group animosity alone. 

A remaining question is whether the observed decline in incivility reflects a genuine change in human behavior or simply the increasing presence of LLM-generated text, which is generally more polite. If a growing share of comments is generated with AI assistance, average toxicity and hostility could mechanically decline even if human-authored comments remain unchanged. To distinguish between these explanations, we repeat the analysis using only comments classified as human-generated. The results, shown in the bottom panel of Figure~\ref{fig:figure5}, closely mirror the earlier findings: hostility and toxicity continue to decline across all partisan groups. This evidence suggests that the reduction in incivility is not merely a compositional effect arising from the addition of more polite AI-generated content. Rather, ChatGPT appears to have altered the broader conversational environment, leading even human-authored comments to become more civil.

Taken together, these findings support the theoretical argument underlying Hypothesis 2. Although ChatGPT increases ideological polarization by encouraging replies to align with the political orientation of the content they address, it simultaneously reduces hostility and toxicity across the community. Generative AI therefore weakens the traditional linkage between ideological divergence and affective polarization, fostering a political discourse that is more polarized in content yet more civil in tone.

\section{Discussion}\label{sec:discussion}
This study offers the first empirical investigation into how the release of ChatGPT, a widely adopted LLM, has influenced political discourse on social media platforms. Drawing on large-scale, detailed Reddit data and a robust measurement of ideological stance derived from congressional speeches, we demonstrate that the release of ChatGPT significantly intensified political polarization among users. Importantly, this increase in polarization does not stem from the creation of more persuasive or ideologically extreme original content. Instead, it appears to originate from how LLM-assisted comments respond to existing posts. By scaling reply-based expression, ChatGPT amplifies users' responses in ways that align with the prevailing narrative of each discussion thread, thereby reinforcing existing positions. As a result, polarization is amplified in same-partisan post--comment interactions, while only modestly moderated in counter-partisan exchanges---effects that are insufficient to offset the overall rise in ideological polarization.

Despite this increase in ideological polarization, our findings reveal a surprising and counterintuitive result---ChatGPT did not exacerbate affective polarization. In fact, indicators of hostility and toxicity declined following ChatGPT's release. This suggests that while users became more ideologically divided, they did so in a more civil manner, possibly due to the structured, measured, and less inflammatory language characteristic of LLM-assisted content. This pattern aligns with recent literature indicating that exposure to balanced or clearly articulated viewpoints can reduce partisan misperceptions and emotional hostility~\citep{voelkel2024megastudy}.

These findings contribute to a growing body of research exploring the societal consequences of generative AI. The findings underscore that the political implications of LLMs are multifaceted: while such tools may amplify ideological signals, they can also support less emotionally charged political exchanges. This dual effect challenges the simplistic narrative that AI tools necessarily degrade the quality of public discourse~\citep{weidinger2021ethical} and instead suggests they may reshape the nature of polarization itself---intensifying ideological differentiation while simultaneously reducing antagonism. More broadly, the results highlight how relatively simple and low-cost technological interventions can foster meaningful improvements in political dialogue, in contrast to large-scale, government-driven regulatory approaches that often risk unintended consequences or public backlash. This suggests that modest, decentralized tools may in practice achieve more sustainable progress in enhancing the quality of democratic discourse than sweeping policy ``big hammers.''

However, several important limitations remain. First, we identify a seasonal pattern of political polarization during the first three months following the release of ChatGPT relative to the corresponding seasonal pattern observed in 2021. As such, the observed polarization pattern reflects a relative deviation from the baseline trend rather than an absolute divergence in political drift. Second, we do not directly observe whether a given comment was generated by LLMs; our identification strategy relies on behavioral shifts in linguistic patterns, which may be subject to measurement errors. Third, our study captures only the early effects of ChatGPT's release. Consequently, our treatment groups may primarily comprise polarization-prone early adopters or technology enthusiasts who are more likely to experiment with a novel tool. As both user familiarity and model capabilities evolve over time, the long-term dynamics of AI-mediated discourse may differ from the initial patterns observed in this study.

Future research should explore how LLMs influence political expression across different platforms and cultural contexts, especially in regions where political sensitivities and AI regulations differ substantially. It is also crucial to examine how design choices---such as content safeguards, prompt engineering, and model fine-tuning---affect the nature and direction of AI's impact on public dialogue. As generative AI tools become increasingly integrated into digital communication, understanding their role in shaping democratic discourse will be essential for both platform governance and public policy.

\clearpage
\onehalfspacing 
\bibliographystyle{econ}
\bibliography{reference}

\clearpage

\makeatletter
\setlength{\@fptop}{0pt}
\makeatother

\begin{figure}
\FIGURE
{\includegraphics[width=0.6\textwidth]{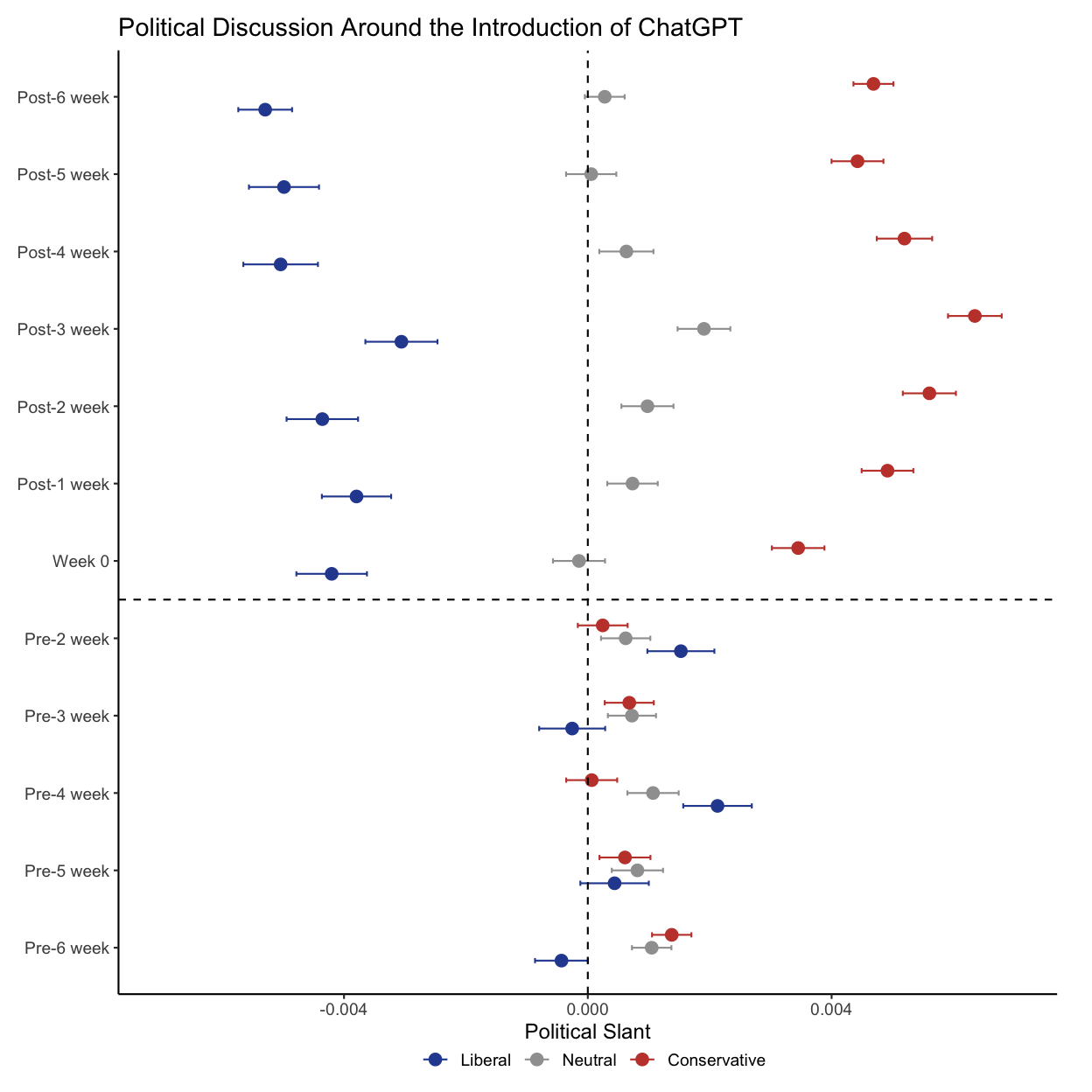}}
{Political Discussion Around the Release of ChatGPT.\label{fig:figure3}}
{This figure presents the results of relative-time regressions estimating the effect of ChatGPT's release on liberal-leaning (blue), neutral (gray), and conservative-leaning (red) authors using Equation~(\ref{eq:relative-time}). Week 0 covers November~28--December~4, 2022, the week containing the launch date of November~30, 2022. Author fixed effects are included in all models. Observations are at the comment level. Error bars show robust standard errors clustered at the author level. The brief right-leaning shift in the third post-launch week is plausibly attributable to the contemporaneous House January~6 Committee referral of Donald Trump for criminal prosecution on December~19, 2022.}
\end{figure}

\begin{figure}
\FIGURE
{\includegraphics[width=1\textwidth]{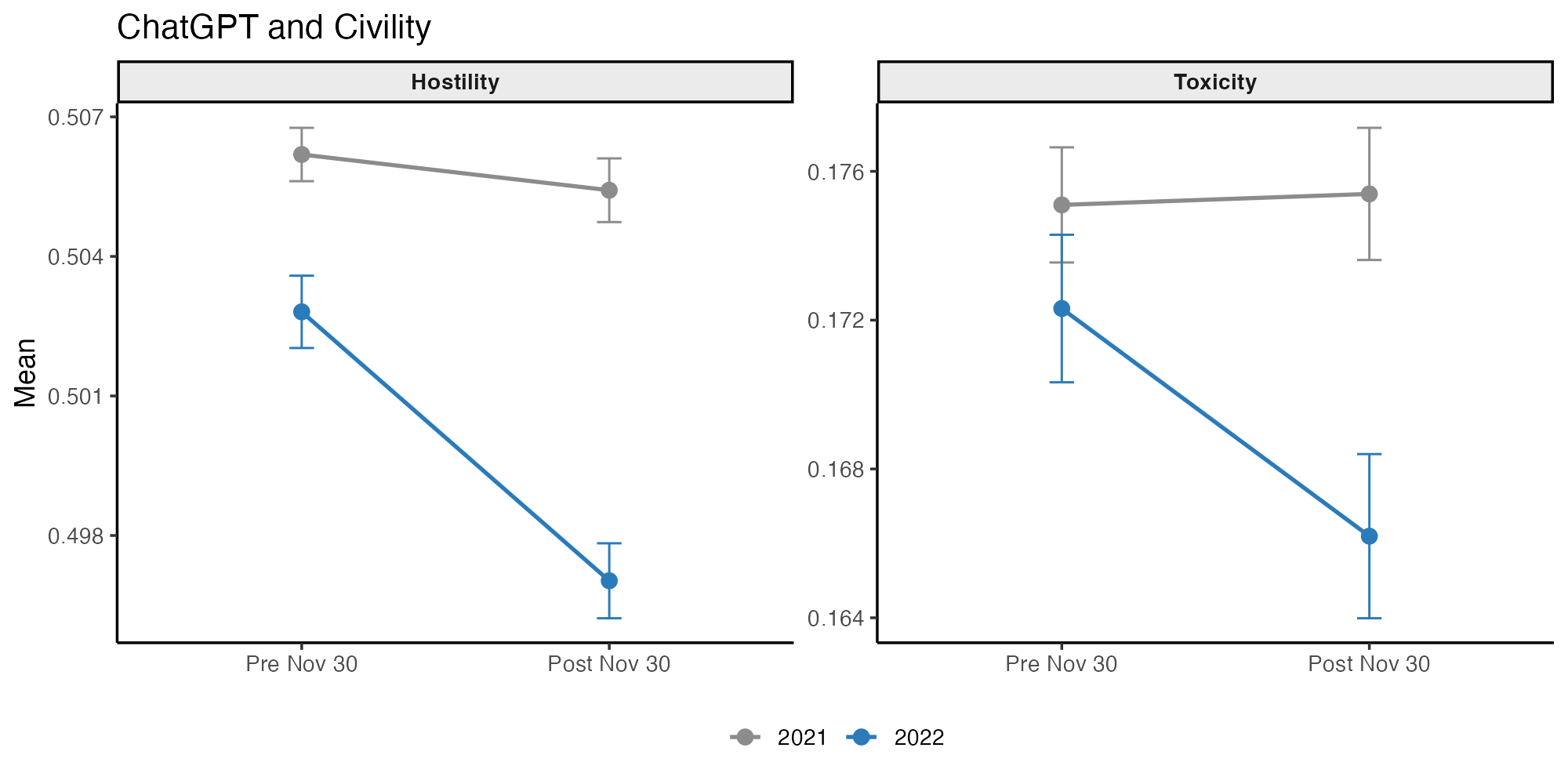}}
{Hostility and toxicity of comments before and after November 30.\label{fig:did_civility}}
{Cell means of comment-level hostility and toxicity, by year (2021  vs. 2022) and period (pre- vs. post-November 30). Points are means; bars are 95\% CIs with standard errors clustered by author.}
\end{figure}

\begin{figure}
\FIGURE
{\includegraphics[width=0.7\textwidth]{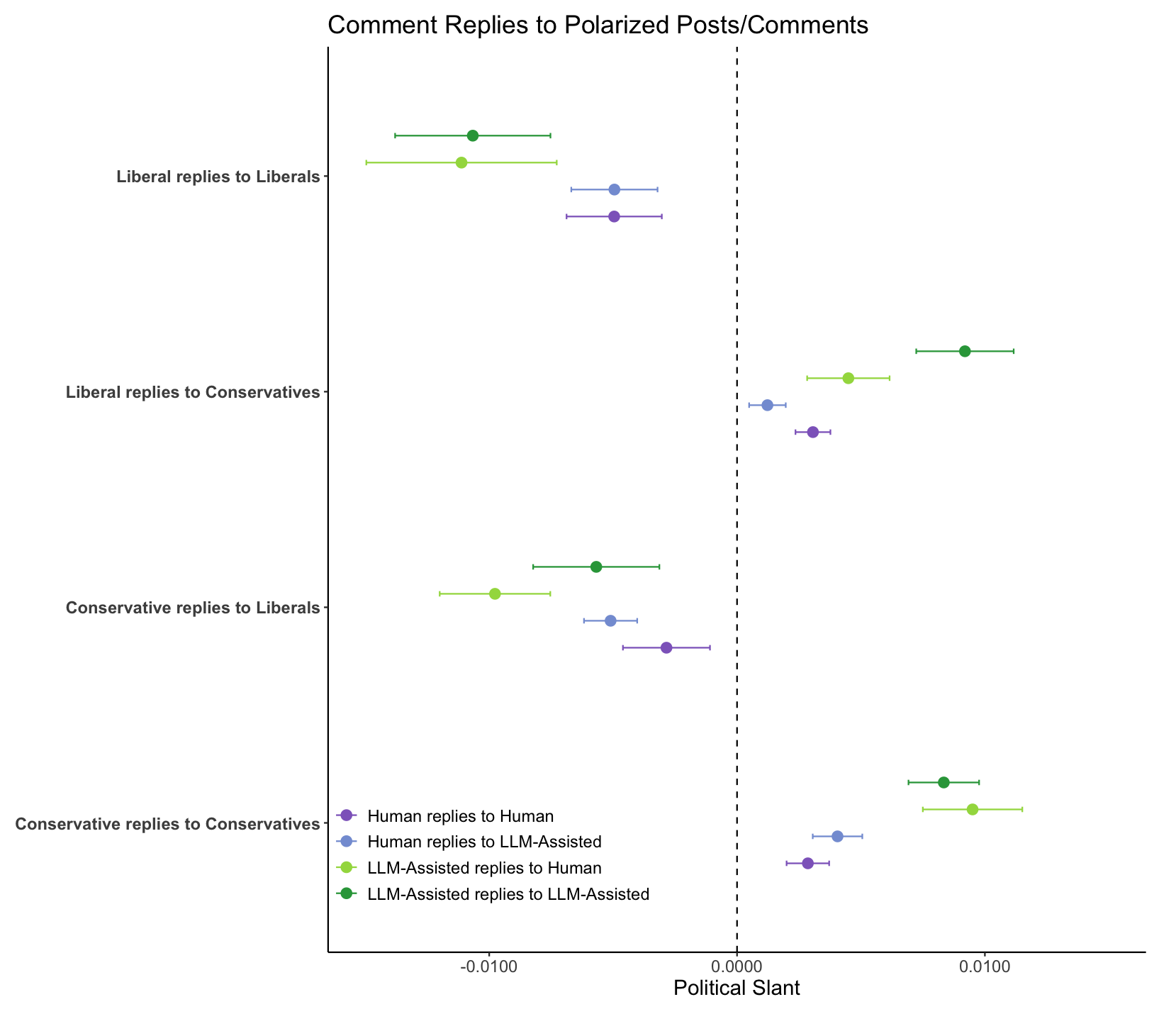}}
{Political Slant Changes Regarding Liberal- versus Conservative-leaning Author's Posts and Comments.\label{fig:figure4}}
{The y-axis shows four groups of replier--thread combinations: (1) liberal-leaning respondents replying to liberal-leaning threads, (2) liberal-leaning respondents replying to conservative-leaning threads, (3) conservative-leaning respondents replying to liberal-leaning threads, and (4) conservative-leaning respondents replying to conservative-leaning threads. The dashed vertical line marks each respondent's average political slant during 2022/09/01--2022/11/29 (3 months before ChatGPT's release), normalized to zero. Points and error bars show the average post-release political slant relative to that baseline. Dark green: LLM-assisted reply to a LLM-assisted thread. Light green: LLM-assisted reply to a human-generated thread. Blue: human-generated reply to an LLM-assisted thread. Purple: human-generated reply to a human-generated thread. Standard errors clustered on author-level.}
\end{figure}

\begin{figure}
\FIGURE
{\includegraphics[width=1\textwidth]{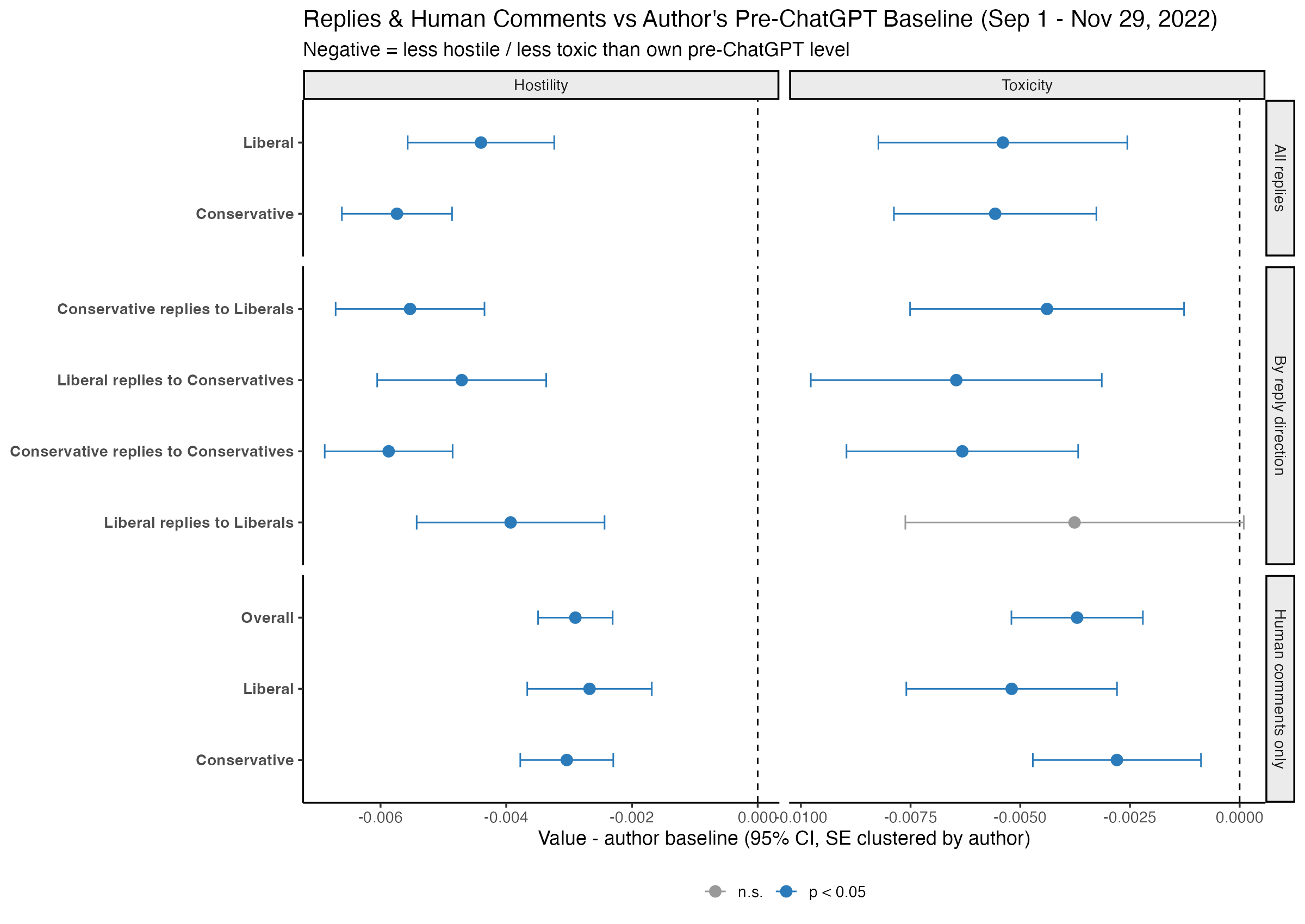}}
{Civility Changes Regarding Liberal- versus Conservative-leaning Author's Posts and Comments.\label{fig:figure5}}
{The figure comprises two columns and three row-panels. The two columns report the two incivility measures, hostility and toxicity. The three row-panels differ in how respondents are grouped along the y-axis. The top panel ("All replies") pools each respondent's post-release replies and groups them by the respondent's own leaning. The middle panel ("By reply direction") splits replies into four replier–thread combinations: (1) liberal-leaning respondents replying to liberal-leaning threads, (2) liberal-leaning respondents replying to conservative-leaning threads, (3) conservative-leaning respondents replying to liberal-leaning threads, and (4) conservative-leaning respondents replying to conservative-leaning threads. The bottom panel ("Human comments only") restricts attention to comments classified as human-generated and reports the overall mean together with the means by respondent leaning. The dashed vertical line marks each respondent's average politeness score during 2022/09/01--2022/11/29 (3 months before ChatGPT's release), normalized to zero. Points and error bars show the average post-release politeness score relative to that baseline.}
\end{figure}
\clearpage

\input{tables/TableS1_Partisan_Phrases}
\input{tables/Table_NormSlant_AllModels.tex}
\input{tables/TableS2_Summary_Statistics}
\input{tables/Table1_Polarized_Political_Discussion}
\input{tables/Overall_Hostility_Toxicity_Changes}
\input{tables/Table4_User_Linguistic_Changes}
\input{tables/Table5_Slant_and_Commenting_Change}
\input{tables/Table6_Hostility_Toxicity_Changes}
\clearpage

\begin{APPENDICES}
\renewcommand{\thefigure}{\thesection\arabic{figure}}
\renewcommand{\thetable}{\thesection\arabic{table}}
\setcounter{figure}{0}
\setcounter{table}{0}

\medskip
\section{Author-level Results}\label{ap:author}

\subsection{Author-Level Slants were also Polarized}
The polarization patterns documented in the main text are estimated using comment-level regressions. While this approach exploits the full granularity of the data, it may overweight the behavior of a small subset of highly prolific authors. To address this concern, we next examine polarization at the author level.

We construct author-day observations by averaging the political slant of all comments posted by a given author on a given day, and estimate regressions using this aggregated measure. This approach assigns equal weight to authors rather than to comments, thereby reducing the influence of extreme posting intensity. We estimate the following author-day level DID regression:
\begin{equation}
\label{eq:author-day-DID}
\text{PoliticalSlant}_{i,t} = \beta_{0} + \beta_{1}\,\text{After\_Nov}_{t} + \beta_{2}\,\text{Treatment}_{t} + \beta_{3}\,\text{After\_Nov}_{t} \times \text{Treatment}_{t} + \gamma_{i} + \varepsilon_{i,t},
\end{equation}
where $i$ indicates an author, $t$ indicates a day, and $\gamma_{i}$ controls for author fixed effects. The results in Table~\ref{tab:table2} are consistent with those in Table~\ref{tab:table1}: after ChatGPT's release, liberal-leaning authors became significantly more liberal-leaning, and conservative-leaning authors became significantly more conservative-leaning. Neutral authors became slightly more liberal-leaning, with the magnitude of the effect being less than 10\% of that for liberal-leaning and conservative-leaning authors. The results are also consistent when we use author-month level data and when we exclude the most prolific authors from the sample, defined as the top 10\% of authors with the largest total number of comments within each political group (Table~\ref{tab:table7} and \ref{tab:tableS3excl}). 

Table~\ref{tab:table12} examines comment-level changes in political slant among neutral authors and those located closest to the ideological center following the release of ChatGPT. Columns (1) and (2) report estimates for the top 20\% and 10\% most centrist liberal-leaning authors, respectively. We find that ChatGPT's release is associated with a modest, statistically insignificant leftward shift in the political discourse of these groups, with magnitudes amounting to roughly 10\% of the effect observed for the full liberal-leaning sample in Table~\ref{tab:table1}. Column (3) reproduces the baseline results from Table~\ref{tab:table1} for comparison. Columns (4) and (5) show that the most centrist conservative-leaning authors similarly exhibit only modest and statistically insignificant shifts in political slant. Taken together, these results indicate that neutral and ideologically centrist authors did not meaningfully contribute to the observed increase in political polarization following ChatGPT's release.

As authors in the same group (i.e., liberal-leaning and conservative-leaning) exhibit diverse political slants and post few opposing-leaning comments, it remains unclear whether the polarization effect is also distributed unevenly. Therefore, we explore whether the polarization effect differs within the same author group. We partition the authors in the same group into three subcategories---Extreme, Regular, and Centrist---based on 33rd and 67th percentile in political slant and examine the ChatGPT effect on each subgroup. Table~\ref{tab:table9} reports the outcomes. We find a statistically pronounced effect on extreme users. Specifically, the ChatGPT effect leads to a 0.015 decrease in the author-level political slant of extreme Liberal-leaning authors and a 0.011 increase in the author-level political slant of extreme Conservative-leaning authors (columns (1) and (4)). Relative to extreme authors, regular and centrist authors show trivial polarized effects, reflected in 0.003 (0.002) and 0.001 (0.001) decreases (increases) in political slant of regular and centrist Liberal-leaning (Conservative-leaning) authors, respectively (columns (2), (3), (5) and (6)).

\subsection{Author-Level Reduced Affective Polarization}
We also conduct additional author-level analyses for the affective polarization pattern. Table~\ref{tab:overall_hostility_toxicity_authorday} repeats the specification in Table~\ref{tab:table2}. Panel A exhibits author-level hostility changes following the launch of ChatGPT and Panel B shows author-level toxicity changes following the launch of ChatGPT. Both panels demonstrate a statistically significant decline in incivility, which is consistent with Table~\ref{tab:table6} in the Main Text.

Besides hostility and toxicity, we conduct a set of emotion analyses to examine whether the adoption of LLMs influences the emotions of users. As reported in Panel A--D of Tables~\ref{tab:table10} and~\ref{tab:table11}, we study the emotional content of comments in the post-ChatGPT period. In general, we do not observe a notable effect of ChatGPT on influencing comments' emotions. By exhaustively investigating four basic emotions, the four panels present a largely consistent result that there is a negligible correlation between LLMs' usage and emotional content, neither for inactive authors who have used LLMs nor for active authors who have not been affected by LLMs. Collectively, Tables~\ref{tab:table10} and~\ref{tab:table11} indicate that the introduction of ChatGPT does not affect the emotional content of users' comments. Although their comments become more biased, exposure to ChatGPT fosters more civil discussion online.

Tables~\ref{tab:table13} and~\ref{tab:table14}, Panel A--F report the author-level emotion changes following the public release of ChatGPT for neutral and middle authors. Tables~\ref{tab:table13} and~\ref{tab:table14}, Panel A reports that the level of author hostility in all author groups has declined significantly following the public release of ChatGPT. Similarly, Panel B indicates a modest reduction in author toxicity, especially for neutral authors and left-leaning conservative authors. These author-level results are consistent with the comment-level results presented in Table~\ref{tab:table6}, suggesting an overall improvement in user civility. Moreover, we study the author's emotional stance in the post-ChatGPT period and present the results in Panel C--F. In with the comment-level emotion findings, in general, we do not observe a remarkable effect of ChatGPT on influencing users' emotions. Collectively, Tables~\ref{tab:table13} and~\ref{tab:table14} indicate that the introduction of ChatGPT does not affect the emotional stance of neutral and middle users. Although neutral authors and the most left-leaning conservative authors shift to liberal positions, exposure to ChatGPT fosters more civil discussion online.
\clearpage

\input{tables/Table2_Author_Level_Polarized}
\input{tables/TableS3_Author_Month_Regression}
\input{tables/SI_TableS3_excl_top10}
\input{tables/Table3_Centrist_Neutral_Leftward}
\input{tables/TableS7_Neutral_Centrist_Effect}
\input{tables/TableS5_Extreme_Regular_Centrist}
\input{tables/Overall_Hostility_Toxicity_AuthorDay}
\input{tables/TableS6_Comment_Emotion_Changes}
\input{tables/TableS8_Author_Emotion_Changes}
\clearpage

\section{Robustness check}
\renewcommand{\thefigure}{\thesection\arabic{figure}}
\renewcommand{\thetable}{\thesection\arabic{table}}
\setcounter{figure}{0}
\setcounter{table}{0}

\subsection{Political Significance of Slant Change}
In Figure~\ref{fig:figure7}, (A) and (B) plot the histograms of political slant of all comments posted by liberal-leaning and conservative-leaning authors prior to ChatGPT's release. In magnitude, (A) shows that a 0.003 decrease in liberal-leaning authors' political slant can push a median-level (5th decile) liberal-leaning comment to the top 3rd decile. (B) illustrates that a 0.003 increase in conservative-leaning authors' political slant can shift a median-level conservative-leaning comment to the top 3rd decile as well. Such results illustrate that the slant changes shown in Table~\ref{tab:table1} are also substantively significant.

In Figure~\ref{fig:hostility_plot}, (A) and (B) depict the histograms of comments' hostility from liberal-leaning and conservative-leaning authors prior to ChatGPT's release, respectively. Similarly, Figure~\ref{fig:toxicity_plot}, (A) and (B) depict the histograms of comments' toxicity from liberal-leaning and conservative-leaning authors prior to ChatGPT's release, respectively.

\subsection{Falsification and Robustness Analyses}
We conduct a series of falsification analyses and robustness checks to ensure that our main finding is not driven by factors other than the release of ChatGPT. For instance, it is possible that there has been an existing trend of political polarization on Reddit, which drives our results, or the polarization was mainly driven by other political events rather than the introduction of ChatGPT, such as the mid-term election on November 8, 2022, which happened 3 weeks before ChatGPT's launch. To address these concerns, we conducted a series of falsification tests by shifting the shock date (ChatGPT's launch) to different times. We report the falsification test results for moving the shock date 1 year, 6 months, and 3 weeks earlier in Table~\ref{tab:table8}, Panels A, B, and C, respectively. None of these results show an existing polarization trend. Specifically, Panels A and B report opposing trends in polarization: liberal-leaning authors became more conservative, and conservative-leaning authors became more liberal before ChatGPT's release. Such results rule out the alternative explanations that either there is a general trend of polarization on Reddit independent of ChatGPT, or that other political events, such as the mid-term election, mainly drive the results. We further stratify authors into active and inactive groups and report consistent results in Table~\ref{tab:falsification_active_inactive}. Together, all these results suggest that the previous findings are unlikely to be driven by alternative factors.

In addition, we identify a short-lived right-leaning shift in user discussions during the three-week post-release period (Figure~\ref{fig:figure3}). This shift likely reflects the contemporaneous external shock of the House January 6 Committee's referral of Donald Trump for criminal prosecution on December 19, 2022. As illustrated in Figure~\ref{fig:figure8}, conservative-leaning expressions became briefly more salient following this event. Importantly, however, the overall degree and direction of polarization remained stable, suggesting that while external political events may perturb the short-term ideological balance, they do not fundamentally affect the long-run polarization dynamics on the platform.

\subsection{Material Significance of Monthly Commenting Change}
The main results in Table~\ref{tab:table5}, Panel B are also materially significant. Figure~\ref{fig:figure9} plots the distribution of monthly comments in each author group in the pre-ChatGPT period. Plot (A) shows that 2.73 more monthly comments can increase the degree of activity of a median-level active liberal-leaning author by over 50\%, which is substantial. Plot (C) suggests a similar result for active conservative-leaning authors.
\clearpage

\begin{figure}
\FIGURE
{\includegraphics[width=0.45\textwidth]{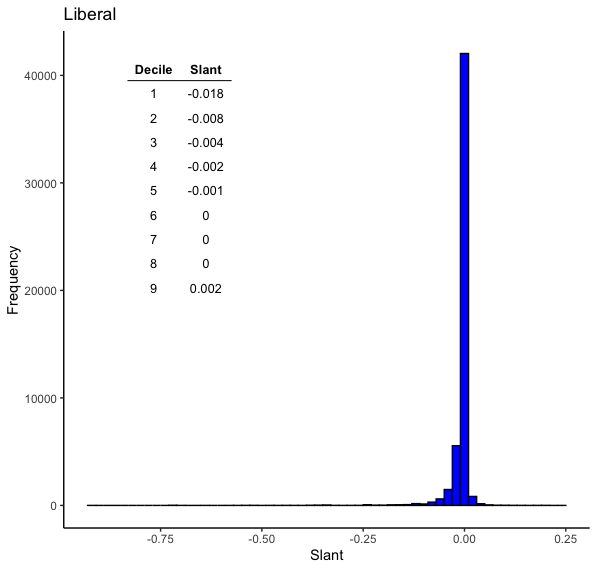}\quad\includegraphics[width=0.45\textwidth]{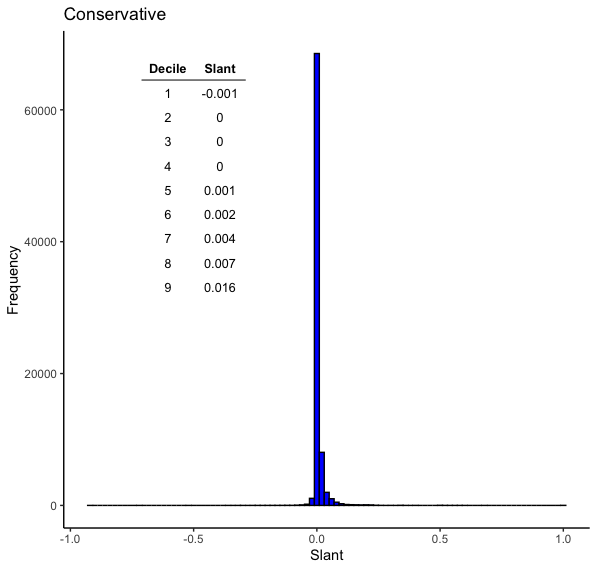}}
{Histogram of Comments' Political Slant on Different Author Groups.\label{fig:figure7}}
{(A) plots the comment-level political-slant histogram for liberal-leaning authors. (B) plots the comment-level political-slant histogram for conservative-leaning authors. Both panels use comments posted prior to ChatGPT's release.}
\end{figure}

\begin{figure}
\FIGURE
{\includegraphics[width=\textwidth]{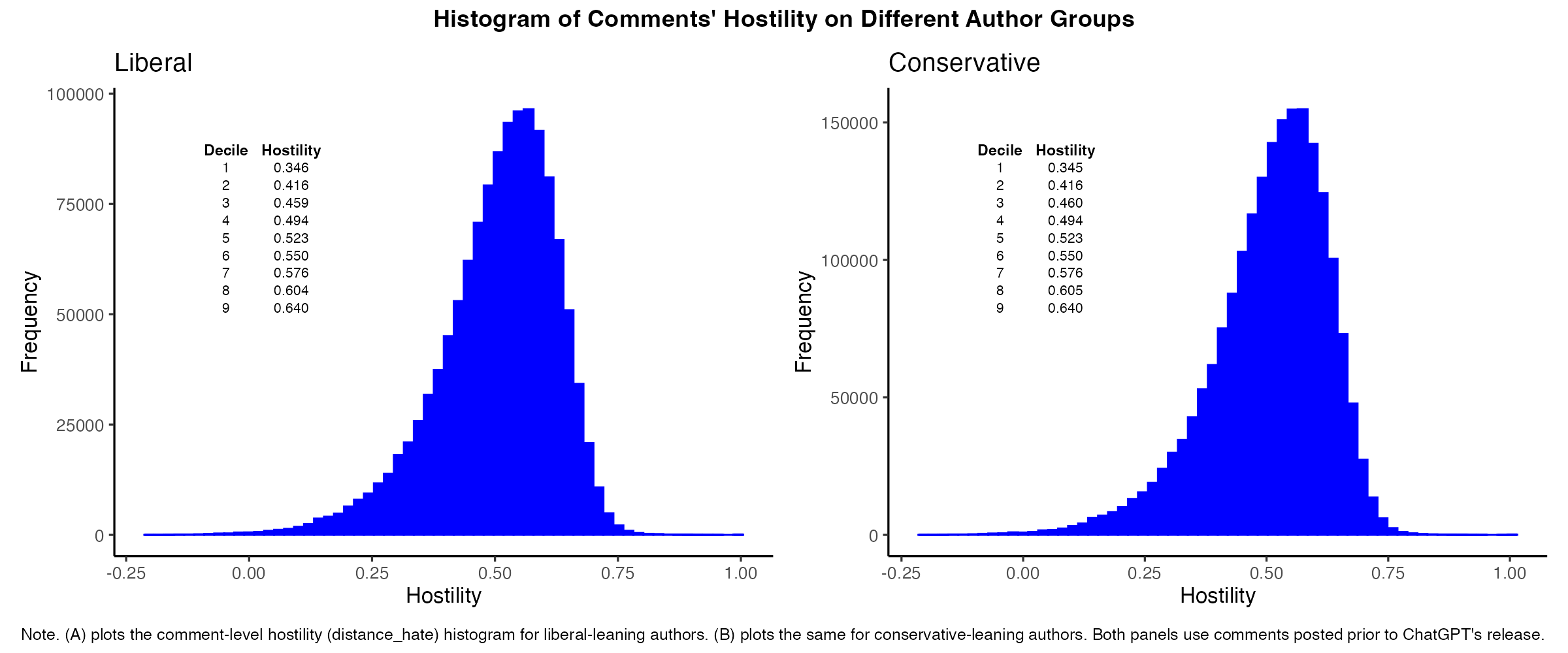}}
{Histogram of Comments' Hostility on Different Author Groups.\label{fig:hostility_plot}}
{(A) plots the comment-level hostility histogram for liberal-leaning authors. (B) plots the comment-level hostility histogram for conservative-leaning authors. Both panels use comments posted prior to ChatGPT's release.}
\end{figure}

\begin{figure}
\FIGURE
{\includegraphics[width=\textwidth]{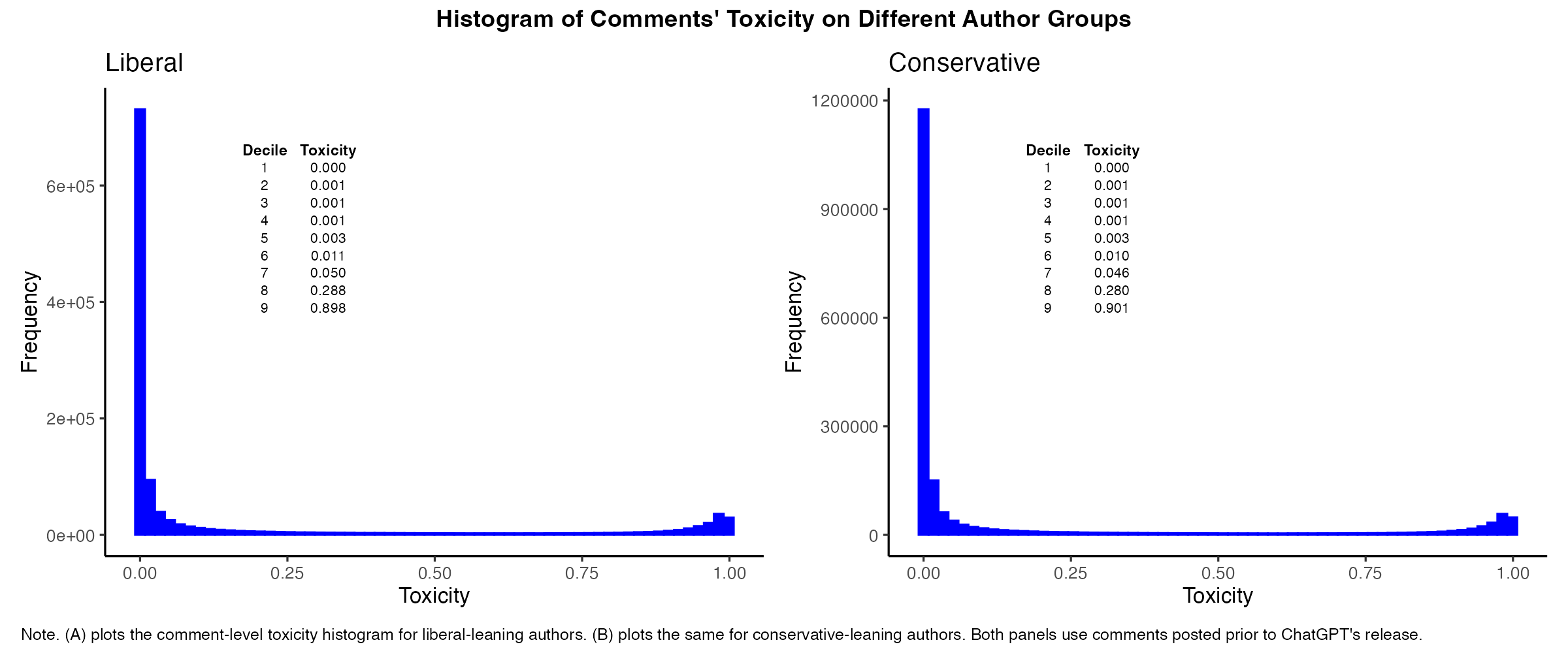}}
{Histogram of Comments' Toxicity on Different Author Groups.\label{fig:toxicity_plot}}
{(A) plots the comment-level toxicity histogram for liberal-leaning authors. (B) plots the comment-level toxicity histogram for conservative-leaning authors. Both panels use comments posted prior to ChatGPT's release.}
\end{figure}

\begin{figure}
\FIGURE
{\includegraphics[width=\textwidth]{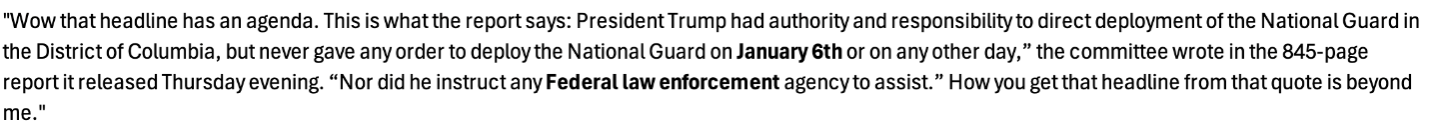}}
{An Example of a Conservative-leaning Comment after December 19, 2022.\label{fig:figure8}}
{Conservative-leaning keywords related to the criminal-prosecution news cycle are highlighted in bold.}
\end{figure}

\begin{figure}
\FIGURE
{\begin{tabular}{cc}
\includegraphics[width=0.45\textwidth]{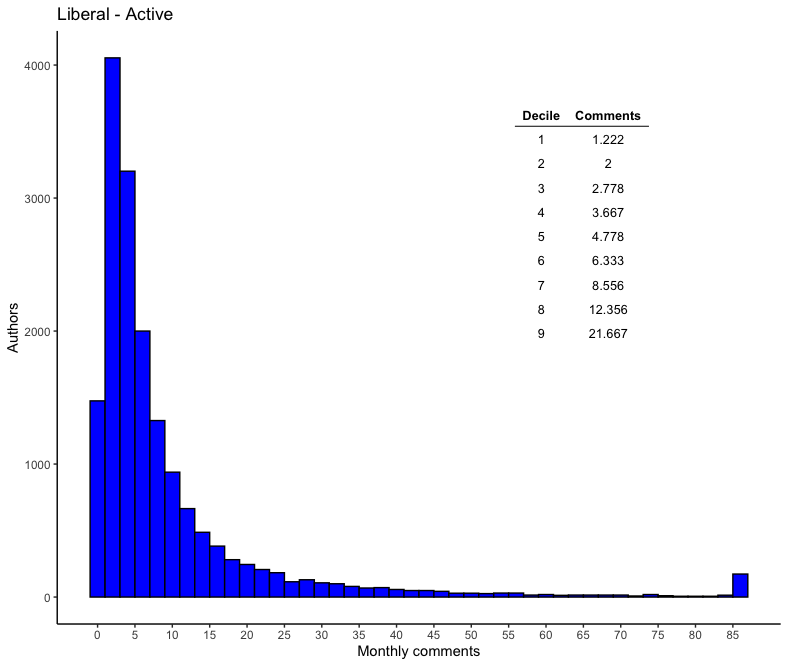} & \includegraphics[width=0.45\textwidth]{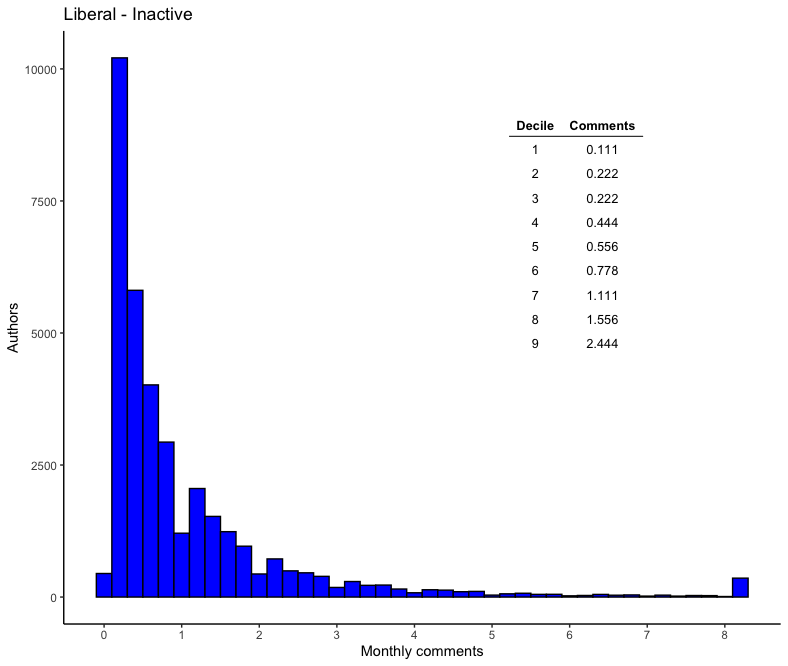} \\[6pt]
\includegraphics[width=0.45\textwidth]{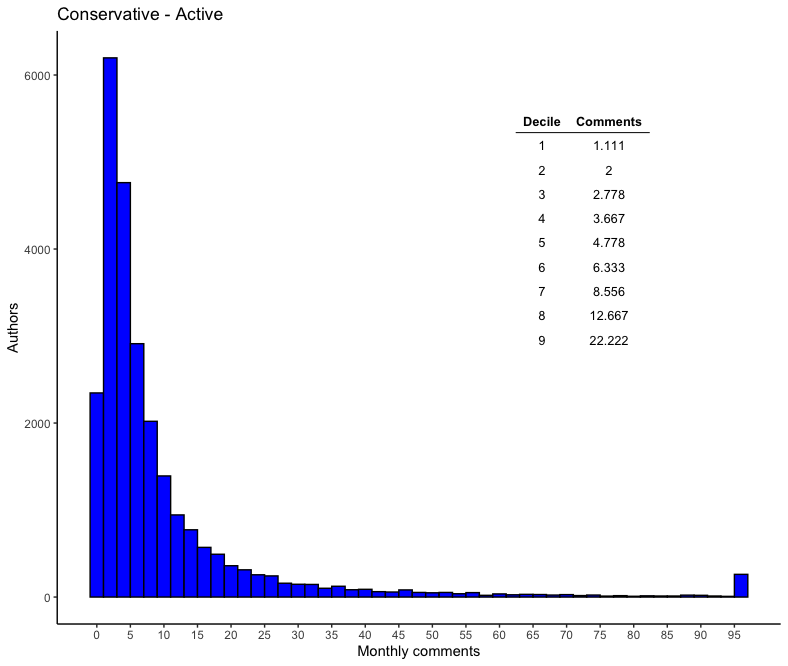} & \includegraphics[width=0.45\textwidth]{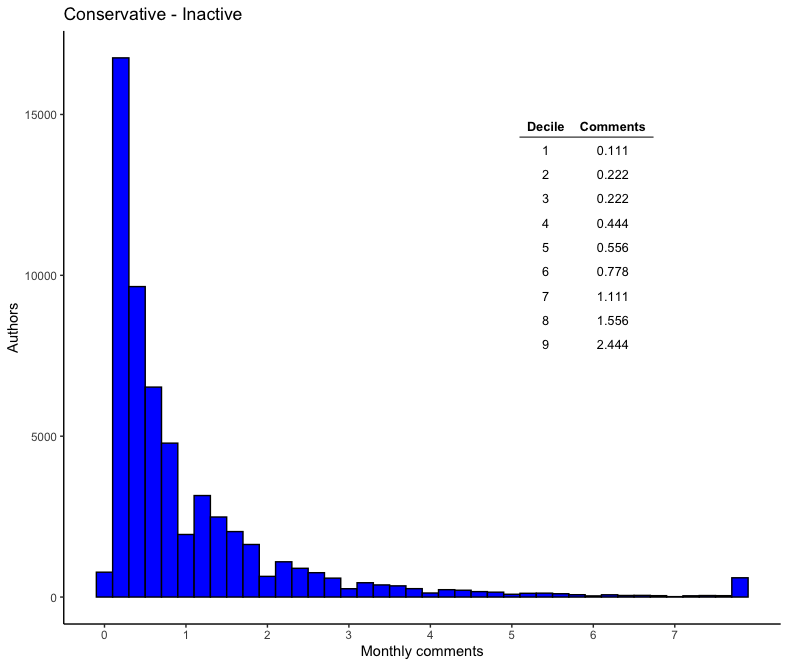}
\end{tabular}}
{Monthly Comments on Different Author Groups.\label{fig:figure9}}
{(A) plots monthly comments posted by active liberal-leaning authors. (B) plots monthly comments posted by inactive liberal-leaning authors. (C) plots monthly comments posted by active conservative-leaning authors. (D) plots monthly comments posted by inactive conservative-leaning authors. Monthly comment counts are winsorized at the 1st and 99th percentiles.}
\end{figure}
\clearpage

\input{tables/TableS4_Falsification_Tests}
\input{tables/TableSX_Falsification_ActiveInactive}
\end{APPENDICES}

\end{document}

%% file: tables/TableS1_Partisan_Phrases.tex
\begin{table}
\TABLE
{Most Partisan Phrases from the 2021 Congressional Record\label{tab:table15}}
{\footnotesize
\begin{tabularx}{\textwidth}{@{}>{\centering\arraybackslash}X>{\centering\arraybackslash}X>{\centering\arraybackslash}X@{}}
\hline
\multicolumn{3}{@{}l@{}}{\up\textit{Panel A: Phrases Used More Frequently by Democrats}} \\ \hline
\multicolumn{3}{@{}l@{}}{\up\textit{Two-Word Phrases}} \\ \hline\up
African American      & sexual assault    & black history       \\
human rights          & black women       & white supremacist   \\
hate crime            & gun violence      & higher education    \\
public health         & sexual harassment & child tax           \\
medical care          & nursing homes     & rights act          \\
justice system        &                   &                     \\ \hline
\multicolumn{3}{@{}l@{}}{\up\textit{Three-Word Phrases}} \\ \hline\up
American rescue plan      & civil rights movement & black lives matter   \\
affordable care act       & public health crisis  & student loan interest \\
voting rights act         & John Lewis voting     & low vaccination rates \\
federal minimum wage      & supreme court ruling  & Federal income tax   \\
criminal justice system   & basic human rights    & child tax credit     \\
public school system      & health care system    &                      \\ \hline
\multicolumn{3}{@{}l@{}}{\up\textit{Panel B: Phrases Used More Frequently by Republicans}} \\ \hline
\multicolumn{3}{@{}l@{}}{\up\textit{Two-Word Phrases}} \\ \hline\up
illegal immigrants    & national security   & illegal immigration  \\
spending bill         & voter id            & national debt        \\
gas prices            & tax rates           & middle east          \\
money laundering      &                     &                      \\ \hline
\multicolumn{3}{@{}l@{}}{\up\textit{Three-Word Phrases}} \\ \hline\up
voter id laws            & Critical Race theory     & school board meeting         \\
federal law enforcement  & national security threat & trillion infrastructure bill \\
Afghan security forces   & capital gains tax        & high gas prices              \\
corporate tax rate       & covid relief bill        & Federal income tax           \\
small business owners    &                          &                              \down\\ \hline
\end{tabularx}}
{}
\end{table}

%% file: tables/Table_NormSlant_AllModels.tex
\begin{table}
\raggedright
\TABLE
{Association of Political Slant on Different ML-Based Measures\label{tab:normslant_allmodels}}
{\footnotesize
\raggedright
\begin{adjustbox}{max width=\textwidth}
\begin{tabular}{@{}lccccccc@{}}
\hline\up
                       & (1)               & (2)                & (3)               & (4)               & (5)              & (6)                & (7)               \\
                       & PoliticalBiasBERT & PolLeaning\_DeBERTa & PolBias\_RoBERTa  & AllSides\_DeBERTa & DeBERTa\_NLI     & DemRep\_DistilBERT & DeBERTaLarge\_NLI \\ \hline\up
Political Slant  & $+0.000263^{***}$ & $+0.000142^{***}$  & $+0.000536^{***}$ & $+0.000713^{***}$ & $+0.000034^{*}$  & $+0.001091^{***}$  & $+0.000415^{***}$ \\
                       & $(0.000013)$      & $(0.000013)$       & $(0.000013)$      & $(0.000013)$      & $(0.000013)$     & $(0.000013)$       & $(0.000013)$      \\
Constant               & $+0.0005^{***}$   & $+0.0005^{***}$    & $+0.0005^{***}$   & $+0.0005^{***}$   & $+0.0005^{***}$  & $+0.0005^{***}$    & $+0.0005^{***}$   \\
                       & $(0.0000)$        & $(0.0000)$         & $(0.0000)$        & $(0.0000)$        & $(0.0000)$       & $(0.0000)$         & $(0.0000)$        \\ \hline
Observations           & 6{,}882{,}091     & 6{,}882{,}091      & 6{,}882{,}091     & 6{,}882{,}091     & 6{,}882{,}091    & 6{,}882{,}091      & 6{,}882{,}091     \\
R-squared              & 0.000057          & 0.000017           & 0.000238          & 0.000422          & 0.000001         & 0.000986           & 0.000143          \down\\ \hline
\end{tabular}
\end{adjustbox}}
{\textit{Notes:} The dependent variable is the comment-level normalized political slant, pooled across liberal, neutral, and conservative authors. Each column reports a separate OLS regression of normalized political slant on the standardized political-slant score from a different ML classifier. Standard errors in parentheses. $^{***}\,p<0.001$, $^{**}\,p<0.01$, $^{*}\,p<0.05$.}
\end{table}

%% file: tables/TableS2_Summary_Statistics.tex
\begin{table}
\TABLE
{Summary Statistics\label{tab:table16}}
{\footnotesize
\raggedright
\begin{adjustbox}{max width=\textwidth}
\begin{tabular}{@{}lcccccccc@{}}
\hline\up
 & N & Mean & Std. Dev. & Min & 25\% & Median & 75\% & Max \\ \hline\up
\emph{Political Slant}      & 6{,}882{,}091 & 0.001  & 0.035  & $-0.924$ & 0      & 0      & 0      & 1      \\
\emph{Comment Length}       & 5{,}604{,}844 & 34.066 & 40.119 & 1        & 11     & 21     & 41     & 461    \\
\emph{ln(Perplexity Score)} & 5{,}604{,}844 & 4.815  & 1.125  & 0.112    & 4.059  & 4.603  & 5.331  & 16.528 \\
\emph{Human Probability}    & 5{,}604{,}844 & 0.286  & 0.305  & 0.0002   & 0.009  & 0.165  & 0.517  & 1      \\
\emph{Hostility}            & 5{,}604{,}844 & 0.504  & 0.121  & $-0.210$ & 0.438  & 0.522  & 0.589  & 1      \\
\emph{Toxicity}             & 5{,}604{,}844 & 0.173  & 0.325  & 0.0003   & 0.001  & 0.003  & 0.112  & 0.999  \\
\emph{Anger Score}          & 5{,}604{,}844 & 0.152  & 0.212  & 0.0004   & 0.016  & 0.051  & 0.194  & 0.995  \\
\emph{Fear Score}           & 5{,}604{,}844 & 0.160  & 0.220  & 0.0001   & 0.016  & 0.057  & 0.211  & 0.992  \\
\emph{Sadness Score}        & 5{,}604{,}844 & 0.036  & 0.127  & 0.0002   & 0.002  & 0.004  & 0.010  & 0.995  \\
\emph{Surprise Score}       & 5{,}604{,}844 & 0.452  & 0.339  & 0.0001   & 0.104  & 0.429  & 0.797  & 0.981  \down\\ \hline
\end{tabular}
\end{adjustbox}}
{}
\end{table}

%% file: tables/Table1_Polarized_Political_Discussion.tex
\begin{table}
\TABLE
{ChatGPT's Release Polarized Political Discussion on Reddit\label{tab:table1}}
{\footnotesize
\raggedright
\begin{tabular}{@{}lccc@{}}
\hline\up
Dependent variable: \emph{Political Slant} & Liberal-leaning authors & Neutral authors & Conservative-leaning authors \\
 & (1) & (2) & (3) \\ \hline\up
After Nov. $\times$ Treatment & $-0.0030^{***}$ & $-0.0003^{***}$ & $0.0026^{***}$  \\
                              & $(0.0001)$      & $(0.0001)$      & $(0.0001)$      \\
Treatment                     & $0.0043^{***}$  & $0.0008^{***}$  & $-0.0020^{***}$ \\
                              & $(0.0001)$      & $(0.0001)$      & $(0.0001)$      \\
After Nov.                    & $0.0029^{***}$  & $0.0002^{**}$   & $-0.0029^{***}$ \\
                              & $(0.0001)$      & $(0.0001)$      & $(0.0001)$      \\ \hline
Author Fixed Effects & Yes & Yes & Yes \\
Observations         & 2{,}179{,}123 & 1{,}231{,}804 & 3{,}471{,}164 \\
R-squared            & 0.0443 & 0.1028 & 0.0396 \down\\ \hline
\end{tabular}}
{\textit{Notes:} This table presents the results of DID regressions estimating the effect of ChatGPT's release on political discussion on Reddit using Equation~(1). Author fixed effects are included in all models. Observations for this regression are at the comment level. Robust standard errors are clustered at the author level and reported in parentheses. $^{***}\,p<0.01$, $^{**}\,p<0.05$, $^{*}\,p<0.1$.}
\end{table}

%% file: tables/Overall_Hostility_Toxicity_Changes.tex
\begin{table}
\TABLE
{Comment Hostility and Toxicity Changes Following the Launch of ChatGPT\label{tab:overall_hostility_toxicity}}
{\footnotesize
\raggedright
\begin{adjustbox}{max width=\textwidth}
\begin{tabular}{@{}lcc@{}}
\hline
\multicolumn{3}{@{}l@{}}{\up\textit{Panel A. Hostility}} \\ \hline\up
Dependent variable: \emph{Hostility} & Liberal-leaning authors & Conservative-leaning authors \\
 & (1) & (2) \\ \hline\up
After Nov. $\times$ Treatment & $-0.0039^{***}$ & $-0.0043^{***}$ \\
                              & $(0.0005)$      & $(0.0004)$      \\
Treatment                     & $-0.0016^{***}$ & $-0.0009^{***}$ \\
                              & $(0.0004)$      & $(0.0003)$      \\
After Nov.                    & $-0.0002$       & $-0.0002$       \\
                              & $(0.0003)$      & $(0.0002)$      \\ \hline
Author Fixed Effects & Yes & Yes \\
Observations         & 2{,}159{,}660 & 3{,}445{,}184 \\
R-squared            & 0.1038 & 0.0999 \\ \hline
\multicolumn{3}{@{}l@{}}{\up\textit{Panel B. Toxicity}} \\ \hline\up
Dependent variable: \emph{Toxicity} & Liberal-leaning authors & Conservative-leaning authors \\
 & (1) & (2) \\ \hline\up
After Nov. $\times$ Treatment & $-0.0022^{*}$ & $-0.0017^{*}$ \\
                              & $(0.0012)$    & $(0.0009)$    \\
Treatment                     & $-0.0003$     & $-0.0006$     \\
                              & $(0.0010)$    & $(0.0008)$    \\
After Nov.                    & $-0.0005$     & $0.0003$      \\
                              & $(0.0008)$    & $(0.0006)$    \\ \hline
Author Fixed Effects & Yes & Yes \\
Observations         & 2{,}159{,}660 & 3{,}445{,}184 \\
R-squared            & 0.1013 & 0.0989 \down\\ \hline
\end{tabular}
\end{adjustbox}}
{\textit{Notes:} These tables present DID regressions estimating the effect of ChatGPT's release on the civility of political discussion, pooling active and inactive authors within each leaning group. Author fixed effects are included in all models. Observations are at the comment level. Robust standard errors are clustered at the author level and reported in parentheses. $^{***}\,p<0.01$, $^{**}\,p<0.05$, $^{*}\,p<0.1$.}
\end{table}

%% file: tables/Table4_User_Linguistic_Changes.tex
\begin{table}
\TABLE
{User Linguistic Changes Following the Launch of ChatGPT\label{tab:table4}}
{\footnotesize
\raggedright
\begin{adjustbox}{max width=\textwidth}
\begin{tabular}{@{}lcccc@{}}
\hline
\multicolumn{5}{@{}l@{}}{\up\textit{Panel A. Comment Length Change}} \\ \hline\up
Dependent variable: \emph{Comment Length} & \multicolumn{2}{c}{Liberal-leaning authors} & \multicolumn{2}{c}{Conservative-leaning authors} \\
\cline{2-3}\cline{4-5}\up
 & Active & Inactive & Active & Inactive \\
 & (1) & (2) & (3) & (4) \\ \hline\up
After Nov. $\times$ Treatment & $0.2187^{*}$    & $3.0641^{***}$  & $-0.0932$       & $2.8601^{***}$  \\
                              & $(0.1245)$      & $(0.2716)$      & $(0.0968)$      & $(0.2073)$      \\
Treatment                     & $-2.5946^{***}$ & $-5.7844^{***}$ & $-2.1600^{***}$ & $-5.3709^{***}$ \\
                              & $(0.0887)$      & $(0.2047)$      & $(0.0685)$      & $(0.1565)$      \\
After Nov.                    & $-0.3534^{***}$ & $-3.2355^{***}$ & $-0.2491^{***}$ & $-2.7178^{***}$ \\
                              & $(0.0762)$      & $(0.1935)$      & $(0.0601)$      & $(0.1483)$      \\ \hline
Author Fixed Effects & Yes & Yes & Yes & Yes \\
Observations         & 1{,}733{,}949 & 425{,}711 & 2{,}762{,}245 & 682{,}939 \\
R-squared            & 0.1921 & 0.2943 & 0.1982 & 0.2969 \\ \hline
\multicolumn{5}{@{}l@{}}{\up\textit{Panel B. Perplexity Score Change}} \\ \hline\up
Dependent variable: \emph{ln(Perplexity Score)} & \multicolumn{2}{c}{Liberal-leaning authors} & \multicolumn{2}{c}{Conservative-leaning authors} \\
\cline{2-3}\cline{4-5}\up
 & Active & Inactive & Active & Inactive \\
 & (1) & (2) & (3) & (4) \\ \hline\up
After Nov. $\times$ Treatment & $-0.0126^{***}$ & $-0.0559^{***}$ & $-0.0078^{***}$ & $-0.0623^{***}$ \\
                              & $(0.0036)$      & $(0.0077)$      & $(0.0028)$      & $(0.0062)$      \\
Treatment                     & $0.0776^{***}$  & $0.1260^{***}$  & $0.0696^{***}$  & $0.1235^{***}$  \\
                              & $(0.0025)$      & $(0.0058)$      & $(0.0020)$      & $(0.0047)$      \\
After Nov.                    & $0.0131^{***}$  & $0.0546^{***}$  & $0.0080^{***}$  & $0.0535^{***}$  \\
                              & $(0.0022)$      & $(0.0055)$      & $(0.0017)$      & $(0.0044)$      \\ \hline
Author Fixed Effects & Yes & Yes & Yes & Yes \\
Observations         & 1{,}733{,}949 & 425{,}711 & 2{,}762{,}245 & 682{,}939 \\
R-squared            & 0.1556 & 0.2263 & 0.1574 & 0.2259 \\ \hline
\multicolumn{5}{@{}l@{}}{\up\textit{Panel C. Human Score Change}} \\ \hline\up
Dependent variable: \emph{Human Probability} & \multicolumn{2}{c}{Liberal-leaning authors} & \multicolumn{2}{c}{Conservative-leaning authors} \\
\cline{2-3}\cline{4-5}\up
 & Active & Inactive & Active & Inactive \\
 & (1) & (2) & (3) & (4) \\ \hline\up
After Nov. $\times$ Treatment & $0.0023^{**}$  & $-0.0102^{***}$ & $0.0049^{***}$ & $-0.0103^{***}$ \\
                              & $(0.0010)$     & $(0.0022)$      & $(0.0008)$     & $(0.0017)$      \\
Treatment                     & $0.0098^{***}$ & $0.0250^{***}$  & $0.0079^{***}$ & $0.0269^{***}$  \\
                              & $(0.0007)$     & $(0.0016)$      & $(0.0006)$     & $(0.0013)$      \\
After Nov.                    & $0.0024^{***}$ & $0.0172^{***}$  & $0.0019^{***}$ & $0.0181^{***}$  \\
                              & $(0.0006)$     & $(0.0015)$      & $(0.0005)$     & $(0.0012)$      \\
Author Fixed Effects & Yes & Yes & Yes & Yes \\
Observations         & 1{,}733{,}949 & 425{,}711 & 2{,}762{,}245 & 682{,}939 \\
R-squared            & 0.0940 & 0.1674 & 0.0933 & 0.1691 \down\\ \hline
\end{tabular}
\end{adjustbox}}
{\textit{Notes:} This table presents the results of DID regressions estimating the effect of ChatGPT's release on comment length (Panel A), perplexity score (Panel B), and human score (Panel C). Author fixed effects are included in all models. Observations for these regressions are at the comment level. Robust standard errors are clustered at the author level and reported in parentheses. $^{***}\,p<0.01$, $^{**}\,p<0.05$, $^{*}\,p<0.1$.}
\end{table}

%% file: tables/Table5_Slant_and_Commenting_Change.tex
\begin{table}
\TABLE
{User Slant and Monthly Commenting Change Following the Launch of ChatGPT\label{tab:table5}}
{\footnotesize
\raggedright
\begin{adjustbox}{max width=\textwidth}
\begin{tabular}{@{}lcccc@{}}
\hline
\multicolumn{5}{@{}l@{}}{\up\textit{Panel A. Slant Change}} \\ \hline\up
Dependent variable: \emph{Political Slant} & \multicolumn{2}{c}{Liberal-leaning authors} & \multicolumn{2}{c}{Conservative-leaning authors} \\
\cline{2-3}\cline{4-5}\up
 & Active & Inactive & Active & Inactive \\
 & (1) & (2) & (3) & (4) \\ \hline\up
After Nov. $\times$ Treatment & $-0.0022^{***}$ & $-0.0082^{***}$ & $0.0019^{***}$  & $0.0072^{***}$  \\
                              & $(0.0001)$      & $(0.0003)$      & $(0.0001)$      & $(0.0002)$      \\
Treatment                     & $0.0035^{***}$  & $0.0095^{***}$  & $-0.0013^{***}$ & $-0.0063^{***}$ \\
                              & $(0.0001)$      & $(0.0002)$      & $(0.0001)$      & $(0.0001)$      \\
After Nov.                    & $0.0021^{***}$  & $0.0084^{***}$  & $-0.0022^{***}$ & $-0.0073^{***}$ \\
                              & $(0.0001)$      & $(0.0002)$      & $(0.0001)$      & $(0.0001)$      \\ \hline
Author Fixed Effects & Yes & Yes & Yes & Yes \\
Observations         & 1{,}733{,}949 & 425{,}711 & 2{,}762{,}245 & 682{,}939 \\
R-squared            & 0.0151 & 0.1598 & 0.0159 & 0.1462 \\ \hline
\multicolumn{5}{@{}l@{}}{\up\textit{Panel B. Commenting Change}} \\ \hline\up
Dependent variable: \emph{Monthly comments} & \multicolumn{2}{c}{Liberal-leaning authors} & \multicolumn{2}{c}{Conservative-leaning authors} \\
\cline{2-3}\cline{4-5}\up
 & Active & Inactive & Active & Inactive \\
 & (1) & (2) & (3) & (4) \\ \hline\up
After Nov. $\times$ Treatment & $2.7295^{***}$  & $-0.0279$       & $2.5995^{***}$  & $0.0461^{***}$  \\
                              & $(0.1727)$      & $(0.0213)$      & $(0.1492)$      & $(0.0175)$      \\
Treatment                     & $-6.4253^{***}$ & $-0.1085^{***}$ & $-6.0424^{***}$ & $-0.1423^{***}$ \\
                              & $(0.1460)$      & $(0.0180)$      & $(0.1261)$      & $(0.0148)$      \\
After Nov.                    & $-5.0758^{***}$ & $-0.2559^{***}$ & $-5.2080^{***}$ & $-0.2845^{***}$ \\
                              & $(0.1221)$      & $(0.0150)$      & $(0.1055)$      & $(0.0124)$      \\ \hline
Author Fixed Effects & Yes & Yes & Yes & Yes \\
Observations         & 235{,}340 & 496{,}916 & 357{,}098 & 809{,}158 \\
R-squared            & 0.3927 & 0.2597 & 0.4267 & 0.2116 \down\\ \hline
\end{tabular}
\end{adjustbox}}
{\textit{Notes:} This table presents the results of DID regressions estimating the effect of ChatGPT's release on different author groups. Author fixed effects are included in all models. Observations for the regression in Panel A are at the comment level, and in Panel B at the author-month level. Robust standard errors are clustered at the author level and reported in parentheses. $^{***}\,p<0.01$, $^{**}\,p<0.05$, $^{*}\,p<0.1$.}
\end{table}

%% file: tables/Table6_Hostility_Toxicity_Changes.tex
\begin{table}
\TABLE
{Comment Hostility and Toxicity Changes Following the Launch of ChatGPT\label{tab:table6}}
{\footnotesize
\raggedright
\begin{adjustbox}{max width=\textwidth}
\begin{tabular}{@{}lcccc@{}}
\hline
\multicolumn{5}{@{}l@{}}{\up\textit{Panel A. Hostility}} \\ \hline\up
Dependent variable: \emph{Hostility} & \multicolumn{2}{c}{Liberal-leaning authors} & \multicolumn{2}{c}{Conservative-leaning authors} \\
\cline{2-3}\cline{4-5}\up
 & Active & Inactive & Active & Inactive \\
 & (1) & (2) & (3) & (4) \\ \hline\up
After Nov. $\times$ Treatment & $-0.0042^{***}$ & $-0.0016^{*}$   & $-0.0049^{***}$ & $-0.0014^{**}$  \\
                              & $(0.0004)$      & $(0.0008)$      & $(0.0003)$      & $(0.0007)$      \\
Treatment                     & $-0.0012^{***}$ & $-0.0041^{***}$ & $-0.0003$       & $-0.0038^{***}$ \\
                              & $(0.0003)$      & $(0.0006)$      & $(0.0002)$      & $(0.0005)$      \\
After Nov.                    & $0.0003$        & $-0.0031^{***}$ & $0.0002$        & $-0.0026^{***}$ \\
                              & $(0.0002)$      & $(0.0006)$      & $(0.0002)$      & $(0.0005)$      \\ \hline
Author Fixed Effects & Yes & Yes & Yes & Yes \\
Observations         & 1{,}733{,}949 & 425{,}711 & 2{,}762{,}245 & 682{,}939 \\
R-squared            & 0.0883 & 0.1659 & 0.0830 & 0.1668 \\ \hline
\multicolumn{5}{@{}l@{}}{\up\textit{Panel B. Toxicity}} \\ \hline\up
Dependent variable: \emph{Toxicity} & \multicolumn{2}{c}{Liberal-leaning authors} & \multicolumn{2}{c}{Conservative-leaning authors} \\
\cline{2-3}\cline{4-5}\up
 & Active & Inactive & Active & Inactive \\
 & (1) & (2) & (3) & (4) \\ \hline\up
After Nov. $\times$ Treatment & $-0.0020^{*}$  & $-0.0025$       & $-0.0017^{**}$ & $-0.0018$       \\
                              & $(0.0011)$     & $(0.0023)$      & $(0.0008)$     & $(0.0019)$      \\
Treatment                     & $0.0003$       & $-0.0036^{**}$  & $-0.0002$      & $-0.0026^{*}$   \\
                              & $(0.0008)$     & $(0.0017)$      & $(0.0006)$     & $(0.0014)$      \\
After Nov.                    & $-0.0005$      & $-0.0010$       & $0.0001$       & $0.0008$        \\
                              & $(0.0007)$     & $(0.0017)$      & $(0.0005)$     & $(0.0013)$      \\ \hline
Author Fixed Effects & Yes & Yes & Yes & Yes \\
Observations         & 1{,}733{,}949 & 425{,}711 & 2{,}762{,}245 & 682{,}939 \\
R-squared            & 0.0848 & 0.1653 & 0.0812 & 0.1651 \down\\ \hline
\end{tabular}
\end{adjustbox}}
{\textit{Notes:} These tables present the results of DID regressions estimating the effect of ChatGPT's release on the civility of political discussion. Author fixed effects are included in all models. Observations for these regressions are at the comment level. Robust standard errors are clustered at the author level and reported in parentheses. $^{***}\,p<0.01$, $^{**}\,p<0.05$, $^{*}\,p<0.1$.}
\end{table}

%% file: tables/Table2_Author_Level_Polarized.tex
\begin{table}
\TABLE
{Author-Level Political Slants Were Also Polarized\label{tab:table2}}
{\footnotesize
\raggedright
\begin{tabular}{@{}lccc@{}}
\hline\up
Dependent variable: \emph{Political Slant} & Liberal-leaning authors & Neutral authors & Conservative-leaning authors \\
 & (1) & (2) & (3) \\ \hline\up
After Nov. $\times$ Treatment & $-0.0038^{***}$ & $-0.0004^{***}$ & $0.0033^{***}$  \\
                              & $(0.0001)$      & $(0.0001)$      & $(0.0001)$      \\
Treatment                     & $0.0049^{***}$  & $0.0010^{***}$  & $-0.0027^{***}$ \\
                              & $(0.0001)$      & $(0.0001)$      & $(0.0001)$      \\
After Nov.                    & $0.0037^{***}$  & $0.0003^{***}$  & $-0.0034^{***}$ \\
                              & $(0.0001)$      & $(0.0001)$      & $(0.0001)$      \\ \hline
Author Fixed Effects & Yes & Yes & Yes \\
Observations         & 923{,}167 & 798{,}896 & 1{,}457{,}083 \\
R-squared            & 0.1170 & 0.1741 & 0.1026 \down\\ \hline
\end{tabular}}
{\textit{Notes:} This table presents the results of DID regressions estimating the effect of ChatGPT's release on political discussion on Reddit using author-day level data. Author fixed effects are included in all models. Robust standard errors are clustered at the author level and reported in parentheses. $^{***}\,p<0.01$, $^{**}\,p<0.05$, $^{*}\,p<0.1$.}
\end{table}

%% file: tables/TableS3_Author_Month_Regression.tex
\begin{table}
\TABLE
{Author-Month Level Regression Results\label{tab:table7}}
{\footnotesize
\raggedright
\begin{tabular}{@{}lccc@{}}
\hline\up
Dependent variable: \emph{Political Slant} & Liberal-leaning authors & Neutral authors & Conservative-leaning authors \\
 & (1) & (2) & (3) \\ \hline\up
After Nov. $\times$ Treatment & $-0.006^{***}$ & $-0.0004^{***}$ & $0.005^{***}$  \\
                              & $(0.0002)$     & $(0.0001)$      & $(0.0002)$     \\
Treatment                     & $0.007^{***}$  & $0.001^{***}$   & $-0.005^{***}$ \\
                              & $(0.0001)$     & $(0.0001)$      & $(0.0001)$     \\
After Nov.                    & $0.006^{***}$  & $0.0003^{**}$   & $-0.005^{***}$ \\
                              & $(0.0001)$     & $(0.0001)$      & $(0.0001)$     \\ \hline
Author Fixed Effects & Yes & Yes & Yes \\
Observations         & 247{,}163 & 487{,}339 & 386{,}986 \\
R-squared            & 0.4122 & 0.3194 & 0.3744 \down\\ \hline
\end{tabular}}
{\textit{Notes:} This table presents the results of DID regressions estimating the effect of ChatGPT's release on political discussion on Reddit using author-month level data. Author fixed effects are included in all models. Robust standard errors are clustered at the author level and reported in parentheses. $^{***}\,p<0.01$, $^{**}\,p<0.05$, $^{*}\,p<0.1$.}
\end{table}

%% file: tables/SI_TableS3_excl_top10.tex
\begin{table}
\TABLE
{Author-Month Level Regression Results, Excluding the Most Prolific 10\% of Authors\label{tab:tableS3excl}}
{\footnotesize
\raggedright
\begin{tabular}{@{}lccc@{}}
\hline\up
Dependent variable: \emph{Political Slant} & Liberal-leaning authors & Neutral authors & Conservative-leaning authors \\
 & (1) & (2) & (3) \\ \hline\up
After Nov. $\times$ Treatment & $-0.007^{***}$ & $-0.0006^{**}$  & $0.006^{***}$  \\
                              & $(0.0003)$     & $(0.0003)$      & $(0.0002)$     \\
Treatment                     & $0.008^{***}$  & $0.001^{***}$   & $-0.005^{***}$ \\
                              & $(0.0002)$     & $(0.0002)$      & $(0.0002)$     \\
After Nov.                    & $0.007^{***}$  & $0.0003^{*}$    & $-0.006^{***}$ \\
                              & $(0.0002)$     & $(0.0002)$      & $(0.0002)$     \\ \hline
Author Fixed Effects & Yes & Yes & Yes \\
Observations         & 197{,}317 & 358{,}635 & 306{,}495 \\
R-squared            & 0.4267 & 0.3850 & 0.3857 \down\\ \hline
\end{tabular}}
{\textit{Notes:} This table replicates Table \ref{tab:table7} after excluding, within each political group, the 10\% of authors with the largest total number of comments. Authors above the group-specific 90th percentile of total comment counts are dropped, and the DID regressions are re-estimated on the remaining author-month observations. Author fixed effects are included in all models. Robust standard errors are clustered at the author level and reported in parentheses. $^{***}\,p<0.01$, $^{**}\,p<0.05$, $^{*}\,p<0.1$.}
\end{table}

%% file: tables/Table3_Centrist_Neutral_Leftward.tex
\begin{table}
\TABLE
{Centrist and Neutral Authors Showed a Slight Leftward Movement\label{tab:table3}}
{\footnotesize
\raggedright
\begin{adjustbox}{max width=\textwidth}
\begin{tabular}{@{}lccccc@{}}
\hline\up
Dependent variable: \emph{Political Slant} & \multicolumn{2}{c}{Liberal-leaning authors} & Neutral authors & \multicolumn{2}{c}{Conservative-leaning authors} \\
\cline{2-3}\cline{5-6}\up
 & 20\% Most Centrist & 10\% Most Centrist &  & 10\% Most Centrist & 20\% Most Centrist \\
 & (1) & (2) & (3) & (4) & (5) \\ \hline\up
After Nov. $\times$ Treatment & $-0.0005^{**}$ & $-0.0004$      & $-0.0004^{***}$ & $-0.001^{**}$   & $-0.0001$       \\
                              & $(0.0002)$     & $(0.0003)$     & $(0.0001)$      & $(0.0002)$      & $(0.0002)$      \\
Treatment                     & $0.0010^{***}$ & $0.0010^{***}$ & $0.0010^{***}$  & $0.001^{***}$   & $0.0005^{***}$  \\
                              & $(0.0001)$     & $(0.0002)$     & $(0.0001)$      & $(0.0002)$      & $(0.0001)$      \\
After Nov.                    & $0.0003^{**}$  & $0.0003$       & $0.0003^{***}$  & $0.0002^{*}$    & $-0.0002$       \\
                              & $(0.0001)$     & $(0.0002)$     & $(0.0001)$      & $(0.0001)$      & $(0.0001)$      \\ \hline
Author Fixed Effects & Yes & Yes & Yes & Yes & Yes \\
Observations         & 244{,}553 & 110{,}788 & 798{,}896 & 176{,}289 & 342{,}972 \\
R-squared            & 0.0410 & 0.0452 & 0.1741 & 0.0528 & 0.0480 \down\\ \hline
\end{tabular}
\end{adjustbox}}
{\textit{Notes:} This table presents the results of difference-in-differences regressions estimating the effect of ChatGPT's release on centrist liberal authors, neutral authors and centrist conservative-leaning authors, respectively. Author fixed effects are included in all models. Observations for this regression are at the author-day level. Robust standard errors are clustered at the author level and reported in parentheses. $^{***}\,p<0.01$, $^{**}\,p<0.05$, $^{*}\,p<0.1$.}
\end{table}

%% file: tables/TableS7_Neutral_Centrist_Effect.tex
\begin{table}
\TABLE
{The Effect of ChatGPT's Release on Neutral and Centrist Authors (Comment-Level)\label{tab:table12}}
{\footnotesize
\raggedright
\begin{adjustbox}{max width=\textwidth}
\begin{tabular}{@{}lccccc@{}}
\hline\up
Dependent variable: \emph{Political Slant} & \multicolumn{2}{c}{Liberal-leaning authors} & Neutral authors & \multicolumn{2}{c}{Conservative-leaning authors} \\
\cline{2-3}\cline{5-6}\up
 & 20\% Most Centrist & 10\% Most Centrist &  & 10\% Most Centrist & 20\% Most Centrist \\
 & (1) & (2) & (3) & (4) & (5) \\ \hline\up
After Nov. $\times$ Treatment & $-0.0002$      & $-0.0002$      & $-0.0003^{***}$ & $-0.0002$      & $0.0001$        \\
                              & $(0.0002)$     & $(0.0002)$     & $(0.0001)$      & $(0.0002)$     & $(0.0001)$      \\
Treatment                     & $0.0010^{***}$ & $0.0009^{***}$ & $0.0008^{***}$  & $0.0006^{***}$ & $0.0004^{***}$  \\
                              & $(0.0001)$     & $(0.0002)$     & $(0.0001)$      & $(0.0001)$     & $(0.0001)$      \\
After Nov.                    & $0.0002^{*}$   & $0.0001$       & $0.0002^{**}$   & $0.0001$       & $-0.0003^{***}$ \\
                              & $(0.0001)$     & $(0.0002)$     & $(0.0001)$      & $(0.0001)$     & $(0.0001)$      \\ \hline
Author Fixed Effects & Yes & Yes & Yes & Yes & Yes \\
Observations         & 592{,}906 & 266{,}111 & 1{,}231{,}804 & 398{,}198 & 795{,}478 \\
R-squared            & 0.0134 & 0.0139 & 0.1028 & 0.0182 & 0.0166 \down\\ \hline
\end{tabular}
\end{adjustbox}}
{\textit{Notes:} This table presents the results of difference-in-differences regressions estimating the effect of ChatGPT's release on neutral and centrist authors. Author fixed effects are included in all models. Observations for this regression are at the comment level. Robust standard errors are clustered at the author level and reported in parentheses. $^{***}\,p<0.01$, $^{**}\,p<0.05$, $^{*}\,p<0.1$.}
\end{table}

%% file: tables/TableS5_Extreme_Regular_Centrist.tex
\begin{table}
\TABLE
{Extreme, Regular and Centrist Authors\label{tab:table9}}
{\footnotesize
\raggedright
\begin{adjustbox}{max width=\textwidth}
\begin{tabular}{@{}lcccccc@{}}
\hline\up
Dependent variable: \emph{Political Slant} & \multicolumn{3}{c}{Liberal-leaning authors} & \multicolumn{3}{c}{Conservative-leaning authors} \\
\cline{2-4}\cline{5-7}\up
 & Extreme & Regular & Centrist & Extreme & Regular & Centrist \\
 & (1) & (2) & (3) & (4) & (5) & (6) \\ \hline\up
After Nov. $\times$ Treatment & $-0.0146^{***}$ & $-0.0025^{***}$ & $-0.0007^{***}$ & $0.0112^{***}$  & $0.0022^{***}$  & $0.0005^{***}$  \\
                              & $(0.0005)$      & $(0.0002)$      & $(0.0001)$      & $(0.0004)$      & $(0.0001)$      & $(0.0001)$      \\
Treatment                     & $0.0173^{***}$  & $0.0039^{***}$  & $0.0014^{***}$  & $-0.0102^{***}$ & $-0.0016^{***}$ & $0.0001^{*}$    \\
                              & $(0.0004)$      & $(0.0001)$      & $(0.0001)$      & $(0.0003)$      & $(0.0001)$      & $(0.0001)$      \\
After Nov.                    & $0.0150^{***}$  & $0.0025^{***}$  & $0.0004^{***}$  & $-0.0115^{***}$ & $-0.0026^{***}$ & $-0.0005^{***}$ \\
                              & $(0.0003)$      & $(0.0001)$      & $(0.0001)$      & $(0.0002)$      & $(0.0001)$      & $(0.0001)$      \\ \hline
Author Fixed Effects & Yes & Yes & Yes & Yes & Yes & Yes \\
Observations         & 303{,}341 & 846{,}210 & 1{,}010{,}109 & 531{,}951 & 1{,}482{,}134 & 1{,}431{,}099 \\
R-squared            & 0.0998 & 0.0166 & 0.0133 & 0.0768 & 0.0141 & 0.0156 \down\\ \hline
\end{tabular}
\end{adjustbox}}
{\textit{Notes:} This table presents the results of difference-in-differences regressions estimating the effect of ChatGPT's release on extreme, regular and centrist authors, respectively. Author fixed effects are included in all models. Observations for this regression are at the comment level. Robust standard errors are clustered at the author level and reported in parentheses. $^{***}\,p<0.01$, $^{**}\,p<0.05$, $^{*}\,p<0.1$.}
\end{table}

%% file: tables/Overall_Hostility_Toxicity_AuthorDay.tex
\begin{table}
\TABLE
{Author-Day Level: Hostility and Toxicity Changes Following the Launch of ChatGPT\label{tab:overall_hostility_toxicity_authorday}}
{\footnotesize
\raggedright
\begin{adjustbox}{max width=\textwidth}
\begin{tabular}{@{}lcc@{}}
\hline
\multicolumn{3}{@{}l@{}}{\up\textit{Panel A. Hostility}} \\ \hline\up
Dependent variable: \emph{Hostility} & Liberal-leaning authors & Conservative-leaning authors \\
 & (1) & (2) \\ \hline\up
After Nov. $\times$ Treatment & $-0.0035^{***}$ & $-0.0037^{***}$ \\
                              & $(0.0005)$      & $(0.0004)$      \\
Treatment                     & $-0.0015^{***}$ & $-0.0008^{***}$ \\
                              & $(0.0004)$      & $(0.0003)$      \\
After Nov.                    & $-0.0009^{***}$ & $-0.0007^{**}$  \\
                              & $(0.0003)$      & $(0.0003)$      \\ \hline
Author Fixed Effects & Yes & Yes \\
Observations         & 923{,}167 & 1{,}457{,}083 \\
R-squared            & 0.1547 & 0.1542 \\ \hline
\multicolumn{3}{@{}l@{}}{\up\textit{Panel B. Toxicity}} \\ \hline\up
Dependent variable: \emph{Toxicity} & Liberal-leaning authors & Conservative-leaning authors \\
 & (1) & (2) \\ \hline\up
After Nov. $\times$ Treatment & $-0.0033^{**}$ & $-0.0033^{***}$ \\
                              & $(0.0014)$     & $(0.0011)$      \\
Treatment                     & $0.0002$       & $0.0010$        \\
                              & $(0.0010)$     & $(0.0008)$      \\
After Nov.                    & $-0.0006$      & $0.0006$        \\
                              & $(0.0009)$     & $(0.0007)$      \\ \hline
Author Fixed Effects & Yes & Yes \\
Observations         & 923{,}167 & 1{,}457{,}083 \\
R-squared            & 0.1562 & 0.1541 \down\\ \hline
\end{tabular}
\end{adjustbox}}
{\textit{Notes:} This table presents the results of DID regressions estimating the effect of ChatGPT's release on incivility using author-day level data. Author fixed effects are included in all models. Robust standard errors are clustered at the author level and reported in parentheses. $^{***}\,p<0.01$, $^{**}\,p<0.05$, $^{*}\,p<0.1$.}
\end{table}

%% file: tables/TableS6_Comment_Emotion_Changes.tex

\begin{table}
\TABLE
{Comment Emotion Changes Following the Launch of ChatGPT\label{tab:table10}}
{\footnotesize
\raggedright
\begin{adjustbox}{max width=\textwidth}
\begin{tabular}{@{}lcccc@{}}

\hline
\multicolumn{5}{@{}l@{}}{\up\textit{Panel A. Anger Score}} \\ \hline\up

Dependent variable: \emph{Anger Score}
& \multicolumn{2}{c}{Liberal-leaning authors}
& \multicolumn{2}{c}{Conservative-leaning authors} \\

\cline{2-3}\cline{4-5}\up
 & Active & Inactive & Active & Inactive \\
 & (1) & (2) & (3) & (4) \\ \hline\up

After Nov. $\times$ Treatment & $-0.0010$       & $-0.0000$       & $-0.0015^{***}$ & $0.0005$        \\
                              & $(0.0007)$      & $(0.0015)$      & $(0.0006)$      & $(0.0012)$      \\
Treatment                     & $-0.0046^{***}$ & $-0.0095^{***}$ & $-0.0052^{***}$ & $-0.0065^{***}$ \\
                              & $(0.0005)$      & $(0.0011)$      & $(0.0004)$      & $(0.0009)$      \\
After Nov.                    & $-0.0008^{*}$   & $-0.0012$       & $0.0003$        & $-0.0018^{**}$  \\
                              & $(0.0004)$      & $(0.0011)$      & $(0.0003)$      & $(0.0009)$      \\ \hline

Author Fixed Effects & Yes & Yes & Yes & Yes \\
Observations         & 1{,}733{,}949 & 425{,}711 & 2{,}762{,}245 & 682{,}939 \\
R-squared            & 0.0459 & 0.1279 & 0.0427 & 0.1289 \\ \hline

\multicolumn{5}{@{}l@{}}{\up\textit{Panel B. Fear Score}} \\ \hline\up

Dependent variable: \emph{Fear Score}
& \multicolumn{2}{c}{Liberal-leaning authors}
& \multicolumn{2}{c}{Conservative-leaning authors} \\

\cline{2-3}\cline{4-5}\up
 & Active & Inactive & Active & Inactive \\
 & (5) & (6) & (7) & (8) \\ \hline\up

After Nov. $\times$ Treatment & $-0.0001$       & $0.0005$        & $-0.0007$       & $0.0024^{*}$    \\
                              & $(0.0007)$      & $(0.0016)$      & $(0.0006)$      & $(0.0013)$      \\
Treatment                     & $-0.0013^{**}$  & $-0.0040^{***}$ & $-0.0010^{**}$  & $-0.0039^{***}$ \\
                              & $(0.0005)$      & $(0.0012)$      & $(0.0004)$      & $(0.0010)$      \\
After Nov.                    & $0.0021^{***}$  & $0.0015$        & $0.0026^{***}$  & $0.0021^{**}$   \\
                              & $(0.0005)$      & $(0.0011)$      & $(0.0004)$      & $(0.0009)$      \\ \hline

Author Fixed Effects & Yes & Yes & Yes & Yes \\
Observations         & 1{,}733{,}949 & 425{,}711 & 2{,}762{,}245 & 682{,}939 \\
R-squared            & 0.0466 & 0.1250 & 0.0433 & 0.1235 \\ \hline

\multicolumn{5}{@{}l@{}}{\up\textit{Panel C. Sadness Score}} \\ \hline\up

Dependent variable: \emph{Sadness Score}
& \multicolumn{2}{c}{Liberal-leaning authors}
& \multicolumn{2}{c}{Conservative-leaning authors} \\

\cline{2-3}\cline{4-5}\up
 & Active & Inactive & Active & Inactive \\
 & (5) & (6) & (7) & (8) \\ \hline\up

After Nov. $\times$ Treatment & $0.0000$        & $0.0003$        & $0.0004$        & $-0.0014^{*}$   \\
                              & $(0.0004)$      & $(0.0010)$      & $(0.0003)$      & $(0.0008)$      \\
Treatment                     & $0.0033^{***}$  & $0.0057^{***}$  & $0.0027^{***}$  & $0.0053^{***}$  \\
                              & $(0.0003)$      & $(0.0007)$      & $(0.0002)$      & $(0.0006)$      \\
After Nov.                    & $-0.0008^{***}$ & $-0.0002$       & $-0.0007^{***}$ & $0.0002$        \\
                              & $(0.0003)$      & $(0.0007)$      & $(0.0002)$      & $(0.0006)$      \\ \hline

Author Fixed Effects & Yes & Yes & Yes & Yes \\
Observations         & 1{,}733{,}949 & 425{,}711 & 2{,}762{,}245 & 682{,}939 \\
R-squared            & 0.0280 & 0.1004 & 0.0271 & 0.1034 \down\\ \hline

\end{tabular}
\end{adjustbox}}
{\textit{Notes:} These tables present the results of difference-in-differences regressions estimating the effect of ChatGPT's release on the emotions of political discussion. Author fixed effects are included in all models. Observations for these regressions are at the comment level. Robust standard errors are clustered at the author level and reported in parentheses. $^{***}\,p<0.01$, $^{**}\,p<0.05$, $^{*}\,p<0.1$.}
\end{table}

\begin{table}
\TABLE
{Comment Emotion Changes Following the Launch of ChatGPT (Continued)\label{tab:table11}}
{\footnotesize
\raggedright
\begin{adjustbox}{max width=\textwidth}
\begin{tabular}{@{}lcccc@{}}

\hline
\multicolumn{5}{@{}l@{}}{\up\textit{Panel D. Surprise Score}} \\ \hline\up

Dependent variable: \emph{Surprise Score}
& \multicolumn{2}{c}{Liberal-leaning authors}
& \multicolumn{2}{c}{Conservative-leaning authors} \\

\cline{2-3}\cline{4-5}\up
 & Active & Inactive & Active & Inactive \\
 & (1) & (2) & (3) & (4) \\ \hline\up

After Nov. $\times$ Treatment & $0.0015$        & $-0.0022$       & $0.0037^{***}$  & $-0.0003$       \\
                              & $(0.0011)$      & $(0.0024)$      & $(0.0009)$      & $(0.0019)$      \\
Treatment                     & $0.0011$        & $0.0049^{***}$  & $0.0015^{**}$   & $0.0028^{*}$    \\
                              & $(0.0008)$      & $(0.0018)$      & $(0.0006)$      & $(0.0015)$      \\
After Nov.                    & $0.0010$        & $0.0036^{**}$   & $-0.0000$       & $0.0020$        \\
                              & $(0.0007)$      & $(0.0017)$      & $(0.0006)$      & $(0.0014)$      \\ \hline

Author Fixed Effects & Yes & Yes & Yes & Yes \\
Observations         & 1{,}733{,}949 & 425{,}711 & 2{,}762{,}245 & 682{,}939 \\
R-squared            & 0.0589 & 0.1377 & 0.0536 & 0.1378 \down\\ \hline

\end{tabular}
\end{adjustbox}}
{\textit{Notes:} Continuation of the previous table. Author fixed effects are included. Observations are at the comment level. Robust standard errors are clustered at the author level and reported in parentheses. $^{***}\,p<0.01$, $^{**}\,p<0.05$, $^{*}\,p<0.1$.}
\end{table}

%% file: tables/TableS8_Author_Emotion_Changes.tex

\begin{table}
\TABLE
{Author Emotion Changes Following the Launch of ChatGPT\label{tab:table13}}
{\footnotesize
\raggedright
\begin{adjustbox}{max width=\textwidth}
\begin{tabular}{@{}lccccc@{}}

\hline
\multicolumn{6}{@{}l@{}}{\up\textit{Panel A: Hostility}} \\ \hline\up
Dependent variable: \emph{Hostility} & \multicolumn{2}{c}{Liberal-leaning authors} & Neutral authors & \multicolumn{2}{c}{Conservative-leaning authors} \\
\cline{2-3}\cline{5-6}\up
 & Right-leaning (top 20\%) & Right-leaning (top 10\%) &  & Left-leaning (top 10\%) & Left-leaning (top 20\%) \\
 & (1) & (2) & (3) & (4) & (5) \\ \hline\up
After Nov. $\times$ Treatment & $-0.0040^{***}$ & $-0.0043^{***}$ & $-0.0128^{***}$ & $-0.0052^{***}$ & $-0.0042^{***}$ \\
                              & $(0.0010)$      & $(0.0015)$      & $(0.0008)$      & $(0.0012)$      & $(0.0009)$      \\
Treatment                     & $-0.0010$       & $-0.0016$       & $0.0056^{***}$  & $0.0001$        & $-0.0002$       \\
                              & $(0.0007)$      & $(0.0011)$      & $(0.0006)$      & $(0.0009)$      & $(0.0006)$      \\
After Nov.                    & $-0.0001$       & $-0.0009$       & $0.0065^{***}$  & $-0.0002$       & $-0.0001$       \\
                              & $(0.0006)$      & $(0.0010)$      & $(0.0006)$      & $(0.0008)$      & $(0.0005)$      \\ \hline
Author Fixed Effects & Yes & Yes & Yes & Yes & Yes \\
Observations         & 244{,}553 & 110{,}788 & 798{,}896 & 176{,}209 & 342{,}972 \\
R-squared            & 0.1424 & 0.1470 & 0.2618 & 0.1410 & 0.1420 \\ \hline

\multicolumn{6}{@{}l@{}}{\up\textit{Panel B: Toxicity}} \\ \hline\up
Dependent variable: \emph{Toxicity} & \multicolumn{2}{c}{Liberal-leaning authors} & Neutral authors & \multicolumn{2}{c}{Conservative-leaning authors} \\
\cline{2-3}\cline{5-6}\up
 & Right-leaning (top 20\%) & Right-leaning (top 10\%) &  & Left-leaning (top 10\%) & Left-leaning (top 20\%) \\
 & (1) & (2) & (3) & (4) & (5) \\ \hline\up
After Nov. $\times$ Treatment & $-0.0003$       & $-0.0034$       & $-0.0105^{***}$ & $-0.0073^{**}$  & $-0.0052^{**}$  \\
                              & $(0.0027)$      & $(0.0040)$      & $(0.0020)$      & $(0.0032)$      & $(0.0023)$      \\
Treatment                     & $-0.0020$       & $-0.0004$       & $0.0022$        & $0.0020$        & $0.0022$        \\
                              & $(0.0020)$      & $(0.0030)$      & $(0.0015)$      & $(0.0024)$      & $(0.0017)$      \\
After Nov.                    & $-0.0017$       & $-0.0006$       & $0.0060^{***}$  & $0.0009$        & $0.0002$        \\
                              & $(0.0017)$      & $(0.0025)$      & $(0.0015)$      & $(0.0021)$      & $(0.0015)$      \\ \hline
Author Fixed Effects & Yes & Yes & Yes & Yes & Yes \\
Observations         & 244{,}553 & 110{,}788 & 798{,}896 & 176{,}209 & 342{,}972 \\
R-squared            & 0.1412 & 0.1409 & 0.3189 & 0.1481 & 0.1454 \\ \hline

\multicolumn{6}{@{}l@{}}{\up\textit{Panel C: Anger}} \\ \hline\up
Dependent variable: \emph{Anger Score} & \multicolumn{2}{c}{Liberal-leaning authors} & Neutral authors & \multicolumn{2}{c}{Conservative-leaning authors} \\
\cline{2-3}\cline{5-6}\up
 & Right-leaning (top 20\%) & Right-leaning (top 10\%) &  & Left-leaning (top 10\%) & Left-leaning (top 20\%) \\
 & (1) & (2) & (3) & (4) & (5) \\ \hline\up
After Nov. $\times$ Treatment & $-0.0005$       & $-0.0038$       & $-0.0069^{***}$ & $-0.0040^{*}$  & $-0.0034^{**}$ \\
                              & $(0.0017)$      & $(0.0025)$      & $(0.0013)$      & $(0.0021)$     & $(0.0015)$     \\
Treatment                     & $-0.0049^{***}$ & $-0.0032^{*}$   & $-0.0014$       & $-0.0026^{*}$  & $-0.0019^{*}$  \\
                              & $(0.0013)$      & $(0.0019)$      & $(0.0010)$      & $(0.0015)$     & $(0.0011)$     \\
After Nov.                    & $-0.0006$       & $0.0010$        & $0.0035^{***}$  & $0.0012$       & $-0.0002$      \\
                              & $(0.0011)$      & $(0.0016)$      & $(0.0009)$      & $(0.0014)$     & $(0.0010)$     \\ \hline
Author Fixed Effects & Yes & Yes & Yes & Yes & Yes \\
Observations         & 244{,}553 & 110{,}788 & 798{,}896 & 176{,}209 & 342{,}972 \\
R-squared            & 0.0893 & 0.0932 & 0.2896 & 0.0992 & 0.0959 \down\\ \hline

\end{tabular}
\end{adjustbox}}
{\textit{Notes:} These tables present the results of difference-in-differences regressions estimating the effect of ChatGPT's release on the emotions of authors. Author fixed effects are included in all models. Observations for these regressions are at the author level. Robust standard errors are clustered at the author level and reported in parentheses. $^{***}\,p<0.01$, $^{**}\,p<0.05$, $^{*}\,p<0.1$.}
\end{table}

\begin{table}
\TABLE
{Author Emotion Changes Following the Launch of ChatGPT (Continued)\label{tab:table14}}
{\footnotesize
\raggedright
\begin{adjustbox}{max width=\textwidth}
\begin{tabular}{@{}lccccc@{}}

\hline
\multicolumn{6}{@{}l@{}}{\up\textit{Panel D: Fear}} \\ \hline\up
Dependent variable: \emph{Fear Score} & \multicolumn{2}{c}{Liberal-leaning authors} & Neutral authors & \multicolumn{2}{c}{Conservative-leaning authors} \\
\cline{2-3}\cline{5-6}\up
 & Right-leaning (top 20\%) & Right-leaning (top 10\%) &  & Left-leaning (top 10\%) & Left-leaning (top 20\%) \\
 & (1) & (2) & (3) & (4) & (5) \\ \hline\up
After Nov. $\times$ Treatment & $0.0007$        & $0.0017$        & $-0.0035^{***}$ & $-0.0034$       & $-0.0015$       \\
                              & $(0.0018)$      & $(0.0026)$      & $(0.0013)$      & $(0.0022)$      & $(0.0015)$      \\
Treatment                     & $-0.0004$       & $-0.0002$       & $-0.0003$       & $-0.0002$       & $-0.0004$       \\
                              & $(0.0013)$      & $(0.0019)$      & $(0.0010)$      & $(0.0015)$      & $(0.0011)$      \\
After Nov.                    & $0.0020^{*}$    & $0.0015$        & $0.0048^{***}$  & $0.0043^{***}$  & $0.0035^{***}$  \\
                              & $(0.0011)$      & $(0.0017)$      & $(0.0010)$      & $(0.0014)$      & $(0.0010)$      \\ \hline
Author Fixed Effects & Yes & Yes & Yes & Yes & Yes \\
Observations         & 244{,}553 & 110{,}788 & 798{,}896 & 176{,}209 & 342{,}972 \\
R-squared            & 0.0921 & 0.0955 & 0.2830 & 0.0930 & 0.0948 \\ \hline

\multicolumn{6}{@{}l@{}}{\up\textit{Panel E: Sadness}} \\ \hline\up
Dependent variable: \emph{Sadness Score} & \multicolumn{2}{c}{Liberal-leaning authors} & Neutral authors & \multicolumn{2}{c}{Conservative-leaning authors} \\
\cline{2-3}\cline{5-6}\up
 & Right-leaning (top 20\%) & Right-leaning (top 10\%) &  & Left-leaning (top 10\%) & Left-leaning (top 20\%) \\
 & (1) & (2) & (3) & (4) & (5) \\ \hline\up
After Nov. $\times$ Treatment & $0.0002$        & $0.0013$        & $0.0034^{***}$  & $0.0022^{*}$    & $0.0018^{**}$   \\
                              & $(0.0010)$      & $(0.0015)$      & $(0.0009)$      & $(0.0013)$      & $(0.0009)$      \\
Treatment                     & $0.0023^{***}$  & $0.0020^{*}$    & $0.0014^{**}$   & $0.0021^{**}$   & $0.0015^{**}$   \\
                              & $(0.0008)$      & $(0.0012)$      & $(0.0007)$      & $(0.0009)$      & $(0.0007)$      \\
After Nov.                    & $-0.0010$       & $-0.0010$       & $-0.0038^{***}$ & $-0.0019^{**}$  & $-0.0022^{***}$ \\
                              & $(0.0006)$      & $(0.0010)$      & $(0.0006)$      & $(0.0008)$      & $(0.0006)$      \\ \hline
Author Fixed Effects & Yes & Yes & Yes & Yes & Yes \\
Observations         & 244{,}553 & 110{,}788 & 798{,}896 & 176{,}209 & 342{,}972 \\
R-squared            & 0.0682 & 0.0733 & 0.2906 & 0.0744 & 0.0756 \\ \hline

\multicolumn{6}{@{}l@{}}{\up\textit{Panel F: Surprise}} \\ \hline\up
Dependent variable: \emph{Surprise Score} & \multicolumn{2}{c}{Liberal-leaning authors} & Neutral authors & \multicolumn{2}{c}{Conservative-leaning authors} \\
\cline{2-3}\cline{5-6}\up
 & Right-leaning (top 20\%) & Right-leaning (top 10\%) &  & Left-leaning (top 10\%) & Left-leaning (top 20\%) \\
 & (1) & (2) & (3) & (4) & (5) \\ \hline\up
After Nov. $\times$ Treatment & $-0.0054^{**}$ & $-0.0044$      & $0.0039^{**}$ & $0.0072^{**}$  & $0.0047^{**}$  \\
                              & $(0.0027)$     & $(0.0041)$     & $(0.0020)$    & $(0.0033)$     & $(0.0023)$     \\
Treatment                     & $0.0049^{**}$  & $0.0063^{**}$  & $0.0000$      & $-0.0034$      & $-0.0014$      \\
                              & $(0.0020)$     & $(0.0030)$     & $(0.0015)$    & $(0.0023)$     & $(0.0017)$     \\
After Nov.                    & $0.0024$       & $0.0020$       & $-0.0003$     & $-0.0010$      & $-0.0001$      \\
                              & $(0.0017)$     & $(0.0025)$     & $(0.0014)$    & $(0.0021)$     & $(0.0015)$     \\ \hline
Author Fixed Effects & Yes & Yes & Yes & Yes & Yes \\
Observations         & 244{,}553 & 110{,}788 & 798{,}896 & 176{,}209 & 342{,}972 \\
R-squared            & 0.1049 & 0.1080 & 0.2982 & 0.1079 & 0.1070 \down\\ \hline

\end{tabular}
\end{adjustbox}}
{\textit{Notes:} Continuation of the previous table. Author fixed effects are included. Observations are at the author level. Robust standard errors are clustered at the author level and reported in parentheses. $^{***}\,p<0.01$, $^{**}\,p<0.05$, $^{*}\,p<0.1$.}
\end{table}

%% file: tables/TableS4_Falsification_Tests.tex
\begin{table}
\TABLE
{Falsification Tests\label{tab:table8}}
{\footnotesize
\raggedright
\begin{adjustbox}{max width=\textwidth}
\begin{tabular}{@{}lccc@{}}

\hline
\multicolumn{4}{@{}l@{}}{\up\textit{Panel A. Move the Treatment Date to One Year Before}} \\ \hline\up
Dependent variable: \emph{Political Slant} & Liberal-leaning authors & Neutral authors & Conservative-leaning authors \\
 & (1) & (2) & (3) \\ \hline\up
After Nov. $\times$ Treatment & $0.0033^{***}$  & $0.0003^{***}$  & $-0.0030^{***}$ \\
                              & $(0.0001)$      & $(0.0001)$      & $(0.0001)$      \\
Treatment                     & $-0.0049^{***}$ & $-0.0012^{***}$ & $0.0019^{***}$  \\
                              & $(0.0001)$      & $(0.0001)$      & $(0.0000)$      \\
After Nov.                    & $-0.0003^{***}$ & $-0.0001^{***}$ & $-0.0001^{*}$   \\
                              & $(0.0000)$      & $(0.0000)$      & $(0.0000)$      \\ \hline
Author Fixed Effects & Yes & Yes & Yes \\
Observations         & 4{,}338{,}597 & 2{,}834{,}298 & 7{,}008{,}890 \\
R-squared            & 0.0281 & 0.0510 & 0.0252 \\ \hline

\multicolumn{4}{@{}l@{}}{\up\textit{Panel B. Move the Treatment Date to Six Months Before}} \\ \hline\up
Dependent variable: \emph{Political Slant} & Liberal-leaning authors & Neutral authors & Conservative-leaning authors \\
 & (1) & (2) & (3) \\ \hline\up
After Nov. $\times$ Treatment & $0.0036^{***}$  & $0.0010^{***}$  & $-0.0020^{***}$ \\
                              & $(0.0001)$      & $(0.0001)$      & $(0.0001)$      \\
Treatment                     & $-0.0004^{***}$ & $-0.0003^{***}$ & $-0.0003^{***}$ \\
                              & $(0.0001)$      & $(0.0001)$      & $(0.0001)$      \\
After Nov.                    & $-0.0033^{***}$ & $-0.0009^{***}$ & $0.0018^{***}$  \\
                              & $(0.0001)$      & $(0.0001)$      & $(0.0000)$      \\ \hline
Author Fixed Effects & Yes & Yes & Yes \\
Observations         & 2{,}750{,}952 & 1{,}324{,}212 & 4{,}333{,}385 \\
R-squared            & 0.0252 & 0.0964 & 0.0278 \\ \hline

\multicolumn{4}{@{}l@{}}{\up\textit{Panel C. Move the Treatment Date to Midterm Election Date}} \\ \hline\up
Dependent variable: \emph{Political Slant} & Liberal-leaning authors & Neutral authors & Conservative-leaning authors \\
 & (1) & (2) & (3) \\ \hline\up
After Nov. $\times$ Treatment & $-0.0009^{***}$ & $-0.0000$       & $-0.0004^{**}$  \\
                              & $(0.0002)$      & $(0.0002)$      & $(0.0002)$      \\
Treatment                     & $0.0051^{***}$  & $0.0006^{***}$  & $-0.0020^{***}$ \\
                              & $(0.0002)$      & $(0.0002)$      & $(0.0001)$      \\
After Nov.                    & $0.0009^{***}$  & $-0.0001$       & $0.0003^{***}$  \\
                              & $(0.0002)$      & $(0.0002)$      & $(0.0001)$      \\ \hline
Author Fixed Effects & Yes & Yes & Yes \\
Observations         & 663{,}585 & 359{,}127 & 1{,}050{,}364 \\
R-squared            & 0.0984 & 0.2054 & 0.0853 \down\\ \hline

\end{tabular}
\end{adjustbox}}
{\textit{Notes:} These tables present the results of a set of falsification tests. Panel A presents the results when moving the date of the treatment to one year earlier, and Panel B reports the results when moving the date of the treatment to six months earlier. Panel C reports the results when using the mid-term election as the treatment. Author fixed effects are included in all models. Observations for these regressions are at the comment level. Robust standard errors are clustered at the author level and reported in parentheses. $^{***}\,p<0.01$, $^{**}\,p<0.05$, $^{*}\,p<0.1$.}
\end{table}

%% file: tables/TableSX_Falsification_ActiveInactive.tex
\begin{table}
\TABLE
{Falsification Tests by Active vs.\ Inactive Authors\label{tab:falsification_active_inactive}}
{\footnotesize
\raggedright
\begin{adjustbox}{max width=\textwidth}
\begin{tabular}{@{}lcccc@{}}

\hline
\multicolumn{5}{@{}l@{}}{\up\textit{Panel A. Move the Treatment Date to Six Months Before}} \\ \hline\up
Dependent variable: \emph{Political Slant}
& \multicolumn{2}{c}{Liberal-leaning authors}
& \multicolumn{2}{c}{Conservative-leaning authors} \\
\cline{2-3}\cline{4-5}\up
 & Active & Inactive & Active & Inactive \\
 & (1) & (2) & (3) & (4) \\ \hline\up

After Nov. $\times$ Treatment & $0.0028^{***}$  & $0.0091^{***}$  & $-0.0014^{***}$ & $-0.0060^{***}$ \\
                              & $(0.0001)$      & $(0.0003)$      & $(0.0001)$      & $(0.0002)$      \\
Treatment                     & $-0.0004^{***}$ & $-0.0005^{**}$  & $-0.0003^{***}$ & $-0.0003^{**}$  \\
                              & $(0.0001)$      & $(0.0002)$      & $(0.0001)$      & $(0.0001)$      \\
After Nov.                    & $-0.0025^{***}$ & $-0.0090^{***}$ & $0.0013^{***}$  & $0.0059^{***}$  \\
                              & $(0.0001)$      & $(0.0002)$      & $(0.0001)$      & $(0.0001)$      \\ \hline

Author Fixed Effects & Yes & Yes & Yes & Yes \\
Observations         & 2{,}252{,}610 & 498{,}342 & 3{,}546{,}907 & 786{,}478 \\
R-squared            & 0.0117 & 0.0821 & 0.0151 & 0.0894 \\ \hline

\multicolumn{5}{@{}l@{}}{\up\textit{Panel B. Move the Treatment Date to Midterm Election Date}} \\ \hline\up
Dependent variable: \emph{Political Slant}
& \multicolumn{2}{c}{Liberal-leaning authors}
& \multicolumn{2}{c}{Conservative-leaning authors} \\
\cline{2-3}\cline{4-5}\up
 & Active & Inactive & Active & Inactive \\
 & (1) & (2) & (3) & (4) \\ \hline\up

After Nov. $\times$ Treatment & $-0.0010^{***}$ & $-0.0001$       & $-0.0004^{**}$  & $-0.0003$       \\
                              & $(0.0003)$      & $(0.0007)$      & $(0.0002)$      & $(0.0005)$      \\
Treatment                     & $0.0043^{***}$  & $0.0099^{***}$  & $-0.0015^{***}$ & $-0.0056^{***}$ \\
                              & $(0.0002)$      & $(0.0006)$      & $(0.0002)$      & $(0.0003)$      \\
After Nov.                    & $0.0010^{***}$  & $0.0003$        & $0.0003^{**}$   & $0.0003$        \\
                              & $(0.0002)$      & $(0.0006)$      & $(0.0001)$      & $(0.0004)$      \\ \hline

Author Fixed Effects & Yes & Yes & Yes & Yes \\
Observations         & 530{,}255 & 133{,}330 & 835{,}731 & 214{,}633 \\
R-squared            & 0.0338 & 0.3126 & 0.0330 & 0.2876 \down\\ \hline

\end{tabular}
\end{adjustbox}}
{\textit{Notes:} This table replicates the falsification tests in Table~S4 separately for active and inactive authors within each partisan leaning. Panel A moves the treatment date to six months before the release of ChatGPT (placebo cutoff: May 30, 2022). Panel B moves the treatment date to the U.S.\ midterm election (placebo cutoff: November 8, 2022). The midterm window is truncated at November 30, 2022 to exclude the actual ChatGPT release. Author fixed effects are included in all models. Observations for these regressions are at the comment level. Robust standard errors are clustered at the author level and reported in parentheses. $^{***}\,p<0.01$, $^{**}\,p<0.05$, $^{*}\,p<0.1$.}
\end{table}

%% file: reference.bib
@article{noy2023experimental,
  title={Experimental evidence on the productivity effects of generative artificial intelligence},
  author={Noy, Shakked and Zhang, Whitney},
  journal={Science},
  volume={381},
  number={6654},
  pages={187--192},
  year={2023},
  publisher={American Association for the Advancement of Science}
}

@article{bastani2025generative,
  title={Generative AI without guardrails can harm learning: Evidence from high school mathematics},
  author={Bastani, Hamsa and Bastani, Osbert and Sungu, Alp and Ge, Haosen and Kabakc{\i}, {\"O}zge and Mariman, Rei},
  journal={Proceedings of the National Academy of Sciences},
  volume={122},
  number={26},
  pages={e2422633122},
  year={2025}, 
  publisher={National Academy of Sciences}
}

@article{zoller2025human,
  title={Human--AI collectives most accurately diagnose clinical vignettes},
  author={Z{\"o}ller, Nikolas and Berger, Julian and Lin, Irving and Fu, Nathan and Komarneni, Jayanth and Barabucci, Gioele and Laskowski, Kyle and Shia, Victor and Harack, Benjamin and Chu, Eugene A and others},
  journal={Proceedings of the National Academy of Sciences},
  volume={122},
  number={24},
  pages={e2426153122},
  year={2025},
  publisher={National Academy of Sciences}
}

@article{brynjolfsson2025generative,
  title={Generative AI at work},
  author={Brynjolfsson, Erik and Li, Danielle and Raymond, Lindsey},
  journal={The Quarterly Journal of Economics},
  volume={140},
  number={2},
  pages={889--942},
  year={2025},
  publisher={Oxford University Press}
}

@article{rutinowski2024self,
  title={The self-perception and political biases of ChatGPT},
  author={Rutinowski, J{\'e}r{\^o}me and Franke, Sven and Endendyk, Jan and Dormuth, Ina and Roidl, Moritz and Pauly, Markus},
  journal={Human Behavior and Emerging Technologies},
  volume={2024},
  number={1},
  pages={7115633},
  year={2024},
  publisher={Wiley Online Library}
}

@article{rozado2023political,
  title={The political biases of ChatGPT},
  author={Rozado, David},
  journal={Social Sciences},
  volume={12},
  number={3},
  pages={148},
  year={2023},
  publisher={MDPI}
}

@article{motoki2024more,
  title={More human than human: measuring ChatGPT political bias},
  author={Motoki, Fabio and Pinho Neto, Valdemar and Rodrigues, Victor},
  journal={Public Choice},
  volume={198},
  number={1},
  pages={3--23},
  year={2024},
  publisher={Springer}
}

@article{waller2021quantifying,
  title={Quantifying social organization and political polarization in online platforms},
  author={Waller, Isaac and Anderson, Ashton},
  journal={Nature},
  volume={600},
  number={7888},
  pages={264--268},
  year={2021},
  publisher={Nature Publishing Group UK London}
}

@article{mosleh2024differences,
  title={Differences in misinformation sharing can lead to politically asymmetric sanctions},
  author={Mosleh, Mohsen and Yang, Qi and Zaman, Tauhid and Pennycook, Gordon and Rand, David G},
  journal={Nature},
  volume={634},
  number={8034},
  pages={609--616},
  year={2024},
  publisher={Nature Publishing Group UK London}
}

@article{wojcieszak2022most,
  title={Most users do not follow political elites on Twitter; those who do show overwhelming preferences for ideological congruity},
  author={Wojcieszak, Magdalena and Casas, Andreu and Yu, Xudong and Nagler, Jonathan and Tucker, Joshua A},
  journal={Science advances},
  volume={8},
  number={39},
  pages={eabn9418},
  year={2022},
  publisher={American Association for the Advancement of Science}
}

@article{bail2018exposure,
  title={Exposure to opposing views on social media can increase political polarization},
  author={Bail, Christopher A and Argyle, Lisa P and Brown, Taylor W and Bumpus, John P and Chen, Haohan and Hunzaker, MB Fallin and Lee, Jaemin and Mann, Marcus and Merhout, Friedolin and Volfovsky, Alexander},
  journal={Proceedings of the National Academy of Sciences},
  volume={115},
  number={37},
  pages={9216--9221},
  year={2018},
  publisher={National Academy of Sciences}
}

@article{flamino2023political,
  title={Political polarization of news media and influencers on Twitter in the 2016 and 2020 US presidential elections},
  author={Flamino, James and Galeazzi, Alessandro and Feldman, Stuart and Macy, Michael W and Cross, Brendan and Zhou, Zhenkun and Serafino, Matteo and Bovet, Alexandre and Makse, Hern{\'a}n A and Szymanski, Boleslaw K},
  journal={Nature Human Behaviour},
  volume={7},
  number={6},
  pages={904--916},
  year={2023},
  publisher={Nature Publishing Group UK London}
}

@article{mamakos2023social,
  title={The social media discourse of engaged partisans is toxic even when politics are irrelevant},
  author={Mamakos, Michalis and Finkel, Eli J},
  journal={PNAS nexus},
  volume={2},
  number={10},
  pages={pgad325},
  year={2023},
  publisher={Oxford University Press US}
}

@article{nyhan2023like,
  title={Like-minded sources on Facebook are prevalent but not polarizing},
  author={Nyhan, Brendan and Settle, Jaime and Thorson, Emily and Wojcieszak, Magdalena and Barber{\'a}, Pablo and Chen, Annie Y and Allcott, Hunt and Brown, Taylor and Crespo-Tenorio, Adriana and Dimmery, Drew and others},
  journal={Nature},
  volume={620},
  number={7972},
  pages={137--144},
  year={2023},
  publisher={Nature Publishing Group UK London}
}

@article{grinberg2019fake,
  title={Fake news on Twitter during the 2016 US presidential election},
  author={Grinberg, Nir and Joseph, Kenneth and Friedland, Lisa and Swire-Thompson, Briony and Lazer, David},
  journal={Science},
  volume={363},
  number={6425},
  pages={374--378},
  year={2019},
  publisher={American Association for the Advancement of Science}
}

@article{siev2024ambivalent,
  title={Ambivalent attitudes promote support for extreme political actions},
  author={Siev, Joseph J and Petty, Richard E},
  journal={Science Advances},
  volume={10},
  number={24},
  pages={eadn2965},
  year={2024},
  publisher={American Association for the Advancement of Science}
}

@article{gentzkow2010drives,
  title={What drives media slant? Evidence from US daily newspapers},
  author={Gentzkow, Matthew and Shapiro, Jesse M},
  journal={Econometrica},
  volume={78},
  number={1},
  pages={35--71},
  year={2010},
  publisher={Wiley Online Library}
}

@article{greenstein2012wikipedia,
  title={Is wikipedia biased?},
  author={Greenstein, Shane and Zhu, Feng},
  journal={American Economic Review},
  volume={102},
  number={3},
  pages={343--348},
  year={2012},
  publisher={American Economic Association}
}

@article{greenstein2016open,
  title={Open content, Linus’ law, and neutral point of view},
  author={Greenstein, Shane and Zhu, Feng},
  journal={Information Systems Research},
  volume={27},
  number={3},
  pages={618--635},
  year={2016},
  publisher={INFORMS}
}

@article{solaiman2019release,
  title={Release strategies and the social impacts of language models},
  author={Solaiman, Irene and Brundage, Miles and Clark, Jack and Askell, Amanda and Herbert-Voss, Ariel and Wu, Jeff and Radford, Alec and Krueger, Gretchen and Kim, Jong Wook and Kreps, Sarah and others},
  journal={arXiv preprint arXiv:1908.09203},
  year={2019}
}

@article{simchon2023computational,
  title={A computational text analysis investigation of the relation between personal and linguistic agency},
  author={Simchon, Almog and Hadar, Britt and Gilead, Michael},
  journal={Communications Psychology},
  volume={1},
  number={1},
  pages={23},
  year={2023},
  publisher={Nature Publishing Group UK London}
}

@article{munoz2024contrasting,
  title={Contrasting linguistic patterns in human and LLM-generated news text},
  author={Mu{\~n}oz-Ortiz, Alberto and G{\'o}mez-Rodr{\'\i}guez, Carlos and Vilares, David},
  journal={Artificial Intelligence Review},
  volume={57},
  number={10},
  pages={265},
  year={2024},
  publisher={Springer}
}

@article{radford2019rewon,
  title={Rewon child, david luan, dario amodei, and ilya sutskever. 2019},
  author={Radford, Alec and Wu, Jeffrey},
  journal={Language models are unsupervised multitask learners. OpenAI blog},
  volume={1},
  number={8},
  pages={9},
  year={2019}
}

@article{naddaf2025ai,
  title={Ai chatbots are sycophants—and it’s harming science},
  author={Naddaf, Miryam},
  journal={Nature},
  volume={647},
  pages={13},
  year={2025}
}

@article{abdurahman2024perils,
  title={Perils and opportunities in using large language models in psychological research},
  author={Abdurahman, Suhaib and Atari, Mohammad and Karimi-Malekabadi, Farzan and Xue, Mona J and Trager, Jackson and Park, Peter S and Golazizian, Preni and Omrani, Ali and Dehghani, Morteza and Harding, Matthew and others},
  journal={PNAS nexus},
  volume={3},
  number={7},
  pages={pgae245},
  year={2024},
  publisher={Proceedings of the National Academy of Sciences}
}

@article{cheng2026sycophantic,
  title={Sycophantic AI decreases prosocial intentions and promotes dependence},
  author={Cheng, Myra and Lee, Cinoo and Khadpe, Pranav and Yu, Sunny and Han, Dyllan and Jurafsky, Dan},
  journal={Science},
  volume={391},
  number={6792},
  pages={eaec8352},
  year={2026},
  publisher={American Association for the Advancement of Science}
}

@article{boxell2024cross,
  title={Cross-country trends in affective polarization},
  author={Boxell, Levi and Gentzkow, Matthew and Shapiro, Jesse M},
  journal={Review of Economics and Statistics},
  volume={106},
  number={2},
  pages={557--565},
  year={2024},
  publisher={MIT Press One Rogers Street, Cambridge, MA 02142-1209, USA journals-info~…}
}

@article{iyengar2012affect,
  title={Affect, not ideology: A social identity perspective on polarization},
  author={Iyengar, Shanto and Sood, Gaurav and Lelkes, Yphtach},
  journal={Public opinion quarterly},
  volume={76},
  number={3},
  pages={405--431},
  year={2012},
  publisher={Oxford University Press US}
}

@article{voelkel2024megastudy,
  title={Megastudy testing 25 treatments to reduce antidemocratic attitudes and partisan animosity},
  author={Voelkel, Jan G and Stagnaro, Michael N and Chu, James Y and Pink, Sophia L and Mernyk, Joseph S and Redekopp, Chrystal and Ghezae, Isaias and Cashman, Matthew and Adjodah, Dhaval and Allen, Levi G and others},
  journal={Science},
  volume={386},
  number={6719},
  pages={eadh4764},
  year={2024},
  publisher={American Association for the Advancement of Science}
}

@article{santoro2022promise,
  title={The promise and pitfalls of cross-partisan conversations for reducing affective polarization: Evidence from randomized experiments},
  author={Santoro, Erik and Broockman, David E},
  journal={Science advances},
  volume={8},
  number={25},
  pages={eabn5515},
  year={2022},
  publisher={American Association for the Advancement of Science}
}

@article{zimmerman2022political,
  title={Political coherence and certainty as drivers of interpersonal liking over and above similarity},
  author={Zimmerman, Federico and Garbulsky, Gerry and Ariely, Dan and Sigman, Mariano and Navajas, Joaquin},
  journal={Science Advances},
  volume={8},
  number={6},
  pages={eabk1909},
  year={2022},
  publisher={American Association for the Advancement of Science}
}

@article{chen2018effect,
  title={The effect of partisanship and political advertising on close family ties},
  author={Chen, M Keith and Rohla, Ryne},
  journal={Science},
  volume={360},
  number={6392},
  pages={1020--1024},
  year={2018},
  publisher={American Association for the Advancement of Science}
}

@article{hohm2024moral,
  title={Do moral values change with the seasons?},
  author={Hohm, Ian and O’Shea, Brian A and Schaller, Mark},
  journal={Proceedings of the National Academy of Sciences},
  volume={121},
  number={33},
  pages={e2313428121},
  year={2024},
  publisher={National Academy of Sciences}
}

@article{goldberg2024regulating,
  title={Regulating privacy online: An economic evaluation of the GDPR},
  author={Goldberg, Samuel G and Johnson, Garrett A and Shriver, Scott K},
  journal={American Economic Journal: Economic Policy},
  volume={16},
  number={1},
  pages={325--358},
  year={2024},
  publisher={American Economic Association 2014 Broadway, Suite 305, Nashville, TN 37203-2425}
}

@article{eichenbaum2024expectations,
  title={Expectations, infections, and economic activity},
  author={Eichenbaum, Martin and Godinho de Matos, Miguel and Lima, Francisco and Rebelo, Sergio and Trabandt, Mathias},
  journal={Journal of Political Economy},
  volume={132},
  number={8},
  pages={2571--2611},
  year={2024},
  publisher={The University of Chicago Press Chicago, IL}
}

@article{burtch2024consequences,
  title={The consequences of generative AI for online knowledge communities},
  author={Burtch, Gordon and Lee, Dokyun and Chen, Zhichen},
  journal={Scientific Reports},
  volume={14},
  number={1},
  pages={10413},
  year={2024},
  publisher={Nature Publishing Group UK London}
}

@article{shan2026examining,
  title={Examining the impact of generative ai on users’ voluntary knowledge contribution: Evidence from a natural experiment on stack overflow},
  author={Shan, Guohou and Qiu, Liangfei},
  journal={Information Systems Research},
  volume={37},
  number={2},
  pages={1021--1041},
  year={2026},
  publisher={INFORMS}
}

@article{bond201261,
  title={A 61-million-person experiment in social influence and political mobilization},
  author={Bond, Robert M and Fariss, Christopher J and Jones, Jason J and Kramer, Adam DI and Marlow, Cameron and Settle, Jaime E and Fowler, James H},
  journal={Nature},
  volume={489},
  number={7415},
  pages={295--298},
  year={2012},
  publisher={Nature Publishing Group UK London}
}

@article{rathje2021out,
  title={Out-group animosity drives engagement on social media},
  author={Rathje, Steve and Van Bavel, Jay J and Van Der Linden, Sander},
  journal={Proceedings of the national academy of sciences},
  volume={118},
  number={26},
  pages={e2024292118},
  year={2021},
  publisher={National Academy of Sciences}
}

@article{gu2026psychological,
  title={Psychological Reactance to the Algorithmic Management of Online Expressions},
  author={Gu, Grace and Yin, Zhitao and Rai, Arun},
  journal={Information Systems Research},
  year={2026},
  publisher={INFORMS}
}

@misc{rai2019editor,
  title={Editor’s comments: Next-generation digital platforms: Toward human--AI hybrids},
  author={Rai, Arun and Constantinides, Panos and Sarker, Saonee},
  journal={MIS quarterly},
  volume={43},
  number={1},
  pages={iii--ix},
  year={2019},
  publisher={Management Information Systems Research Center, University of Minnesota}
}

@article{he2025platform,
  title={Platform governance with algorithm-based content moderation: An empirical study on Reddit},
  author={He, Qinglai and Hong, Yili and Raghu, TS},
  journal={Information Systems Research},
  volume={36},
  number={2},
  pages={1078--1095},
  year={2025},
  publisher={INFORMS}
}

@article{gao2026does,
  title={Does social bot help socialize? Evidence from a microblogging platform},
  author={Gao, Yang and Zhang, Maggie Mengqing and Lysyakov, Mikhail},
  journal={Information Systems Research},
  volume={37},
  number={1},
  pages={416--433},
  year={2026},
  publisher={INFORMS}
}

@article{xue2026can,
  title={Can ChatGPT kill user-generated Q\&A platforms?},
  author={Xue, Junzhi and Wang, Lizheng and Zheng, Jinyang and Li, Yongjun and Tan, Yong},
  journal={Information Systems Research},
  year={2026},
  publisher={INFORMS}
}

@article{bai2022training,
  title={Training a helpful and harmless assistant with reinforcement learning from human feedback},
  author={Bai, Yuntao and Jones, Andy and Ndousse, Kamal and Askell, Amanda and Chen, Anna and DasSarma, Nova and Drain, Dawn and Fort, Stanislav and Ganguli, Deep and Henighan, Tom and others},
  journal={arXiv preprint arXiv:2204.05862},
  year={2022}
}

@article{ouyang2022training,
  title={Training language models to follow instructions with human feedback},
  author={Ouyang, Long and Wu, Jeffrey and Jiang, Xu and Almeida, Diogo and Wainwright, Carroll and Mishkin, Pamela and Zhang, Chong and Agarwal, Sandhini and Slama, Katarina and Ray, Alex and others},
  journal={Advances in neural information processing systems},
  volume={35},
  pages={27730--27744},
  year={2022}
}

@article{guess2023reshares,
  title={Reshares on social media amplify political news but do not detectably affect beliefs or opinions},
  author={Guess, Andrew M and Malhotra, Neil and Pan, Jennifer and Barber{\'a}, Pablo and Allcott, Hunt and Brown, Taylor and Crespo-Tenorio, Adriana and Dimmery, Drew and Freelon, Deen and Gentzkow, Matthew and others},
  journal={Science},
  volume={381},
  number={6656},
  pages={404--408},
  year={2023},
  publisher={American Association for the Advancement of Science}
}

@article{cinelli2021echo,
  title={The echo chamber effect on social media},
  author={Cinelli, Matteo and De Francisci Morales, Gianmarco and Galeazzi, Alessandro and Quattrociocchi, Walter and Starnini, Michele},
  journal={Proceedings of the national academy of sciences},
  volume={118},
  number={9},
  pages={e2023301118},
  year={2021},
  publisher={National Academy of Sciences}
}

@article{bakshy2015exposure,
  title={Exposure to ideologically diverse news and opinion on Facebook},
  author={Bakshy, Eytan and Messing, Solomon and Adamic, Lada A},
  journal={Science},
  volume={348},
  number={6239},
  pages={1130--1132},
  year={2015},
  publisher={American Association for the Advancement of Science}
}

@inproceedings{biswas2025political,
  title={Political elites in the attention economy: Visibility over civility and credibility?},
  author={Biswas, Ahana and Lin, Yu-Ru and Tai, Yuehong Cassandra and Desmarais, Bruce A},
  booktitle={Proceedings of the International AAAI Conference on Web and Social Media},
  volume={19},
  pages={241--258},
  year={2025}
}

@article{arhin2021ground,
  title={Ground-Truth, Whose Truth?--Examining the Challenges with Annotating Toxic Text Datasets},
  author={Arhin, Kofi and Baldini, Ioana and Wei, Dennis and Ramamurthy, Karthikeyan Natesan and Singh, Moninder},
  journal={arXiv preprint arXiv:2112.03529},
  year={2021}
}

@article{hebbelstrup2024offline,
  title={The offline roots of online hostility: Adult and childhood administrative records correlate with individual-level hostility on Twitter},
  author={Hebbelstrup Rye Rasmussen, Stig and Bor, Alexander and Petersen, Michael Bang},
  journal={Proceedings of the National Academy of Sciences},
  volume={121},
  number={44},
  pages={e2412277121},
  year={2024},
  publisher={National Academy of Sciences}
}

@article{brady2017emotion,
  title={Emotion shapes the diffusion of moralized content in social networks},
  author={Brady, William J and Wills, Julian A and Jost, John T and Tucker, Joshua A and Van Bavel, Jay J},
  journal={Proceedings of the National Academy of Sciences},
  volume={114},
  number={28},
  pages={7313--7318},
  year={2017},
  publisher={National Academy of Sciences}
}

@article{morag2025narratives,
  title={Narratives and valuations},
  author={Morag, Dor and Loewenstein, George},
  journal={Management Science},
  volume={71},
  number={6},
  pages={5376--5395},
  year={2025},
  publisher={INFORMS}
}

@article{del2024large,
  title={Large language models reduce public knowledge sharing on online Q\&A platforms},
  author={del Rio-Chanona, R Maria and Laurentsyeva, Nadzeya and Wachs, Johannes},
  journal={PNAS nexus},
  volume={3},
  number={9},
  pages={pgae400},
  year={2024},
  publisher={Oxford University Press US}
}

@article{weidinger2021ethical,
  title={Ethical and social risks of harm from language models},
  author={Weidinger, Laura and Mellor, John and Rauh, Maribeth and Griffin, Conor and Uesato, Jonathan and Huang, Po-Sen and Cheng, Myra and Glaese, Mia and Balle, Borja and Kasirzadeh, Atoosa and others},
  journal={arXiv preprint arXiv:2112.04359},
  year={2021}
}

@article{frimer2023incivility,
  title={Incivility is rising among American politicians on Twitter},
  author={Frimer, Jeremy A and Aujla, Harinder and Feinberg, Matthew and Skitka, Linda J and Aquino, Karl and Eichstaedt, Johannes C and Willer, Robb},
  journal={Social Psychological and Personality Science},
  volume={14},
  number={2},
  pages={259--269},
  year={2023},
  publisher={Sage Publications Sage CA: Los Angeles, CA}
}

@article{festinger1962cognitive,
  title={Cognitive dissonance},
  author={Festinger, Leon},
  journal={Scientific American},
  volume={207},
  number={4},
  pages={93--106},
  year={1962},
  publisher={JSTOR}
}

@article{turner1979social,
  title={Social comparison and group interest in ingroup favouritism},
  author={Turner, John C and Brown, Rupert J and Tajfel, Henri},
  journal={European journal of social psychology},
  volume={9},
  number={2},
  pages={187--204},
  year={1979},
  publisher={Wiley Online Library}
}

@book{mutz2006hearing,
  title={Hearing the other side: Deliberative versus participatory democracy},
  author={Mutz, Diana C},
  year={2006},
  publisher={Cambridge University Press}
}

@article{lord1979biased,
  title={Biased assimilation and attitude polarization: The effects of prior theories on subsequently considered evidence.},
  author={Lord, Charles G and Ross, Lee and Lepper, Mark R},
  journal={Journal of personality and social psychology},
  volume={37},
  number={11},
  pages={2098},
  year={1979},
  publisher={American Psychological Association}
}

@article{fiorina2008political,
  title={Political polarization in the American public},
  author={Fiorina, Morris P and Abrams, Samuel J},
  journal={Annu. Rev. Polit. Sci.},
  volume={11},
  number={1},
  pages={563--588},
  year={2008},
  publisher={Annual Reviews}
}

@article{sharma2023towards,
  title={Towards understanding sycophancy in language models},
  author={Sharma, Mrinank and Tong, Meg and Korbak, Tomasz and Duvenaud, David and Askell, Amanda and Bowman, Samuel R and Cheng, Newton and Durmus, Esin and Hatfield-Dodds, Zac and Johnston, Scott R and others},
  journal={arXiv preprint arXiv:2310.13548},
  year={2023}
}

@article{shapira2026rlhf,
  title={How RLHF Amplifies Sycophancy},
  author={Shapira, Itai and Benade, Gerdus and Procaccia, Ariel D},
  journal={arXiv preprint arXiv:2602.01002},
  year={2026}
}

@book{brown1987politeness,
  title={Politeness: Some universals in language usage},
  author={Brown, Penelope and Levinson, Stephen C},
  volume={4},
  year={1987},
  publisher={Cambridge university press}
}

@article{mutz2007effects,
  title={Effects of “in-your-face” television discourse on perceptions of a legitimate opposition},
  author={Mutz, Diana C},
  journal={American Political Science Review},
  volume={101},
  number={4},
  pages={621--635},
  year={2007},
  publisher={Cambridge University Press}
}
